\documentclass[]{aa}  

\usepackage{txfonts}
\usepackage{graphicx}
\usepackage{gensymb}    % For degree symbol
\usepackage[dvipsnames]{xcolor}
\definecolor{bblue}{RGB}{70, 130, 180}
\usepackage[colorlinks=true,linkcolor=bblue,citecolor=bblue,urlcolor=bblue]{hyperref}
\usepackage{xspace}
\usepackage{siunitx}    % for S option in table to align decimals
\usepackage{multirow}
\usepackage{hhline}
\usepackage{upgreek}  % For upright Greek letters
\usepackage{pifont} % for Xs and tick marks
\newcommand{\arii}{[\ion{Ar}{ii}]\xspace}
\newcommand{\ariii}{[\ion{Ar}{iii}]\xspace}
\newcommand{\neii}{[\ion{Ne}{ii}]\xspace}
\newcommand{\neiii}{[\ion{Ne}{iii}]\xspace}
\newcommand{\siii}{[\ion{S}{iii}]\xspace}

\newcommand{\clii}{[\ion{Cl}{ii}]\xspace}
\newcommand{\htwo}{H$_2$\xspace}
\newcommand{\halpha}{H$\alpha$\xspace}
\newcommand{\hii}{\ion{H}{ii}\xspace}

\newcommand{\pahfit}{{\small PAHFIT}\xspace}
\newcommand{\ofive}{Obs$_{\rm r1.5,h0.5}$\xspace}
\newcommand{\osix}{Obs$_{\rm r1.5,h1.0}$\xspace}
\newcommand{\oseven}{Obs$_{\rm r4.7,h0.5}$\xspace}
\newcommand{\oeight}{Obs$_{\rm r4.7,h1.0}$\xspace}
\newcommand{\olowh}{Obs$_{\rm h0.5}$\xspace}
\newcommand{\ohighh}{Obs$_{\rm h1.0}$\xspace}
\newcommand{\upmicron}{$\upmu$m\xspace}

\usepackage{svg}
\newcommand{\orcid}[1]{\href{https://orcid.org/#1}{\includegraphics[width=8pt]{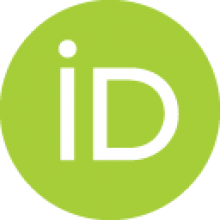}}}

\begin{document}

   \title{Survival of very small carbonaceous dust grains in the inner-CGM of NGC~891 from JWST/MIRI MRS}

   \author{
   {J\'er\'emy~Chastenet}\inst{\ref{UGent}}\orcid{0000-0002-5235-5589}\thanks{jeremy.chastenet@ugent.be} \and
    {Ilse~De~Looze}\inst{\ref{UGent}}\orcid{0000-0001-9419-6355}\thanks{PI: ilse.delooze@ugent.be} \and
    {Karl~D.~Gordon}\inst{\ref{STScI},\ref{UGent}}\orcid{0000-0001-5340-6774} \and
    {Suzanne~C.~Madden}\inst{\ref{CEA}}\orcid{0000-0003-3229-2899} \and
    {Monica~Rela\~no}\inst{\ref{DeptGranada}, \ref{UGranada}}\orcid{0000-0003-1682-1148} \and
    {Karin~M.~Sandstrom}\inst{\ref{UCSD}}\orcid{0000-0002-4378-8534} \and
    {Dries~Van~De~Putte}\inst{\ref{UOntario}}\orcid{} \and
    {Maarten~Baes}\inst{\ref{UGent}}\orcid{0000-0002-3930-2757} \and
    {Simone~Bianchi}\inst{\ref{INAF}}\orcid{0000-0002-9384-846X} \and
    {Alberto~D.~Bolatto}\inst{\ref{UMaryland}}\orcid{0000-0002-5480-5686} \and
    {Viviana~Casasola}\inst{\ref{INAFRadio}}\orcid{0000-0002-3879-6038} \and
    {Daniel~A.~Dale}\inst{\ref{UWyoming}}\orcid{0000-0002-5782-9093} \and
    {Sara~Duval}\inst{\ref{UToledo}}\orcid{0000-0003-0014-0508} \and
    {Jacopo~Fritz}\inst{\ref{UMexico}}\orcid{0000-0002-7042-1965} \and
    {Fr\'ed\'eric~Galliano}\inst{\ref{CEA}}\orcid{0000-0002-4414-3367} \and
    {Simon~C.~O.~Glover}\inst{\ref{UHeidelbergAstro}}\orcid{0000-0001-6708-1317} \and
    {Stavroula Katsioli}\inst{\ref{Athens1}, \ref{Athens2}}\orcid{0000-0001-7095-8933} \and
    {Vasileios Katsis}\inst{\ref{Athens1}, \ref{UAixMarseille}} \and
    {Florian~Kirchschlager}\inst{\ref{UGent}}\orcid{0000-0002-3036-0184} \and
    {Ralf S.\ Klessen}\inst{\ref{UHeidelbergAstro},\ref{IWR}}\orcid{0000-0002-0560-3172} \and
    {Rebecca~C.~Levy}\inst{\ref{STScI}}\orcid{0000-0003-2508-2586} \and
    {Sharon~E.~Meidt}\inst{\ref{UGent}}\orcid{0000-0002-6118-4048} \and
    {Aleksandr~V.~Mosenkov}\inst{\ref{UBrigham}}\orcid{0000-0001-6079-7355} \and
    {Kentaro~Nagamine}\inst{\ref{UOsaka}, \ref{UOsakaTJR},  \ref{KavliTokyo}, \ref{UNevada}, \ref{NCfA}}\orcid{0000-0001-7457-8487} \and
    {Lara~Pantoni}\inst{\ref{UGent}}\orcid{0000-0003-2666-5759} \and
    {Helena~M.~Richie}\inst{\ref{UPittsburg}}\orcid{0000-0001-6325-9317} \and
    {Stefanie Walch}\inst{\ref{UCologne}}\orcid{0000-0001-6941-7638} \and
    {Thomas~G.~Williams}\inst{\ref{Oxford}}\orcid{0000-0002-0786-7307}\and
    {Emanuele M. Xilouris}\inst{\ref{Athens1}}
}

\institute{
    % Sterrenkundig Observatorium, Universiteit Gent, Krijgslaan 281-S9, 9000 Gent, Belgium
    Department of Physics and Astronomy, Universiteit Gent, Proeftuinstraat 86 N3, B-9000 Ghent, Belgium
    \label{UGent} \and
    Space Telescope Science Institute, 3700 San Martin Drive, Baltimore, MD 21218, USA
    \label{STScI} \and
    Université Paris-Saclay, Université Paris Cité, CEA, CNRS, AIM, 91191, Gif-sur-Yvette, France
    \label{CEA} \and
    Dept. Fisica Teorica y del Cosmos, E-18071 Granada, Spain
    \label{DeptGranada} \and
    Instituto Universitario Carlos I de Fisica Teorica y Computacional, Universidad de Granada, E-18071 Granada, Spain
    \label{UGranada} \and
    Department of Astronomy \& Astrophysics,  University of California, San Diego, 9500 Gilman Drive, La Jolla, CA 92093, USA
    \label{UCSD} \and
    Department of Physics \& Astronomy, The University of Western Ontario, London ON N6A 3K7, Canada
    \label{UOntario} \and
    INAF – Osservatorio Astrofisico di Arcetri, Largo E. Fermi 5, 50125 Firenze, Italy
    \label{INAF} \and
    Department of Astronomy and Joint Space-Science Institute, University of Maryland, College Park, MD 20742, USA
    \label{UMaryland} \and
    INAF - Istituto di Radioastronomia, Via Piero Gobetti 101, 40129 Bologna, Italy
    \label{INAFRadio} \and
    Department of Physics and Astronomy, University of Wyoming, Laramie, WY 82071, USA
    \label{UWyoming} \and
    Ritter Astrophysical Research Center, University of Toledo, Toledo, OH 43606, USA
    \label{UToledo} \and
    Instituto de Radioastronomía y Astrofísica, Universidad Nacional Autónoma de México, Morelia, Michoacán 58089, Mexico
    \label{UMexico} \and
    Universit\"{a}t Heidelberg, Zentrum f\"{u}r Astronomie, Institut f\"{u}r Theoretische Astrophysik, Albert-Ueberle-Stra{\ss}e 2, D-69120 Heidelberg, Germany
    \label{UHeidelbergAstro} \and
    National Observatory of Athens, IAASARS, Ioannou Metaxa and Vasileos Pavlou GR-15236 Athens, Greece 
    \label{Athens1} \and
    Faculty of Physics, Department of Astrophysics, Astronomy \& Mechanics, University of Athens, Panepistimiopolis, GR-15784 Zografos, Athens, Greece
    \label{Athens2} \and
    Physique des Interactions Ioniques et Mol\'eculaires, CNRS, Aix Marseille Université, Marseille, France
    \label{UAixMarseille} \and
    Universit\"{a}t Heidelberg, Interdisziplin\"{a}res Zentrum f\"{u}r Wissenschaftliches Rechnen, Im Neuenheimer Feld 225, 69120 Heidelberg, Germany
    \label{IWR} \and
    Department of Physics and Astronomy, N283 ESC, Brigham Young University, Provo, UT 84602, USA
    \label{UBrigham} \and
    Theoretical Astrophysics, Department of Earth and Space Science, Osaka University, 1-1 Machikaneyama, Toyonaka, Osaka 560-0043, Japan
    \label{UOsaka} \and
    Theoretical Joint Research (TJR), Graduate School of Science, Osaka University, 1-1 Machikaneyama, Toyonaka, Osaka 560-0043, Japan
    \label{UOsakaTJR} \and
    Kavli-IPMU (WPI), University of Tokyo, 5-1-5 Kashiwanoha, Kashiwa, Chiba, 277-8583, Japan
    \label{KavliTokyo} \and
    Department of Physics \& Astronomy, University of Nevada, Las Vegas, 4505 S. Maryland Pkwy, Las Vegas, NV 89154-4002, USA
    \label{UNevada} \and
    Nevada Center for Astrophysics, University of Nevada, Las Vegas, 4505 S. Maryland Pkwy, Las Vegas, NV 89154-4002, USA
    \label{NCfA} \and
    Physics and Astronomy Department, University of Pittsburgh, 3941 O'Hara St, Pittsburgh, PA 15260
    \label{UPittsburg} \and
    Universit\"{a}t zu K\"{o}ln, I. Physikalisches Institut, Z\"{u}lpicher Str. 77, 50937 K\"{o}ln, Germany
    \label{UCologne} \and
    Sub-department of Astrophysics, Department of Physics, University of Oxford, Keble Road, Oxford OX1 3RH, UK 
    \label{Oxford}
}
   \date{}

    \abstract{
    We present new spectroscopic observations of the inner circumgalactic medium (CGM) of NGC~891 taken with the Mid-Infrared Imager/Medium Resolution Spectroscopy instrument onboard JWST, in four positions: two near the bulge and two at galactocentric radii ($r$) of $\sim 1.5, 4.7$~kpc. Each pair of pointings has one position along the minor axis ($h$) at $\sim 0.5~$kpc and one at $\sim 1$~kpc away from the mid-plane. We analyse both 1D spectra and 3D cubes using the dust emission model PAHFIT to extract properties of typical mid-IR features. These spectra reveal that the earlier reported mid-IR emission out to 4 kpc is dominated by the emission of polycyclic aromatic hydrocarbons (PAHs), and not hot dust continuum, providing direct evidence of the survival of PAHs in the inner CGM of NGC\,891. Comparing PAH band ratios with other environments (Orion, M51), it is obvious that the 11.2\,\upmicron PAH feature---and not the usual 7.7\,\upmicron---dominates in NGC\,891, which seems to imply the presence of more neutral, large PAHs in the CGM. Overall, PAH-to-continuum ratios show little variations with scale-height and radius in NGC\,891, which suggests little PAH processing. However, we do see a decrease in the PAH feature strengths with the \neiii/\neii ratio, pointing at elevated dust processing with increased radiation field hardness. We also confirm a tight correlation between \htwo and PAH features, suggesting that both tracers must be co-spatial and, hence, implying that PAH emission predominantly arises from cool dense parts of cloudlets entrained in galactic outflows. Finally, we report the clear detection of a previously unidentified PAH feature at 16.72~\upmicron.
  }


   \maketitle
%

%-------------------------------------------------------------------
%-------------------------------------------------------------------

\section{Introduction}
The nature of dust in the interstellar medium (ISM) of galaxies is an intense area of study, because of the impact dust grains have on the evolution of their host galaxy. 
It is widely accepted that interstellar dust consists of carbonaceous and silicate-rich material, with grains spanning a range of sizes, from a few nanometres to micrometre-sized grains \citep{Field1975, Allamandola1989b, WD2001, Kemper2004, Min2008}.
Small grains are stochastically heated and their emission is primarily in the mid-infrared (mid-IR) regime, while emission from larger grains lies in the far-infrared and is well modelled as a (modified) blackbody.
Several dust models are available, calibrated primarily to reproduce Galactic emission, extinction, depletion observations, and laboratory measurements \citep[e.g.][]{DL2007, THEMIS, HD2023}. 
Using these models, several studies investigated the lifecycle of dust grains in external galaxies, leading to some crucial insights into their formation and destruction mechanisms \citep[e.g.][]{RemyRuyer2013, Davies2017, Aniano2020, Galliano2021, Chastenet2025}.

 In external galaxies, the total dust mass scales positively with stellar mass and global star formation \citep[e.g.][]{Cortese2012, Casasola2020, DeLooze2020}, and the same behaviour is observed on resolved scales \citep[e.g.][]{Leroy2008, Hsieh2017, Lin2019, Casasola2022, Chastenet2025}. These relations fall well within other scaling laws explaining galactic evolution, such as the fundamental relation between mass, star formation, and metallicity \citep{Mannucci2010, Maiolino2019}, Kennicutt-Schmidt and gas-stellar mass relations \citep{KennicuttEvans12KSReview, Saintonge2022}, and the dust-to-metal ratio with metallicity \citep{Peroux2020}. 

The grains at the smallest end of the carbonaceous grain size distribution are responsible for prominent mid-infrared broad features.
The nomenclature of the species responsible for these features is debated as growing evidence suggests that some mid-IR features originate from other bonds than aromatic bonds, leading to modelling and naming adjustments \citep[][]{DL2007, THEMIS, HD2023, THEMIS2, Pendleton2025}.
In this paper, most of the detected emission features are attributed to aromatic bonds (except for one olefinic bond), and for readability purposes, we use the term `PAHs' (Polycyclic Aromatic Hydrocarbons) to describe the species responsible for the mid-IR features.
These features are characteristic of C=C, C--H vibrational modes in aromatic and aliphatic material, at 3.3, 6.2, 7.7, 8.6, 11.2,\footnote{This refers to the same feature sometimes noted 11.3~\upmicron in the literature.} 12.7, and 17~\upmicron. 
The 3.3~\upmicron feature is associated with small neutral particles \citep{Maragkoudakis2020, Kerkeni2022}, while the 7.7 and 11.2~\upmicron features are associated with larger ionized and neutral particles, respectively \citep{Bauschlicher2008, Bauschlicher2009, Draine2021}.
These bright emission features are used to trace the properties and evolution of dust, and the local conditions of the ISM itself.
Theoretical works suggest that the destruction of PAHs can occur in hot, high energy gas, through photo-ionisation or grain-grain collisions  \citep{Micelotta2010gas, Micelotta2010shocks, Bocchio2012, Montillaud2013}, as well as in intense/hard radiation fields \citep{Madden2006, Gordon2008} in \hii~regions \citep{Chastenet2019, Egorov2023, Sutter2024}. The harsh environment in these media leads to the destruction of these particles, rendered unstable by the gain/loss of electrons or hydrogen atoms.
 
Positive scaling relations between PAH fraction and CO emission \citep{Cortzen2019, Chown2021, Leroy2023, Whitcomb2023, ShivaeiBoogaard2024, Chown2025PHANGS} suggest that these grains are well mixed with molecular gas, but also well mixed with diffuse neutral gas \citep{Hensley2022, Sutter2024}. This co-spatiality is invoked to argue for the growth of hydrocarbon nanoparticles in the ISM.
Similar positive correlations are seen with metallicity \citep{Engelbracht2005, Draine2007, Galliano2008, RemyRuyer2015, Galliano2021}, suggesting a formation route dependent on available dust and metals.
 
Most of the above information about interstellar dust has been inferred from studies of galactic disks. There have been, however, several studies focused on grains outside the main disks of nearby galaxies, in their circumgalactic medium (CGM). 
Early evidence of dusty material in the CGM was found through absorption and reddening of background sources through galaxy haloes \citep[at redshift $z\sim0.3$:][see also \citealt{Peek2015}, \citealt{McCleary2026}]{Menard2010}. Recent space-based infrared telescopes had sufficient increases in sensitivity and resolution to directly trace the infrared emission in the CGM of nearby edge-on systems.
For example, \textit{Spitzer} observations of edge-on galaxies {NGC~5529}, {NGC~5775}, and {NGC~5907} found strong evidence of dust features dominating the mid-IR emission of their haloes, out to several kiloparsecs \cite[][]{Lee2001, IrwinMadden2006, Irwin2007}. 
Recent JWST observations of M82 show very conspicuous, mid-IR-bright emission away from the mid-plane \citep{Bolatto2024, Fisher2025}, confirming the presence of the responsible species in the lower-CGM, in the form of filaments and arc-like layers of large bubbles.
Despite other observational evidence of the CGM being filled with ionised gas and X-ray emission, hydrocarbon nanoparticles appear to be surviving in this harsh environment (S. Cronin et al., in prep). 
Their mere presence raises questions about how they came to be in the halo, and is suggestive of outflows carrying them, although the launching mechanisms of these outflows remain debated. 

In this paper, we focus on the edge-on galaxy NGC~891, which has been the target of several works studying the CGM of nearby galaxies. For example,  \cite{Bocchio2016} and \citet{Yoon2021} measured the scale-height of the dusty thick disk using \textit{Herschel} observations \citep[$\sim 0.3~$kpc for the thin disk, and $\sim 1.44~$kpc for the thick disk; see also][]{Whaley2009}.
\footnote{In this study, we use the term `(lower-)CGM', although that denomination will vary depending on the perspective of the study. In our case, it is meant as a separation from the main (thin) disk, although scale-height measurements might sometimes encompass our pointing positions.}
In the mid-IR, \citet{Rand2008} found very clear spectroscopic evidence of dust features in the first kiloparsecs away from the mid-plane of NGC~891. Heavy elements were also identified in absorption by \cite{Bregman2013} within 5~kpc of the plane. 
Recent deep, high angular resolution JWST observations of NGC~891 confidently detected PAHs out to 4~kpc from the mid-plane using deep MIRI imaging \citep{GO2180}, also in the forms of arcs and bubbles. 
Zoom-in and tall-box simulations are good tools to estimate the impact of disk activity on the outflow properties in the lower CGM \citep[e.g.][]{Walch2015, Girichidis2016, Emerick2019, Schneider2020, McCallum2024, TanFielding2024}.
Star formation activity and supernovae are believed to be important actors in launching outflows. 
Empirically, \citet{Heckman2002} found a threshold value of $\rm \Sigma_{SFR}\sim 0.1~M_\odot~yr^{-1}~kpc^{-2}$ enough to launch localised outflows, where dust grains can be entrained with the ejected gas and travel out of the thick disk. While this value is reached in the starburst M82, the disk of NGC~891 shows a lower star formation rate surface density \citep[0.03~$\rm M_\odot~yr^{-1}~kpc^{-2}$,][]{Yoon2021}.
The work by \citet[][see also \citealt{Strickland2004}]{RossaDettmar2003}, however, finds a lower energy injection threshold that allows for extraplanar material ($\rm SFR\sim 0.001~M_\odot~yr^{-1}~kpc^{-2}$), and the work of \citet{Yoon2021} suggests that the SFR of NGC~891 is enough to launch material off the mid-plane.
In that same paper, the authors report a super-bubble and dust spur that they attribute to feedback processes within the disk. Additionally, it is possible that NGC~891 had a more intense SFR in the past, leading to outflows \citep[see][and references therein]{Whaley2009}. The newly observed structures combined with the past literature lead us to believe that NGC~891 indeed exhibits outflows. Now, while we very clearly detect escaping filaments, it is unclear whether these will develop in full-blown winds or fall back onto the disk.

Note that recent work from the DUVET survey suggests that the specific star formation rate (SFR/M$_\star$) is a better predictor for outflows than star formation rate \citep[][]{Reichardt2025}.
The detection of PAHs in the lower-CGM of galaxies raises questions about the survival mechanisms that allow such small grains to endure the hot, high-velocity winds and maintain integrity long enough to be detected at these distances from the disk.

In this study, we use JWST MIRI/MRS spectroscopy in four positions outside the mid-plane of NGC~891 to investigate the variation of PAH properties in the lower-CGM, coupled with \htwo and Ne emission lines. By tracking these variations, we aim to identify trends with galactocentric radius and/or distance from the mid-plane. These trends would be a way to shed light on how hydrocarbon dust evolves in the CGM, and give us clues on the formation and destruction mechanisms of nano- and micron-sized grains.
We present these new data in Section~\ref{sec:data}, followed by a qualitative description of 1D spectra between these four pointings in Section~\ref{sec:qualitativedescription}, where we also highlight similarities and differences with other measurements. In Section~\ref{sec:dustinCGM}, we apply a infrared spectrum model to 1D spectra and 3D cubes to track the intensity of mid-IR features in the CGM, combined with gas emission lines, leading to a discussion of grain survival in Section~\ref{sec:discussion}.

\section{Data acquisition and reduction}
\label{sec:data}
\subsection{MIRI~MRS spectroscopy}
\begin{figure*}
    \centering
    \includegraphics[width=0.68\textwidth, angle=-90, clip, trim={2.8cm 0 4cm 0}]{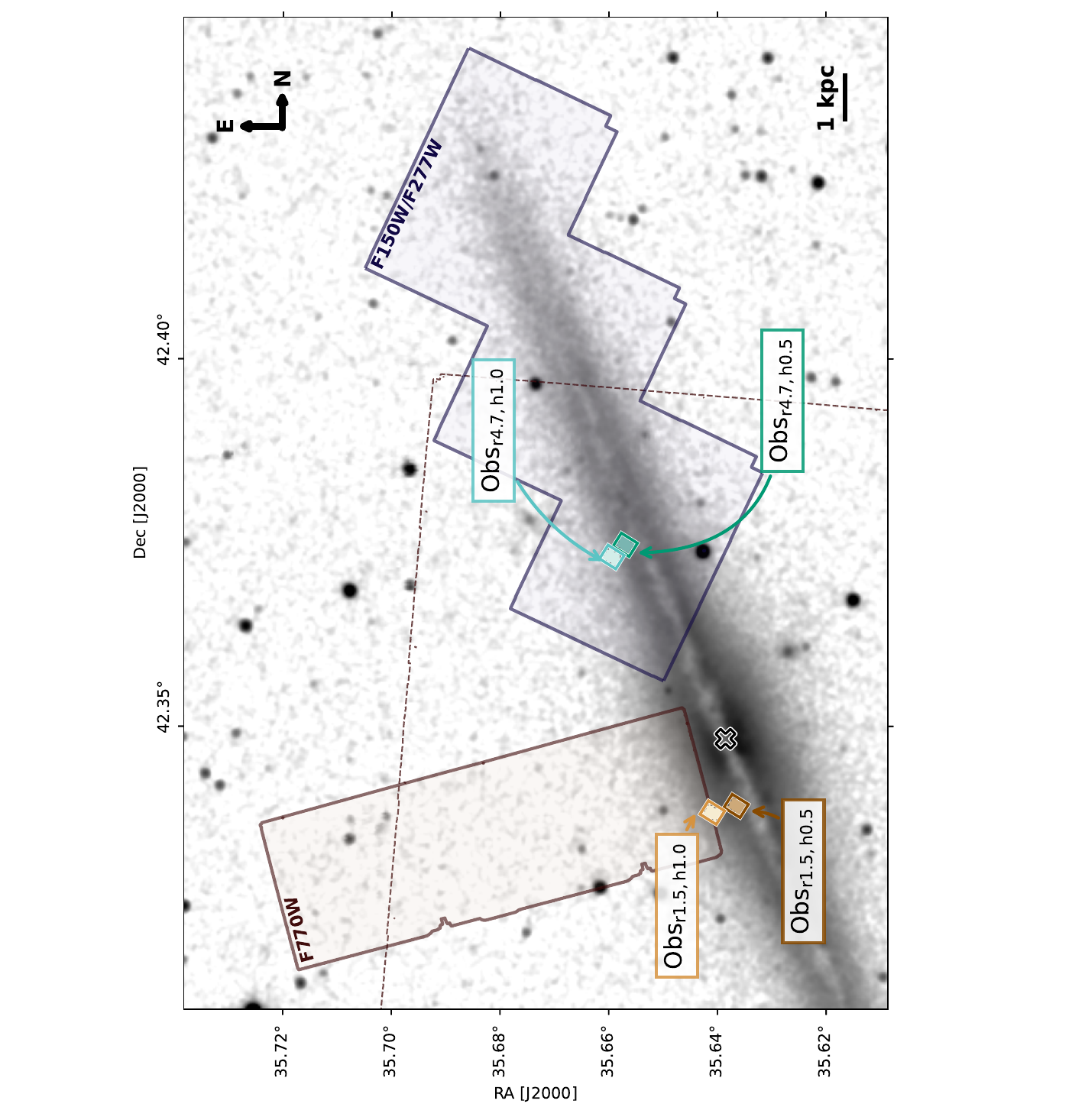}
    \caption{2MASS J-band image of NGC~891 with the four {MIRI/MRS} positions presented in this paper: 
    \ofive and \osix lie near the bulge, and \oseven and \oeight at $r\sim4.7~$kpc, where each pair has pointings at $h\sim0.5~$kpc and $h\sim1.0~$kpc.
    Additional footprints of available JWST data are shown: MIRI imaging is shown in dark red/brown, with the strip available in F770W \citep[see][]{GO2180}. The dark blue footprint shows the NIRCam coverage for F150W/F277W \citep{GO2180}.
    The dashed line shows the footprint of GO~2433 (PI: Howk) data available in MIRI F560W/F770W. The cross marks the centre of the galaxy.}
    \label{fig:mosaic}
\end{figure*}

The spectroscopic data presented in this paper were taken as part of the JWST~GO~2180 (PI: De~Looze), focused on the edge-on galaxy NGC~891. The first observations of this program are presented in \citet{GO2180}.
Here, we present four MIRI~MRS observations\footnote{Note that the labelling in the public entry of the program is no longer accurate, due to a change in the pointing coordinates: the pointings are at 0.5 and 1~kpc, and not 1 and 2~kpc away from the plane.} at two galactocentric radii, $r\sim1.5, 4.7$~kpc, and two distances from the plane, $h\sim0.5, 1$~kpc, in pairs. 
These positions are listed in Table~\ref{tab:obs_info} (we provide information of observations 5 through 8 as referred to in the public form, but use their $(r,~h)$ in the text) and shown in Fig.~\ref{fig:mosaic}, along with other JWST data available from GO~2180 and 2433.
The observations are taken in all 12 channels, with a {\sc slowr}{\small 1} readout pattern, 10 groups/integration, 2 integrations/exposure for 1 exposure, with a standard {\small 4}{\sc -point} dither. All observations were taken 18--19 September 2023.
The total observing time for each pointing is $\sim 6020~$s.

We use the pipeline version {\tt jwst 1.17.1} Build 11.2 to reduce the spectroscopic data, with CRDS context 1322. 
We turn on the {\tt jump.find\_shower} function in the {\tt detector1} step, and the {\tt straylight.clean\_showers} in the {\tt spec2} step, which are greatly beneficial to reduce spurious bumps in the final spectra. 
The {\tt residual\_fringe} step is also turned on as it improves the quality of the final products.
Every other parameter is kept as the default one. 
The background is removed in each frame pixel-by-pixel, instead of using a master background.
For each pointing, the pipeline outputs 4 different 3D cubes (one for each Channel), which spatially overlap, with the smallest intersection being the area of Channel~1, and contiguous in wavelength between 5 and 25~\upmicron.

\renewcommand{\arraystretch}{1.4}
\begin{table*}[]
    \centering
    \caption{
    Key information for the four main observations and background pointing.}
    \begin{tabular}{l|l|ll|ccccccc}
        \textbf{Observation} & \textbf{R.A., Dec} & \textbf{$h$ [kpc]}\tablefootmark{a} & \textbf{$r$ [kpc]}\tablefootmark{b} & \htwo\tablefootmark{c} & \neii & \neiii & \siii & \arii & \ariii & \clii\\
        \hline
        Obs. 5 & (02h22m32.7222s,  & $\sim 0.5$ & $\sim 1.5$\tablefootmark{d} & $S(5)$ & \ding{51} & \ding{51} & \ding{51} & \ding{51} & $\sim$ & $\sim$ \\
        & 42\degree20$'$20.33$''$) & & & & & & & \\
        Obs. 6 & (02h22m33.7740s,  & $\sim 1.0$ & $\sim 1.5$\tablefootmark{d} & $\sim$ $S(3)$ & \ding{51} & \ding{51} & \ding{51} & \ding{51} & $\sim$ & \ding{55} \\
        & 42\degree20$'$17.05$''$) & & & & & & & \\
        Obs. 7 & (02h22m37.6347s,  & $\sim 0.5$ & $\sim 4.7$ & $S(3)$ & \ding{51} & \ding{51} & \ding{51} & \ding{51} & $\sim$ & $\sim$ \\
        & 42\degree22$'$28.38$''$) & & & & & & & \\
        Obs. 8 & (02h22m38.2322s,  & $\sim 1.0$ & $\sim 4.7$ & $S(3)$ & \ding{51} & \ding{51} & \ding{51} & \ding{51} & \ding{55} & \ding{55} \\
        & 42\degree22$'$22.31$''$) & & & & & & & \\
        Bkg & (02h24m39.1970s,  & -- & -- & & & & & & & \\
        & 42\degree11$'$38.07$''$) & & & & & & & 
    \end{tabular}
    \label{tab:obs_info}
    \tablefoot{We list observations 5 through 8 and refer to their $(r,~h)$ in the text. The $\sim$ symbol indicates a tentative detection.
                \tablefoottext{a}{Distance from mid-plane.}
                \tablefoottext{b}{Galactocentric radius.}
                \tablefoottext{c}{We confidently detect \htwo~$S(1)$ up to the \htwo~$S(x)$ line indicated in the entry.}
                \tablefoottext{d}{Southward of galactic centre.}}
\end{table*}

\subsection{Extracting 1D spectra}
\subsubsection{Extraction}
\label{sec:1dextraction}
For each region, we extracted a 1D spectrum using the same stitching algorithm as the PDRs4All project.
Briefly, the stitching is done using Channel~2 {\sc long} as the reference to measure the offset between overlapping wavelengths. Matching is done using a sliding weighted average. More details are available in Section~2.2 of \citet[][see also \citealt{vdPutte2025}]{vdPutte2024}, and the PDRs4All routine {\tt extract\_templates}.\footnote{\url{https://github.com/pdrs4all/pdrs4all}}
The extraction region is given by the field-of-view of Channel~1, the smallest overlapping region between all Channels.

\begin{figure*}
    \centering
    \includegraphics[width=\textwidth]{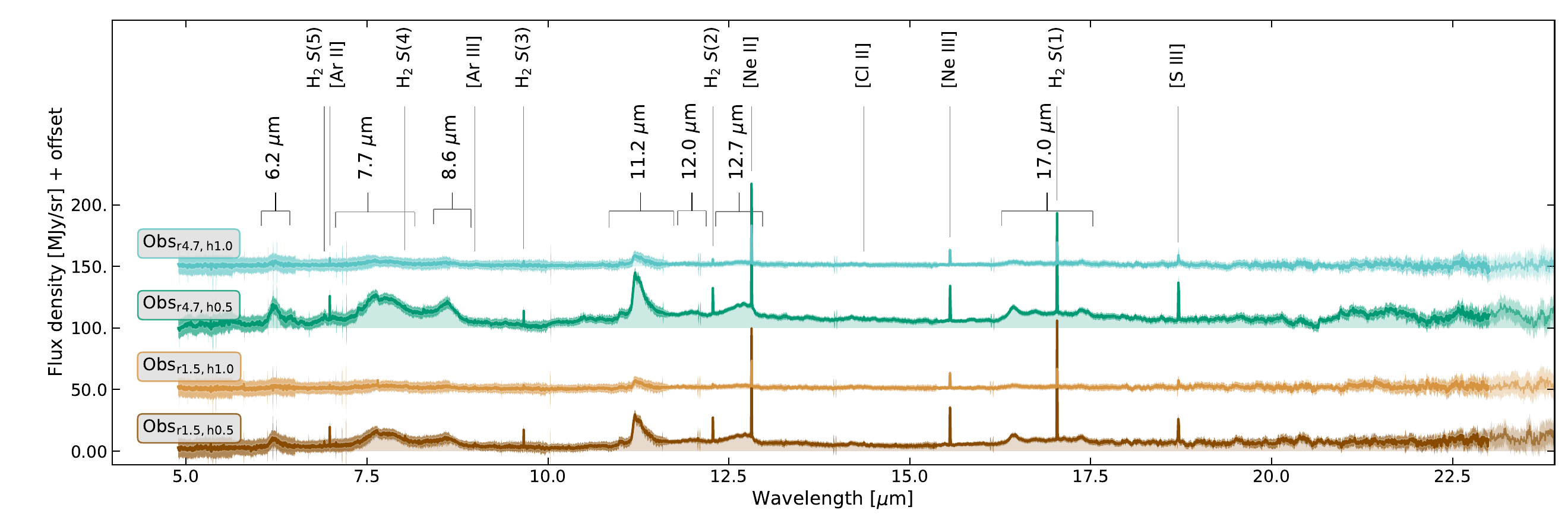}
    \caption{Average spectra for the four observations, with 50~MJy/sr~offsets for  (the bottom of the shaded area shows the 0-point in each case). We label mid-IR features and emission lines of interest. \ofive and \oseven show the cleanest features and highest intensities for the identified lines. While showing a lower S/N, \osix and \oeight still exhibit some features and the brightest emission lines. The wavelength range $\uplambda \geq 23~\upmu$m is not shown as it is not used at this stage.}
    \label{fig:all1Dspectra}
\end{figure*}

In Fig.~\ref{fig:all1Dspectra} we show all four extracted 1D spectra from our observations, with an offset for better visualisation. We show the spectra only up to $\lambda \sim 23~$\upmicron, for clarity. At longer wavelengths, the spectra are very noisy (and still show potential fringing, with a steep rise past 25~\upmicron), and this analysis is focused on the portion 6~\upmicron$ \leqslant \lambda \leqslant $ 20~\upmicron, where the carbonaceous dust features are present.

We present a detailed inspection of these spectra in Section~\ref{sec:qualitativedescription}.
Briefly, we clearly detect the 6.2, 7.7, 8.6, 11.2, 12.0, 12.7, 17~\upmicron mid-IR features attributed to hydrocarbon nanoparticles, and some fainter ones in \ofive and \oseven.
We also clearly identify \htwo~$S(1)-S(5)$ (up to $S(2)$ in the lower-S/N \osix and \oeight; $S(0)$ lies in the noisy part of the spectrum that is not considered in this paper, and we do not detect it), as well as the bright \neii and \neiii lines at 12.8135 and 15.5550~\upmicron (all are rest wavelengths) respectively; the \siii lines at 18.709~\upmicron; the \arii and \ariii lines at 6.9853 and 8.9914~\upmicron, respectively. We also detect the \clii line at 14.3678~\upmicron, in \ofive and \oseven; we note that it is close to the [\ion{Ne}{v}] line at 14.32~\upmicron, but the spectral resolution is high enough ($\sim 0.005~$\upmicron) to confidently distinguish both emission lines.
We point out the possible detection of the [\ion{Fe}{ii}] line at 5.34~\upmicron in \ofive and \oseven, although uncertain given the strong noise shortwards of 5.5~\upmicron. A dedicated paper will focus on the line detections and investigate their variations.

Using the error computed by the pipeline, we find overall S/N of 1.99, 0.52, 2.82, 0.52 for \ofive, \osix, \oseven, and \oeight, respectively, for wavelengths between 5 and 20~\upmicron, at native resolution, and S/N of 2.21, 0.57, 3.14, 0.58 when considering the brightest PAH features only.
Later in this study (see Section~\ref{sec:pahfit1D}), we resample the output cubes both spatially and spectrally to increase our S/N. Doing so, we find new S/N of 6.22, 1.60, 8.55, 1.60 between 5 and 20~\upmicron, and 6.51, 1.67, 8.96, 1.68 for the PAH features.

\subsubsection{Redshift correction}
We observe variable line shifts among the different spectra.
We note that \ofive and \osix seem to have a similar shift, while \oseven and \oeight show a different one. Both pairs of observations were made at different galactocentric radii, and the difference in shifts is likely due to the rotational velocity of the galactic disk. 
There have been several studies of the differential lags of gas phases with distance from the disk \citep[e.g.][]{Levy2019, Lu2024}.
For the purpose of the analysis in this paper, we simply perform an ad-hoc correction so that all spectra are corrected to the rest-frame. 
The velocity shift due to galaxy motion is expected, but not easy to attribute to any particular radius in the galaxy. 
A detailed measure of the width of the emission lines might give us information on the depth probed by our measurements, in this edge-on case. A more detailed decomposition of the emission lines and their information (including CGM kinematics) is not within the scope of this paper and will be presented in the future.

We use emission lines and their reference wavelengths from the theoretical list used in PDRs4All\footnote{\url{https://hebergement.universite-paris-saclay.fr/edartois/jwst_line_list.html}} \citep[used in][]{Chown2024, Peeters2024, vdPutte2024} to measure redshifts. 
Some of the faintest/tentative lines, e.g. \htwo~$S(4)$ and \ariii, show large discrepancies with respect to the median of only the brighter ones, and we do not consider them for correction purposes.
We extract the final corrections as the mean values of the redshifts for only the brightest lines, visible in all 1D~spectra, i.e. \htwo~$S(1)$, \neii, \neiii, \siii, \htwo~$S(2)$, \arii.
The final redshift used for \ofive and \osix is $1.993\times10^{-3}$, and $1.259\times10^{-3}$ for \oseven and \oeight. We do not find a significant enough difference between observations at the same galactocentric radius to warrant different corrections. While other works find differential lags of gas in the CGM, e.g. \citep{Levy2018, Lu2024}, these appear of the order of $20-30$~km~s$^{-1}$, lower than the achievable MIRI~MRS resolution.

\section{Qualitative description and model creation}
\label{sec:qualitativedescription}
Some well-known mid-IR features and complexes can be identified in all spectra, with varying intensities and clarity.
Similarly to the emission lines, \ofive and \oseven show the clearest features.

\subsection{Intra-comparison: between 6 and 18~\texorpdfstring{$\mu$m}{}}
\label{sec:qualitativedescriptionIntra}
\begin{figure*}
    \centering
    \includegraphics[width=1.0\linewidth]{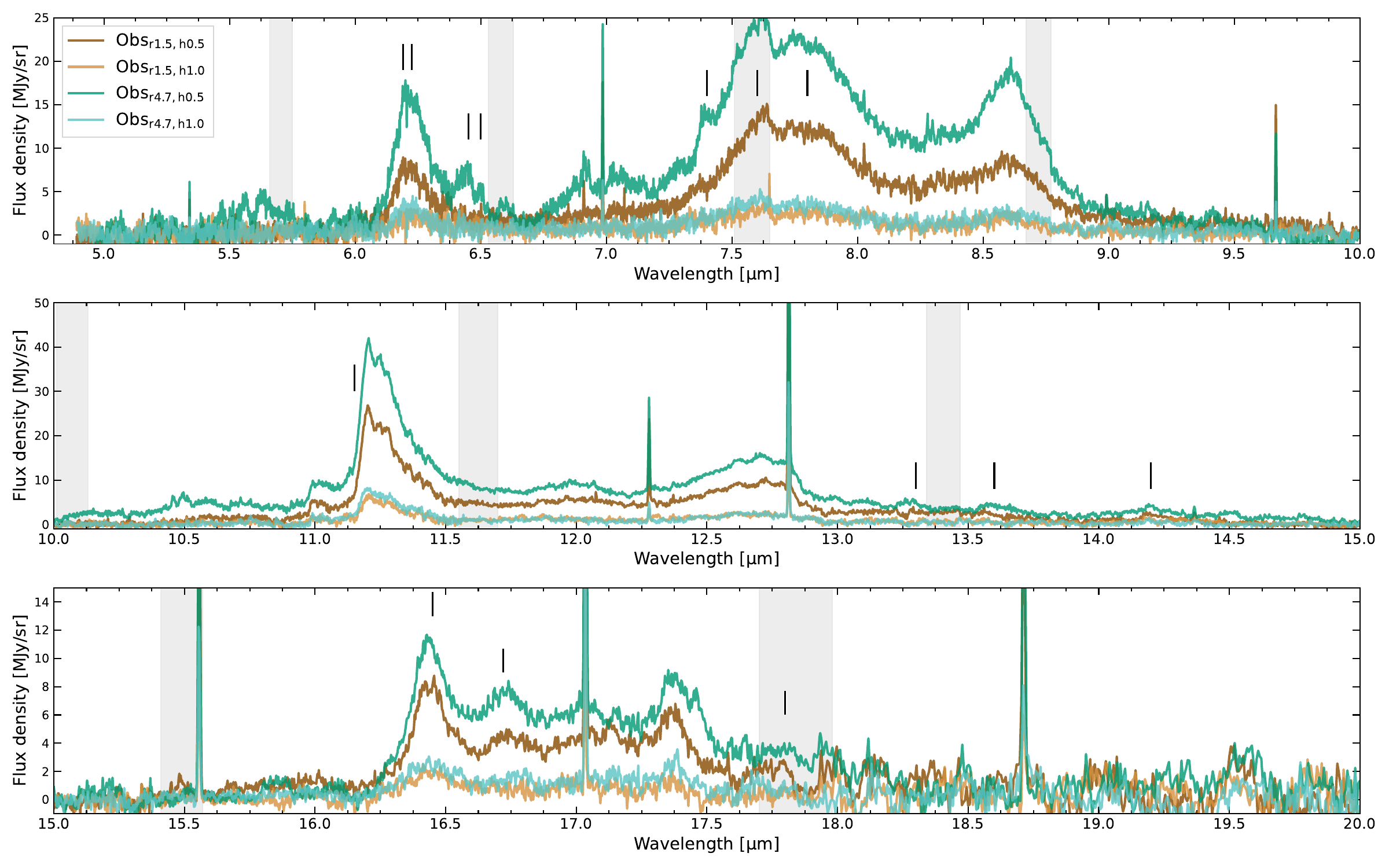}
    \caption{All spectra, continuum-subtracted using the best-fit stellar and hot dust components from 1D \pahfit results. While \oseven (green) shows notably higher flux densities, the shapes of the mid-IR features are in good agreement with \ofive (brown). \osix (orange) and \oeight (blue) show a clear decrease in mid-IR emission, but the brightest features persist (e.g. 7.7, 11.2, 17~\upmicron). The shaded grey regions mark the stitching wavelengths between channels. Some of the sub-features, at 6.19, 6.23, 6.45, 6.50, 7.40, 7.60, 7.80, 11.15, 13.30, 13.60, 14.20, 16.45, 16.72, and 17.8~\upmicron, discussed in Section~\ref{sec:qualitativedescriptionIntra} are marked by black vertical lines.}
    \label{fig:allObsSpectraContSub}
\end{figure*}

Here we describe the differences and similarities between the four pointings in the lower-CGM of NGC~891. 
Figure~\ref{fig:allObsSpectraContSub} shows the four spectra, where the continuum removed in each case is the sum of the stellar and hot dust components (Section~\ref{sec:pahfit1D}).

\paragraph{6.2~{$\upmu$m} feature ---}
The 6.2~\upmicron emission is associated with C=C stretch. We see that observations display global similarities in pairs: \olowh show a prominent 6.2~\upmicron peak, while \ohighh show a moderate bump, when all are continuum-subtracted. 
\ofive and \oseven show a smaller peak at 6.45~\upmicron, with a potential additional one at $\sim 6.5$~\upmicron in \oseven only. 
Other observations show a rather rapid decrease on the red side on the feature, without any clear additional peaks.
\oseven shows a relatively narrow peak around 6.9~\upmicron, absent in the other positions in NGC~891. It is also not seen in other spectra we compare to (see Section~\ref{sec:qualitativedescriptionExtra}) and is likely an artifact.

\noindent Given the noise in all observations, we attempt to extract information from the spectra after convolution to a coarser spectral resolution. Details of this approach are given in Appendix~\ref{sec:app:convolvedSpectra}. 
Convolving the 1D spectra with a Gaussian of $\sigma=5$ (pixels), we tentatively identify two peaks in \ofive and \oeight. 
The 6.2~\upmicron complex is made of two sub-features from olefinic and aromatic stretch modes \citep[6.19 and 6.25~\upmicron, respectively;][]{Pendleton2025}. In our case, we find a peak at 6.19~\upmicron but the second peak at $\sim6.22-6.23~$\upmicron. \oseven shows a broad peak rather than two separate peaks, possibly hinting at the mixture of the two features from stretching modes.

\paragraph{7.7~$\upmu$m complex ---}
The 7.7~\upmicron complex is attributed to large cations and represents a mixture of C=C stretching and C--H in-plane bending \citep{Bregman1989, Peeters2002}, making up several sub-features. 
In our case, \olowh show very similar shapes, which would be modelled by the same number of components, although \oseven might display three distinct features a little more clearly than \ofive, and a clear 7.4~\upmicron bump.  
Both \ohighh show a very large bump that is just above the continuum, without any clear decomposition into sub-features.

\noindent Work by \citet{Shannon2019} identified a shift of the peak wavelength in this feature depending on the global size distribution of grains emitting in this spectral range. Namely, a smaller (in terms of number of carbon atoms) grain population peaks at 7.6~\upmicron while a larger population peaks near 7.8~\upmicron. 
Following the approach described in Appendix~\ref{sec:app:convolvedSpectra}, we do not find a significant wavelength shift between our four convolved spectra. The high-S/N \ofive and \oseven peak around 7.620 and 7.625~\upmicron, a very small difference. Including \osix and \oeight shows a larger range, from 7.615 to 7.629~\upmicron, but the peak position is much more uncertain.

\paragraph{8.6~$\upmu$m feature ---}
This feature also rises from a mixture of C--H bending and C=C stretching \citep[e.g.][]{Bauschlicher2008}.
For this feature, we once again see similarities in pairs: \olowh show a clear, well defined feature while \ohighh seem to be just above the continuum level.
There is potentially a systematic difference in peak wavelength between convolved spectra, ranging from $\sim 8.56$ to 8.61~\upmicron (see Appendix~\ref{sec:app:convolvedSpectra}).

\paragraph{11.2~$\upmu$m complex ---}
The multi-peaked 11.2~\upmicron complex arises from C--H out-of-plane bending modes \citep{Hony2001, Bauschlicher2008}.
All our observations show an increase in flux starting at $\sim 10.95$~\upmicron, with a small feature that was also identified in the Orion PDR \citep{Chown2024}.
\ohighh look identical, with a similar second rise at $\sim 11.15~$\upmicron, peaking at 11.2~\upmicron and slowly decreasing over the range of the complex.
\olowh also show a very similar 11.2~\upmicron feature. The rise between 11.15 and 11.2~\upmicron in \oseven seems to be slightly steeper and shows three distinct bumps a little more clearly between 11.2 and 11.3~\upmicron, compared to \ofive. Additionally, \ofive shows a shallower, broader red wing.

\noindent The peak wavelength of the 11.2~\upmicron feature is very stable between all our observations, even when convolved to coarser resolution (less than 0.5\% difference). Convolving results in blending the multiple peaks, but all spectra respond similarly to degrading the spectral resolution (see Appendix~\ref{sec:app:convolvedSpectra}).

\paragraph{12--15~$\upmu$m range ---}
This range is (theoretically) filled with numerous features that pertain to different C--H bending modes \citep{Hony2001}. 
First, we see that the 12~\upmicron feature looks virtually absent in the lower S/N spectra of \ohighh, while it is weakly detected in the other two. 
The 12.7~\upmicron complex appears more visible in all spectra, with potential sub-features identified in \oseven \citep[similarly to what is observed in Orion;][]{Chown2024}.
Additional bumps are somewhat visible in the highest S/N spectrum around 13.3, 13.6, and 14.2~\upmicron, some of which can be found in other targets \citep{Chown2024}.

\paragraph{17~$\upmu$m complex ---}
The 17~\upmicron complex has been attributed to C--C--C bending modes, likely dominated by large molecules \citep{Peeters2004}.
In the 17~\upmicron region, we see a very similar shape between \ofive and \oseven, with all features identifiable in one seen in the other as well, for a total of at least four features between 16.2 and 17.6~\upmicron. \ohighh show much less bright features, although we can identify by eye a small contribution of the 16.4 and 17.4~\upmicron features in both, while the bump in-between disappears in \oeight. 
The 17.8~\upmicron is (moderately) identifiable in all observations but \oeight. 

Overall, \ofive and \oseven are very similar and present the same (sub-)features, while \osix and \oeight sometimes miss a clear identification of a few features. We do not notice any clear shifts of the central wavelengths of these features between the four spectra presented here, but widths may vary slightly.

\subsection{Comparison with other mid-IR spectra}
\label{sec:qualitativedescriptionExtra}
\begin{figure*}
    \centering
    \includegraphics[width=\linewidth]{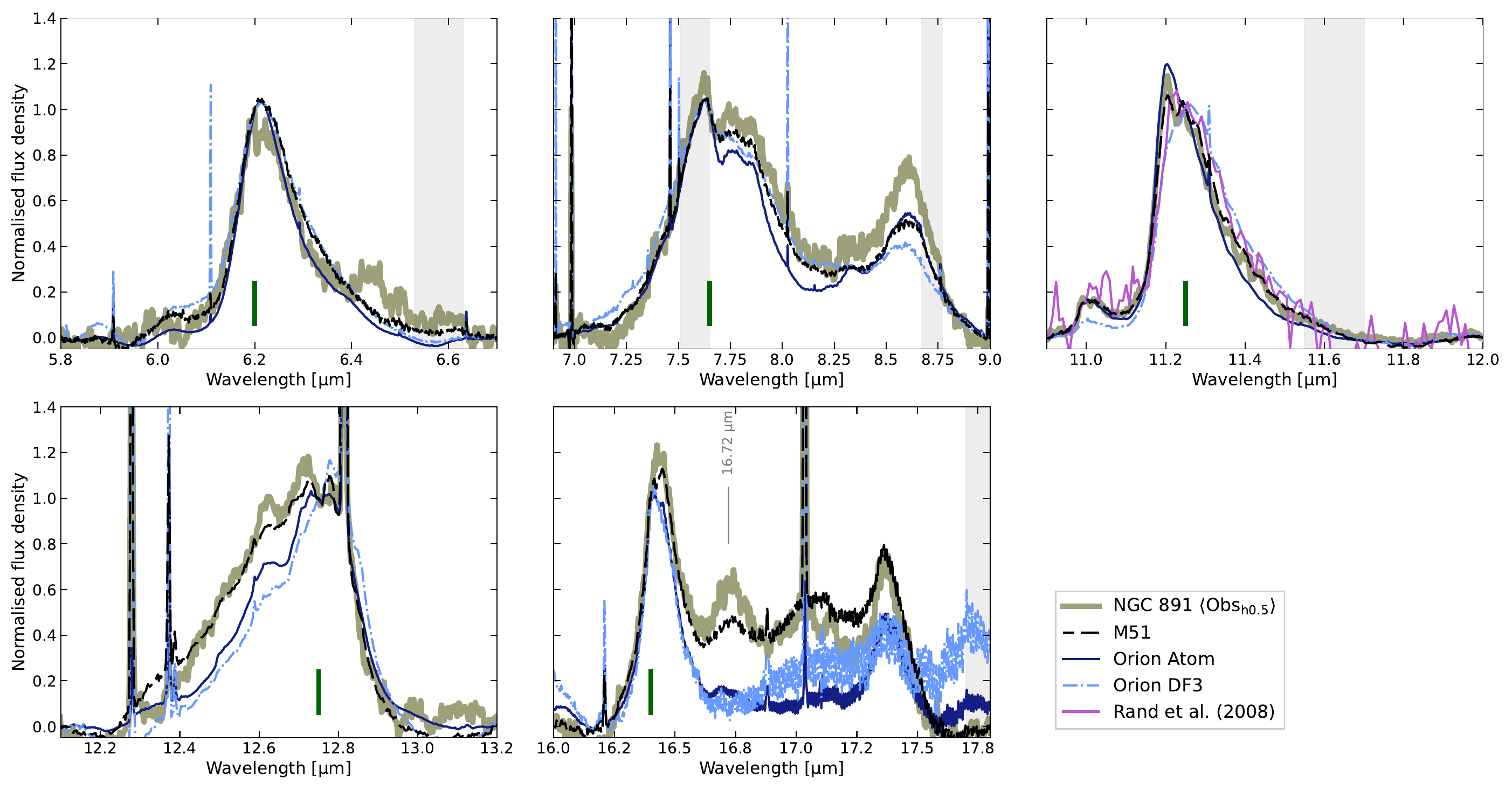}
    \caption{Portions of mid-IR spectra focused on a few typical PAH features. We show an average of the observations \ofive and \oseven spectra for NGC~891 in olive-green colour, the `atomic' and `DF3' templates from the PDRs4All program in solid and dash-dot blue lines, and an average of the three pointings in M51 in dash black. We add the average spectrum of the east pointing from \citet{Rand2008} in the 11.2~\upmicron panel (it is too noisy in other panels).  The shaded grey regions mark the stitching wavelengths between channels. The dark green lines mark the normalising wavelengths.
    }
    \label{fig:specs_comp}
\end{figure*}

We compare our data with the spectra of other sources recently available in the literature, namely the Orion star forming region, representative of a photodissociation region, and those of M51, focused on star forming region in a normal galaxy. Through this exercise, we aim to identify potential systematic shifts in the peak wavelength of PAH features, tracing different grain populations in these various environments. 
In Fig.~\ref{fig:specs_comp}, we show an average of our \ofive and \oseven, the higher-S/N spectra with the clearest features. 
In blue lines, we show the mid-IR atomic (atom., solid dark blue) and the dissociation front 3 (DF3, dot-dash blue) spectra from the Orion photo-dissociation region \citep[from PDRs4All;][]{Chown2024}. We choose the Atomic template as we consider it a ``reference'' PDR spectrum, and the DF3 as it shows striking differences with other spectra.
In Appendix~\ref{app:compwithPDRs}, Fig.~\ref{fig:app:wPDRs}, we show the average spectrum of \olowh of NGC~891 and all PDR templates, `atomic', `\hii', and the three `DF' spectra.
We also add to Fig.~\ref{fig:specs_comp} recent observations from M51 (GO~3435; Dale, Sandstrom et al., in prep). With dashed black lines, we show an average of their three available MRS data, pointed at star-forming regions. 
For a qualitative comparison, and because each spectrum probes a very different environment, we remove the local continuum in each panel and normalise the spectra. We simply fit a line using the surface brightness at the wavelength bounds of each panel, remove a linear component to that portion of the spectrum, and normalise them to their respective surface brightness at a common wavelength in each portion shown in the panels. This limits the comparison to the shape of the features only.

Overall, the displayed spectra show a lot of similarities.
In the 6.2~\upmicron feature, associated with cations \citep[][]{Allamandola1999}, the NGC~891 and Orion Bar Atomic spectra are similar on the red wing, while the Orion Bar DF3 spectrum shows the broadest profile, with NGC~891 following the same steep blue rise. The M51 spectrum is intermediate, closer to the PDR~Atomic blue rise but following the DF3 spectrum on the red wing. All peak at the same wavelength, although our data do not show a smooth, clean bump, as mentioned in Section~\ref{sec:qualitativedescriptionIntra} and Appendix~\ref{sec:app:convolvedSpectra}. 
This could be an indication of different relative contributions from the olefinic and aromatic bonds in the complex.
The apparent bump around 6.02~\upmicron in the M51 and Orion spectra \citep[C=O modes;][]{Allamandola1989b}, could tentatively be identified at slightly shorter wavelength in NGC~891, at 6.0~\upmicron, but it is rather faint.

The spectral region that includes the 7.7~\upmicron feature also displays strong resemblance amongst the sources. The blue rise is once again very similar between all spectra, slightly broader in DF3.
The width of the complex is the same in NGC~891 and M51, both on the blue and red sides, while the Atom. spectrum is significantly narrower, DF3 being the one with the broadest of the Orion profiles.
All spectra also display multiple peaks within the 7.7~\upmicron complex, although those seen in the Orion~Atomic data appear the clearest. Their relative intensity is very different between the pair NGC~891/M51 and the Atomic spectrum, with the DF3 spectrum being intermediate.
Based on the work of \citet{Shannon2019}, this difference could be attributed to a different size distribution of the grains responsible for the 7.7~\upmicron emission (see also Section~\ref{sec:qualitativedescriptionIntra}).
This would imply a higher processing level (fewer large grains) in the case of the Orion spectrum.
However, we also notice the higher surface brightness in the transition between the 7.7~\upmicron complex to the 8.6~\upmicron feature, which could suggest a stronger plateau in our data, M51, and DF3.
If this difference around 8.2~\upmicron is real (and not only due to normalisation effects), this could indicate again a lower level of grain processing in the CGM of NGC~891, as the transition from class C (with broad emission between the 7.7 and 8.6~\upmicron features) to class B\footnote{Classes of PAH features aim to form broad families based on the shape of the feature. In the case of the 7.7~\upmicron region, the main difference is in the peak wavelength of the feature, with Class~B peaking at $\sim 7.9~$\upmicron and a clear separate 8.6~\upmicron feature, while Class~C shows a single, broad feature peaking at $\sim 8.2~$\upmicron.} emission features \citep[][]{Peeters2002, Shannon2019} has been attributed to the first levels of processing of small grains.
The 8.6~\upmicron feature appears very similar between NGC~891, M51, and the Atomic spectra, while DF3 shows a noticeable fainter feature. The steep 7.7~\upmicron red wing but similar 8.6~\upmicron bump in the Atom. spectrum, and the opposite behaviour in DF3, and their respective resemblance with NGC~891 is interesting, but limited due to the subtraction and normalisation needed here for comparison. 

Between 11.0 and 11.2~\upmicron, the match is excellent between the NGC~891, M51, and Orion~Atomic spectra on the blue side, including the 11.0~\upmicron feature and the steep rise to the whole complex. 
The red wings are slightly different, with NGC~891's slope being between M51 and the Atomic ones (possibly closer to the PDR~\hii, Fig.~\ref{fig:app:wPDRs}). There appear to be several bumps in the red wing of the NGC~891 spectrum, and although we do not know if these are to be associated with PAH features, the M51 data could present them as well. 
The DF3 spectrum in that region shows a noticeable shift, making it closer to a Class~B profile \citep{Chown2024}, as it peaks at longer wavelengths and shows no strong 11.2~\upmicron sub-feature.
We also show the east pointing from \citet{Rand2008} in this panel. The \textit{Spitzer}/IRS data show a fairly similar shape, including what seems to be matching the steepness of the red-wing slope of the MRS data, but the difference in resolution makes it hard to draw more conclusions.

Our data in the 12.7~\upmicron feature region align really well with M51. The blue rise looks very similar in both cases, with a small deviation shortwards of 12.4~\upmicron, and the red slope shows the same steepness. We also detect three similar peaks in these two spectra. 
This resemblance contrasts with the data from the PDR templates: both the atomic and DF3 spectra show a different blue side of the feature, and only the atomic template shows a similar red wing. Recent work by \citet{Khan2025} focuses on a detailed decomposition of the 12.7~\upmicron feature in the PDRs4All data, and finds it is best fitted with six components. In our case, we do not use six sub-features (see next Section~\ref{sec:dustinCGM_PAHFITmodel}).

In the 17~\upmicron region, the 16.4~\upmicron feature appears very similar between both the NGC~891 and the M51 spectra, in width and intensity. Later in this paper, we investigate the conspicuous yet unknown 16.72~\upmicron bump (Section~\ref{sec:16.72bump}). The 17~\upmicron bump appears different between both examples, wider and rising more in the M51 data, possibly because of a strong continuum.

To summarise, we find that the strong features discussed here show little to no variation in their peak wavelengths (e.g. at 6.2, 11.2, 12.7~\upmicron). As sometimes pointed out, the number of peaks may vary. These may be real but may also be due to lower/higher S/N (e.g. at 7.7~\upmicron, or the red wing of 11.2~\upmicron feature). Overall, the most noticeable differences are those seen in the width of the features, pertaining to different slopes on the blue and/or red side(s).

The differences noted here reflect various bending and stretching modes: C=C in the 6.2~\upmicron feature, C--H stretching in-plane in the 7.7~\upmicron feature and out-of-plane in the 11.2~\upmicron features. 
The hint of smaller grains dominating the 7.7~\upmicron feature in Orion compared to NGC~891 could indicate a more advanced processing stage in the PDR. This could be linked to a more intense radiation field, given the immediate vicinity of the star in the PDR. Although the CGM of NGC~891 would be subject to high energy photons, the continuous exposure to a softer but more intense and more consistent radiation field in the PDR might lead to enhanced PAH processing. The steeper red slope of the 11.2~\upmicron feature of Orion might also indicate a higher level of processing there, although we note a lack of clear evidence from laboratory measurements.
These differences rely on the local continuum subtraction and relatively small shifts in wavelength, which make the above interpretation mostly tentative.
In Section~\ref{sec:dustinCGM}, we discuss these differences more from a band ratio perspective.

\subsection{Mid-IR features model with {\small PAHFIT}}
\label{sec:dustinCGM_PAHFITmodel}
We use \pahfit \citep[][]{Smith2007, vdPutte2025} to model the dust features in our spectra and extract their intensities.
The \pahfit tool uses a model file with input on the central wavelength and width of the profiles of the features (characterised as Drude profiles) that need to be modelled. 
To create the model, as an observational constraint, we use an average of the \ofive and \oseven, as they both show higher S/N than the other two. We do so as we expect (unknown) variation between these positions, but do not assume either to be the ``true'' standard spectrum.
In this analysis, we constrain ourselves to wavelengths up to $\lambda = 20$~\upmicron. As mentioned earlier, none of the spectra show reliable data past this threshold, and we wish to focus on the PAH features at shorter wavelengths. 

In this study, we create our own model file by adjusting the one optimised for the PDRs4All consortium and NGC~7023 \citep{vdPutte2025}, and remove any lines and features that we do not find in our data from the starter model file. Naturally, the Orion spectra show many more emission lines and more dust (sub-)features than our data \citep[][]{Chown2024, vdPutte2024, vdPutte2025}, and we keep only the emission lines that can be identified by eye in the high S/N spectra.
For the dust features, we first overlap our averaged spectrum with the templates given by the PDRs4All consortium for the ionised, molecular, and dissociation fronts regions. We inspect these data carefully to remove the features that do not show up in the new NGC~891 data. 
From this empirical model, we proceed iteratively: we run \pahfit on our average 1D spectrum, and visually inspect the results to (i) adjust central wavelengths and widths, (ii) add features that clearly improve the quality of the fit, and (iii) remove features that turn out not to improve the quality of the fit.

Our final model is comprised of the following large complexes, with several sub-features meant to reproduce the shapes of the profiles, without verified counterparts from other spectra or laboratory data:
\vspace{-0.1cm}
\begin{itemize}
\setlength\itemsep{0.05cm}
    \item 6.2~\upmicron: made of four sub-features, two forming the main asymmetrical emission profile, and two additional small bumps in the red tail;
    \item 7.7~\upmicron: made of five sub-features, four of which are meant to reproduce the main emission, including a wide broad-winged feature;
    \item 11.2~\upmicron: made of three peaked features for the main emission, and a collection of ad-hoc, non-physical features to reproduce the very asymmetrical profile on the red side;
    \item 12.7~\upmicron: made of a broad, slow blue rise and three narrow peaks;
    \item 17~\upmicron: made of three distinct bumps;
\end{itemize}
and more ``isolated'' features (i.e. not belonging to a known, broader complex) at 5.99, 6.91, 7.04, 8.30, 8.60, 10.60, 11.01, 11.95, 13.30, 13.57, 14.20, 16.4~\upmicron (modelled by three components to best reproduce the asymmetric red wing), and 16.72~\upmicron.
Note that the 10.60 and 16.72~\upmicron bumps appear conspicuous enough to warrant an input that helps the model, but they are not tabulated in \citet{Chown2024}. Aside from these particular points, all other entries for $\lambda < 17$~\upmicron in our model can be considered to have an identified or tentative equivalent in the PDRs4All census. 
The final model is tabulated in Appendix~\ref{sec:app:finalmodel}, with central wavelength and width of all features.

\subsection{Applying PAHFIT}
\label{sec:pahfit1D}
\begin{figure*}
    \centering
    \includegraphics[width=\textwidth, clip, trim={2cm 1cm 1cm 2.1cm}]{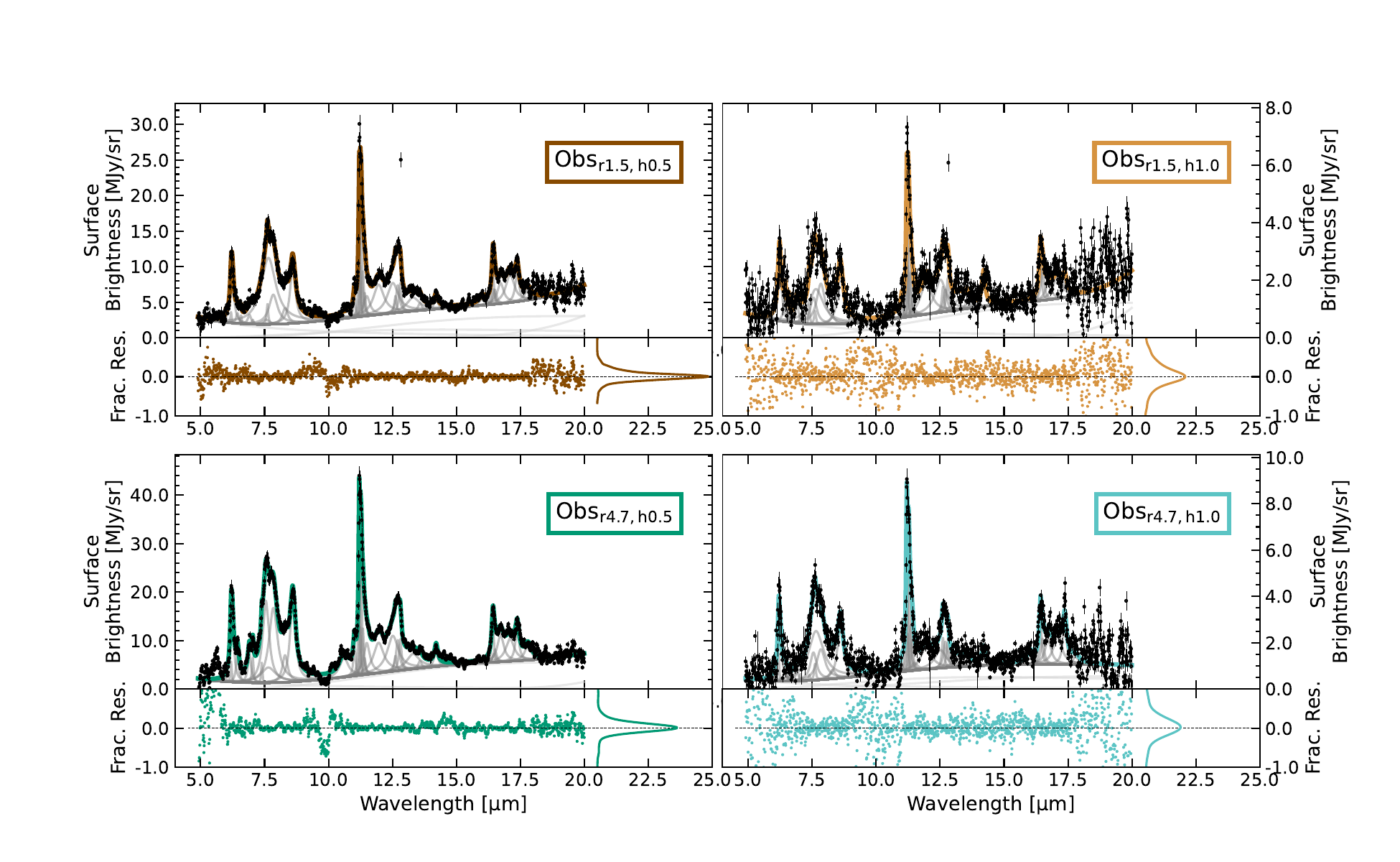}
    \caption{Best fits for each 1D spectrum with our spectral model described in Section~\ref{sec:dustinCGM_PAHFITmodel} using \pahfit (with emission lines removed for better comparison of dust features). 
    Overall, the model based on an average of \ofive and \oseven fits well all observations, and adjusts to the low-emission spectra \osix and \oeight. Note the difference in continuum between \ofive and \oseven, near the disk, \osix and \oeight away from the disk. Although \oseven is at higher galactocentric radius, it shows more pronounced PAH features than \ofive, closer to the bulge.
    We also show the wavelength-dependent residuals corresponding to each fit, and their histogram. As expected, the fit appears better for the high-S/N spectra of \ofive and \oseven.
    }
    \label{fig:pahfit_allObs}
\end{figure*}

We first apply our spectral model to 1D spectra. As mentioned earlier, to mitigate the relatively low S/N in our observations, and since we do not expect to detect very faint and narrow sub-features, we regrid to coarser spatial and spectral pixels our extracted spectra. We use the {\tt calwebb\_spec3.cube\_build} function with a target $11\times13$ pixels cube with $\Delta\uplambda = 0.01$~\upmicron. 
This spectral resampling does not matter in terms of emission line fitting as we do not use the results from \pahfit to investigate the gas properties.
In Fig.~\ref{fig:pahfit_allObs} we show the fit to all 1D spectra using the model described above. We remove emission lines for a better view of the dust features.\footnote{We note a few discrepant points in \ofive and 6 at $\sim 11.2$ and $\sim 12.8~\upmu$m, from removing emission lines for display.}
Similarly to Fig.~\ref{fig:all1Dspectra}, we clearly see the difference in intensity and S/N between the observations close to the disk (\ofive and \oseven) and those at 1~kpc (\osix and \oeight). 

Under each panel of Fig.~\ref{fig:pahfit_allObs}, we show the fractional residuals as a function of wavelength. All behave fairly similarly, are centred on 0, although the fits of \osix and \oeight show a wider distribution of residuals.
The dip just before 10~\upmicron visible in \oseven (although it is somewhat noticeable in all cases) aligns with a dip in the data. Despite its closeness with known silicate absorption, we do not believe it is tracing that feature because of the low $A_{\rm V}$ associated with the diffuse CGM. Instead, we believe this drop in surface brightness is an artifact, potentially related to the stitching.

The continuum is modelled as a combination of a 3000~K blackbody accounting for starlight, and three modified blackbodies at 50, 150, and 200~K accounting for host dust emission (all shown in light gray in Fig.~\ref{fig:pahfit_allObs}).
We find that the low-temperature dust continuum has the highest intensity in observations \ofive, \osix, and \oseven compared to the hotter temperatures, which have a much lower contribution of several orders of magnitude. 

\begin{figure*}
    \centering
    \includegraphics[width=1.0\textwidth, clip, trim={2cm 1.25cm 1cm 0}]{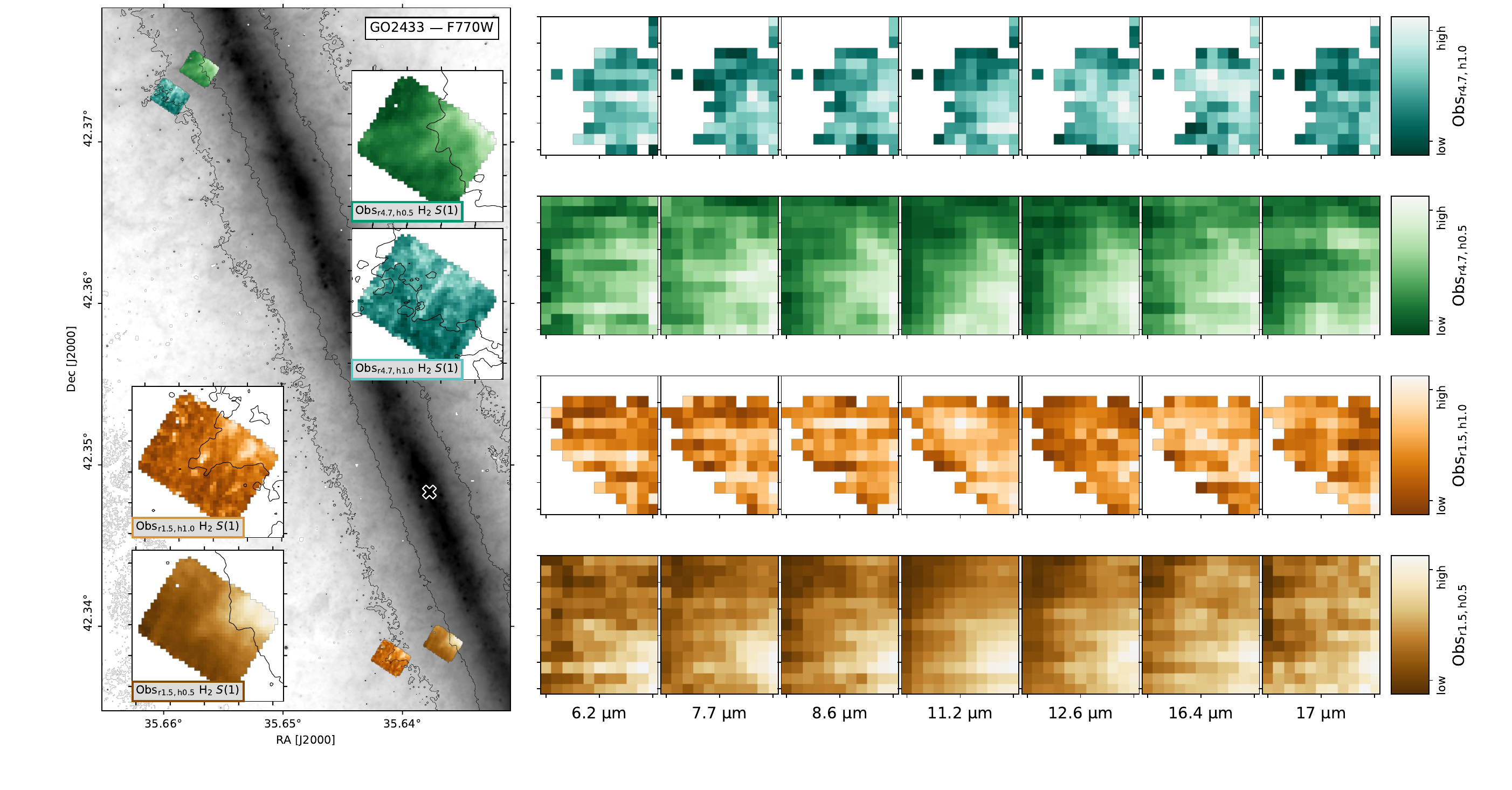}
    \caption{\textit{Left:} GO~2433's F770W map of NGC~891, with the \htwo~$S(1)$ maps for the four MRS pointings in our study, in colour. The contours in the background and inset images are F770W surface brightnesses. The white cross marks the centre of the galaxy.
    \textit{Right:} Total intensity maps for a few key dust features in all observations. The maps are obtained by fitting a 1D model to each pixel independently, and summing the intensities of all sub-features belonging to the same complex. Because the 11.2~\upmicron feature dominates the emission, each map is shown with its own scaling; we focus on the variations within each maps rather than the absolute intensity values.
    Non-fit pixels, skipped by \pahfit, are shown in white.}
    \label{fig:pahfitcube_allObs}
\end{figure*}

We also wish to apply the same model to the 3D cubes, and perform fits pixel-by-pixel (spatially). 
Each spatial pixel is thus fit independently with \pahfit. In Fig.~\ref{fig:pahfitcube_allObs} we show the results of the cube fits for a few key features. In the case of complexes, these maps are computed by summing all sub-features belonging to that complex. The white pixels are those deemed too uncertain to be fit.\footnote{A spectrum in the cube is skipped in the fit when there are more than 20\% pixels along the wavelength axis with negative values.}

At this stage, \pahfit does not output errors associated with the fits directly. To have a general idea of the quality of the fits, we simply run \pahfit on the same spectra with added noise. 
We use the uncertainty cubes produced by the JWST pipeline to create noise which is directly added to the spectra. In both the 1D and 3D cases (in the latter, noise is added in each spaxel), we repeat the fitting procedure 100 times with different offsets drawn from a random distribution that matches the properties of the noise.
We measure the standard deviation in the values we are interested in (total intensity, band ratio) and quote this as error.

\subsection{The 16.72~\texorpdfstring{$\mu$m}{} bump}
\label{sec:16.72bump}
We find a rather conspicuous bump in the 17~\upmicron complex, so far not attributed to a species or bond to the best of our knowledge. Previous works focused on the 15--20~\upmicron range observed with ISO and \textit{Spitzer} have drawn the list of features associated with PAHs. Usually, the ``17~\upmicron complex'' is split between the 16.4, 17.4, 17.8~\upmicron features and a $\sim$15--$18~\upmu$m plateau, with additional features on each side at 15.8 and 18.9~\upmicron \citep[e.g.][]{Boersma2010, Peeters2012, Shannon2015}. The plateau can be fit as a broad component peaking around 17.2~\upmicron. 

We point out that this feature appears to be identifiable in other spectra.
We find it in the MIRI spectra of three pointings in the M51 galaxy (GO~3435, PIs: Sandstrom, Dale, in prep).
It appears slightly visible in some of the PDRs4All's Orion spectra \citep{vdPutte2025}. It is also noticeable in some earlier works using \textit{Spitzer}/IRS \citep[e.g.][]{BernardSalas2009, Rand2011}.
In the M51 and Orion spectra, the continuum at $\lambda \geq 15~$\upmicron is much stronger than in our spectra. 
Finally, as we found no warning from the pipeline official releases, we consider this feature to be real and attempt to extract information for it.

Both our high S/N 1D spectra of \ofive and \oseven diverge from standard spectra seen in the literature and instead can be best fit with two components at 16.72 and 17.09~\upmicron. 
It is visible in the bottom right panel of Fig.~\ref{fig:specs_comp}, for the average emission of \ofive and \oseven combined. We clearly see that the bump is large enough to warrant an additional feature that helps fitting the data better. Although less visible in the low S/N spectra of \osix and \oeight, that feature remains present. 

\begin{figure}
    \centering
    \includegraphics[width=1.0\linewidth]{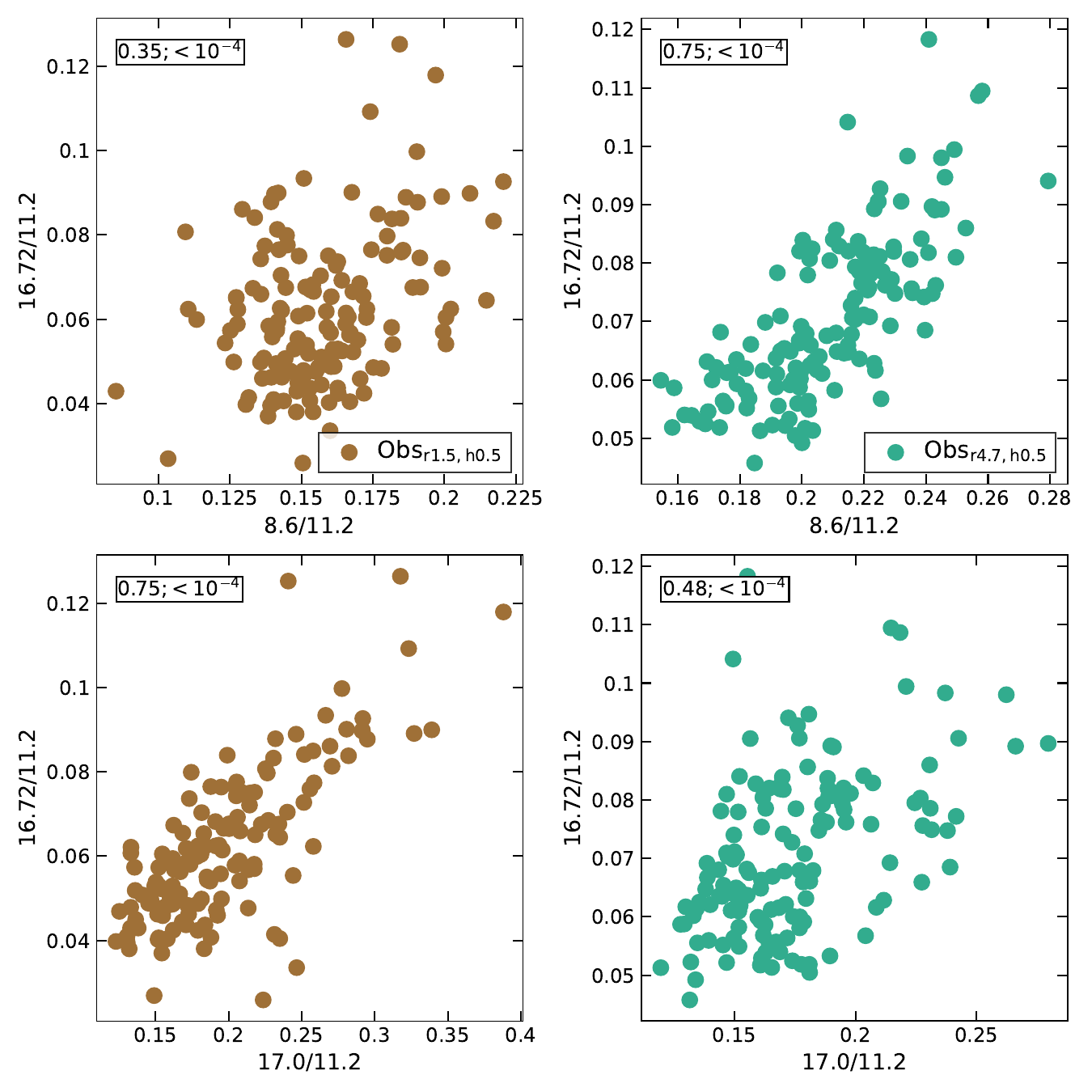}
    \caption{Examples of correlation between the 16.72~\upmicron feature and the 8.6 and 17~\upmicron features, normalised by the 11.2~\upmicron emission, showing the highest $\rho$ values in \ofive and \oseven. The labels indicate the Pearson coefficient, $\rho$, and $p$-value.}
    \label{fig:16.72corr}
\end{figure}

We investigate the possible correlation of this feature with other known features with a known PAH population attributed for their emission. 
Using the results from fitting the 3D cubes, we look into the variations of the 16.72~\upmicron emission pixel-by-pixel normalised by continuum emission and by the total emission from different features.
First, we can report that these correlations for \osix and \oeight show high $p$-values that indicate no significant correlations on a pixel scale.
We find different trends between \ofive and \oseven. In \ofive, normalising the 16.72~\upmicron emission by a local continuum, or the emission in the 7.7 or 11.2~\upmicron features leads to the highest Pearson coefficient, $\rho$, when correlated with the (similarly) normalised 17~\upmicron. We also find that the 16.72~\upmicron/Norm. vs 8.6~\upmicron/Norm. correlations shows reasonably high $\rho$ values to be significant. 
Correlation with the 17~\upmicron emission (in our case, the sum of $\sim 17.09, 17.4, 17.8~$\upmicron emission) could also be a modelling result, as all these features are close in wavelength and variations in one will inevitably propagate to the others.
In \oseven, the correlation with 8.6~\upmicron is consistently highest, closely followed by high $\rho$ values for correlations with the 7.7~\upmicron features.
We show a few of these correlations in Fig.~\ref{fig:16.72corr}.

Spatial variations (e.g. as shown in Fig.~\ref{fig:pahfitcube_allObs}) are somewhat difficult to interpret because of the limited (reliable) spatial coverage of only \ofive and \oseven. 
The features that seem to correlate with 16.72~\upmicron are usually attributed to ionised PAHs \citep[][]{Maragkoudakis2025}. 
Here, without being completely conclusive, we suggest that this feature could be linked to the same populations (see Appendix~\ref{app:16.72feature}).
Higher S/N observations such as these of M51 or the Orion Bar will certainly bring clearer results.

\section{The evolution of dust in the CGM}
\label{sec:dustinCGM}
\subsection{Evolution with $r$ and $h$}
As mentioned before, when constructing the model, some entries are added only to improve the fit and reproduce a known asymmetric shape that is not reproduced by a Drude profile (for example, the very asymmetric 11.2~\upmicron feature in our case). They are not (laboratory-)identified peaks and are purely practical, mitigating the lack of asymmetric profiles in the current version of \pahfit. Therefore, we do not study the sub-features making up each complex individually, but rather sum up all features attributed to a complex to get a total value for each large complex/feature at 6.2, 7.7, 11.2, 12.7, 16.4, and 17~\upmicron, including ad-hoc features, if any. 
These ad-hoc features contribute to a small percentage of the total emission within a complex. 

We see in Fig.~\ref{fig:allObsSpectraContSub} that \oseven consistently shows the highest surface brightness values for all dust features, followed by \ofive. Since they both lie at the same distance from the disk, $h\sim 0.5~$kpc, but at different radii, it could indicate an increase of the mid-IR emission with increasing radius.
(We note that \osix and \oeight follow the same trend.)
However, the higher flux densities in \oseven are more likely a result of the pointing choice than a gradient with galactocentric radius: \oseven and \oeight are positioned to align with an identified filament escaping the disk, therefore not quite sampling the same ``diffuse'' environment as \ofive and \osix, but with known enhanced emission.
Work by \citet{Rand2008}, which includes data distributed along the disk of NGC~891 in five positions, show no consistent trends with radius $r$. 
We therefore believe that the higher values found in \oseven and \oeight compared to \ofive and \osix are not due to an increase in galactocentric radius. That said, other works have found a gradient with galactocentric radius \citep[e.g.][]{Popescu2011}, but we argue that the main difference between our pointings is not linked to that gradient.

\begin{figure}
    \centering
    \includegraphics[width=1.0\linewidth]{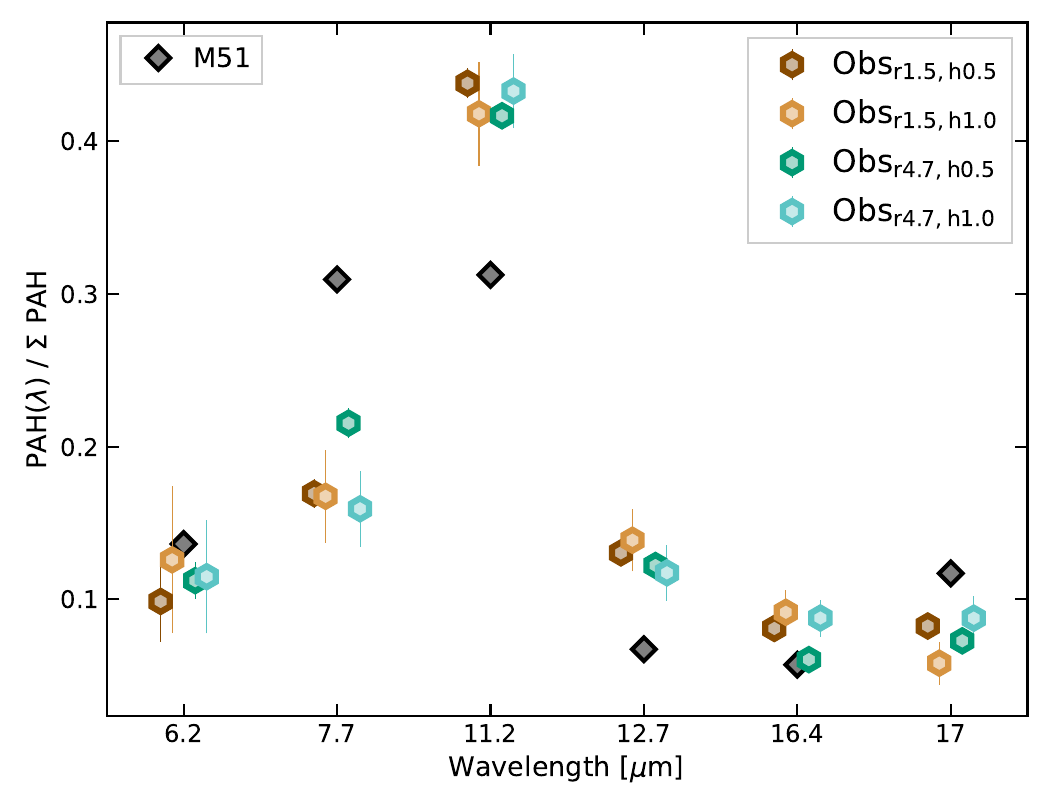}
    \caption{Total intensity in several PAH features normalised by the total intensity in all features for each observation, as well as M51, for reference.}
    \label{fig:variations_r_h_1d}
\end{figure}

This is confirmed in Fig.~\ref{fig:variations_r_h_1d}, where we show the total intensity in several mid-IR features, normalised by the total intensity in all features. 
Most observations show similar behaviours to one another for all features. At 7.7~\upmicron, \oseven shows a $\sim 30$\% difference with the others but does not stand out otherwise. We also see that the 11.2~\upmicron complex clearly emits most of power in all observations.
There is a mild trend identifiable between \ofive and \oseven. We see that \oseven emits more in 6.2 and 7.7~\upmicron complexes compared to \ofive, which emits more at 11.2, 12.7, 16.4 and 17~\upmicron. 
This could be a sign of a different population in both positions, where \oseven has more ionised and smaller grains compared to \ofive \citep{Draine2021}.
Comparing other observations in pairs does not conclusively yield similar or other trends. 
For reference, we calculate the same values for the M51 average spectrum described above.\footnote{For the M51 spectrum, we use the \pahfit model optimised for Orion and NGC~7023 \citep{Misselt2025, vdPutte2025}.} These observations for M51 are closer to the emission from the disk than the NGC~891 observations taken in the CGM. With M51, we see that the 7.7 and 11.2~\upmicron features are contributing to the total intensity to the same level, a behaviour identified in the past \citep{Smith2007}. Indeed, the 7.7~\upmicron complex is often expected to be the brightest. 
The difference between these two galaxies indicate that, for the same PAH emission, there are more grains contributing to the 7.7~\upmicron feature (i.e. large, ionised grains) in the M51 spectrum than in NGC~891, although we see no similar discrepancy between M51 and NGC~891 in the 6.2~\upmicron feature. 
Conversely, there are more 11.2~\upmicron-emitting grains (neutral, large) in NGC~891, which is consistent with the values for the 12.7 and 16.4~\upmicron features (but not 17~\upmicron).
The shapes of the 7.7 (ionised, large) or 11.2~\upmicron between NGC~891 and M51 do not differ significantly either (Section~\ref{sec:qualitativedescriptionExtra}), and do not point at sub-features being responsible for this difference.

We also see a decrease of mid-IR emission with increasing distance from the disk, $h$, regardless of the radius, as evidenced by the systematically lower fluxes measured in the features of \osix and \oeight compared to those of \ofive and \oseven (Fig.~\ref{fig:allObsSpectraContSub}).
This behaviour was already observed by \citet{Rand2008} who observed three pointings at the same radius $r$ on east, disk, and west positions with the \textit{Spitzer}/IRS instrument.
They found that the 11.2~\upmicron and 17.4~\upmicron features (equivalent to the three peaks in our 17~\upmicron) both decrease with increasing distance from the disk. 

\begin{figure}
    \centering
    \includegraphics[width=1.0\linewidth]{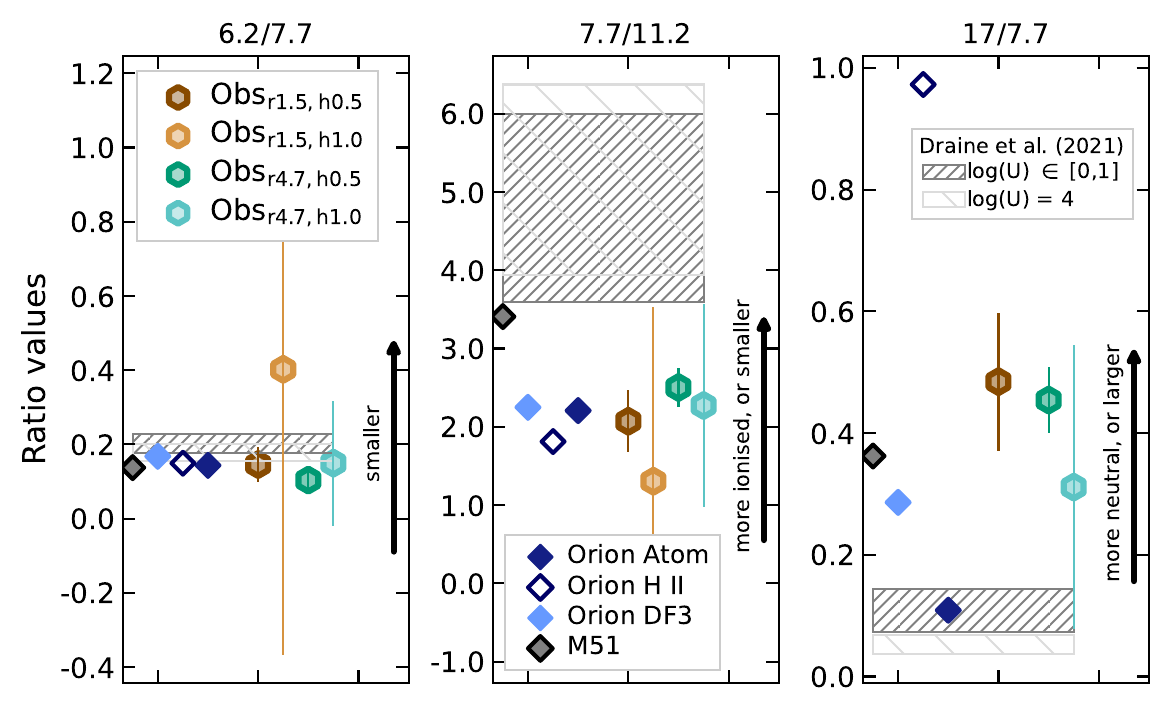}
    \caption{A few key band ratios extracted from our 1D spectra, using clip-values as in \citet{Draine2021}. 
    We also show the ratios extracted from the M51 average spectrum (black diamond), the PDRs4All templates (Atomic, \hii, DF3; shades of blue diamonds), and a few emission spectra from \citet[][``D2021'', grey boxes]{Draine2021}: we use the mMMP model with all sizes (`sma', `std', `lrg') and ionisation (`lo', `st', `hi') for radiation fields intensities from 1 to 10 together, and $10^4$.
    The horizontal offsets are only for legibility.}
    \label{fig:bandratios_comp}
\end{figure}

Figure~\ref{fig:bandratios_comp} shows some of the common mid-IR band ratios accessible with our data. 
These ratios are computed using the same approach as in \citet{Draine2021}: we use the same clipping points, remove a simple continuum between these points (that line is calculated on a smoothed spectrum, to avoid the clip points falling on spurious values, leading to too many negative values after subtraction), and integrate the flux within the bounds. 
We also use the values from the \citet{Draine2021} emission spectra, using the mMMP radiation field \citep[modified spectrum from][]{Mathis1983}, with intensity scaling from 1 to 10, as well as $10^4$---closer to the radiation field intensity in Orion---, and all grain sizes (`sma', `std', `lrg'), and ionisations (`lo' `st', `hi'). Finally, we use the PDRs4All templates and report the same ratios with the same technique.

Putting aside \osix, which is very noisy (sometimes leading to negative clipped values after local continuum removal), we do not observe particularly large variations in each ratio, between pointings.
We find that the 7.7/11.2 ratio, considered to be a tracer of PAH ionisation, as 7.7~\upmicron traces large ionised PAHs and 11.2~\upmicron traces large, more neutral PAHs \citep[][also the 6.2/11.2 ratio has been used as a ionisation tracer, see \citealt{Boersma2014}]{Maragkoudakis2020, Draine2021, Maragkoudakis2022}, slightly increases with galactocentric radius, looking at \ofive (brown) and \oseven (green).
This suggests that the grain population is more ionised further away from the bulge. This means that the larger stellar content at small radii is not affecting the PAH population as much as the environment at larger radii.
This is opposite to what \citet{Laine2010} found in the disk of NGC~4565, where they notice a decrease of the IRAC~8/IRS~11.3~\upmicron from $r\sim 5$ to $r\sim 15~$kpc.
We also look at the 17/7.7 ratio, as a tracer of the global PAH size \citep{RichieHensley2025}, although note that \citet{Maragkoudakis2023} find that the best size estimate comes from the 3.3/11.2 ratio. In our case, their recommendations would suggest using the emission in the 15--20~\upmicron region and the largest wavelength difference for the other band. Given the more robust estimate at 7.7~\upmicron, compared to 6.2~\upmicron, we opt for that former band.
We find overall larger grains (traced by the 17~\upmicron emission) near the bulge of the galaxy.

Overall, our ratios match those from PDRs4All. 
If we focus only on the higher S/N values from \ofive and \oseven, we find that the ionised region (\hii) of the Orion PDR, marked by the open-diamond symbol, agrees best with our positions near the disk in Fig.~\ref{fig:bandratios_comp}. 
That spectrum, however, is not trivial to interpret as the \hii template probes PDRs in the background of the real \hii region in Orion, which induces strong line of sight mixing.

\subsection{Mid-IR features and gas tracers}
\label{sec:pah_gas}
The correlation between mid-IR features with gas tracers, and for our case with \htwo, is well established in the disks of nearby galaxies \citep[e.g.][]{Roussel2007, Hensley2022, Leroy2023, Chown2025PHANGS}, and recent JWST observations found a similar correlation in the CGM \citep{Villanueva2025,Lopez2026}.
Here we investigate the relation between these features and \htwo emission lines, tracing warm molecular gas, and Ne lines.

To fit the line emission, we define line-free wavelength ranges on either side of the \htwo line and we fit a N-th order polynomial to subtract the underlying continuum emission. In most cases, a first-order polynomial provides a good fit to the underlying continuum. In some cases, when the line is located on a PAH feature, we opt for a second- or third-order polynomial to better reproduce the non-linear behaviour in the underlying dust emission. Then, we keep the parameters of this N-th order polynomial fixed, and we fit a single Gaussian component to model the \htwo line emission. The best-fit parameters for the N-th order polynomial and the Gaussian line profile are determined based on a Non-Linear Least-Squares Fitting minimization routine in Python ({\sc lmfit}). The line flux in each spaxel is calculated as the flux integrated over the best-fit Gaussian profile.

\begin{figure}
    \centering
    \includegraphics[width=\linewidth]{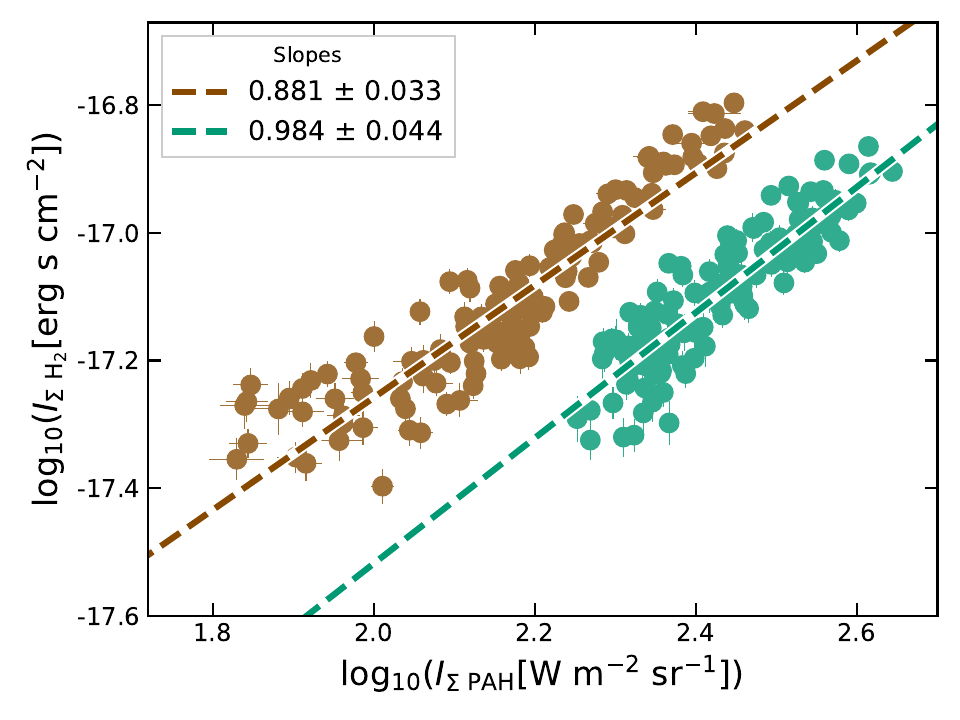}
    \includegraphics[width=\linewidth]{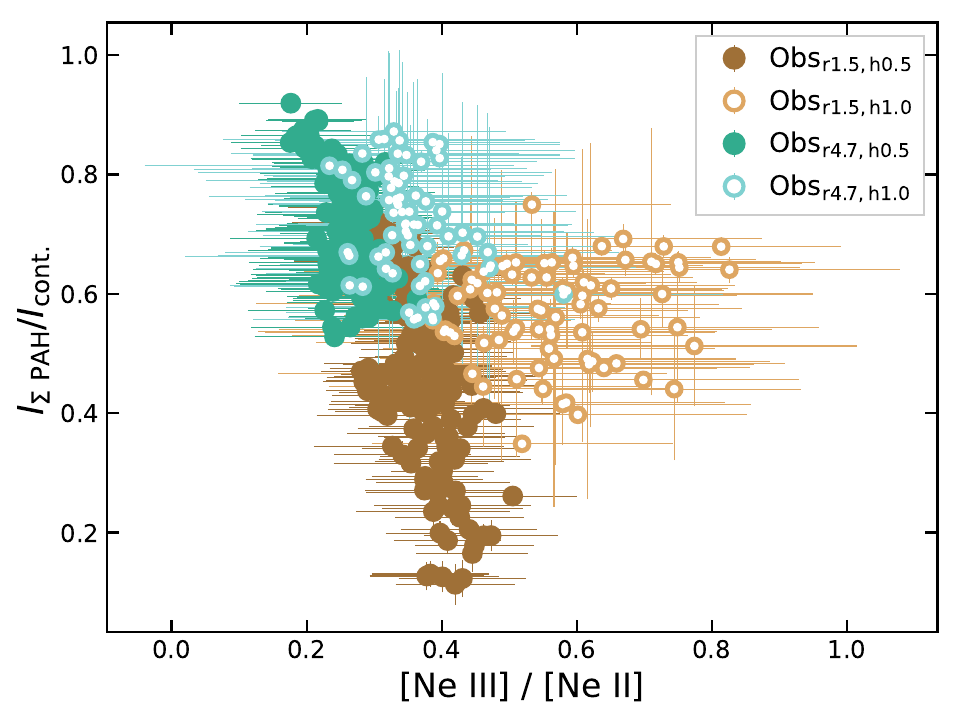}
    \caption{\textit{Top:} Total intensity in the \htwo~$S(1)+S(2)$ lines from Gaussian fitting as a function of the total intensity in the mid-IR features from \pahfit, for \ofive in brown and \oseven in green. The slopes are indicated in the legend. \textit{Bottom:} Total intensity in the mid-IR features normalised to the best-fit dust continuum from \pahfit, as a function of the \neiii/\neii ratio in all four observations. }
    \label{fig:pah_vs_gas}
\end{figure}

In the top panel of Fig.~\ref{fig:pah_vs_gas}, we show the pixel-by-pixel correlation between the total emission in \htwo~$S(1)$ and $S(2)$ as a function of the total emission in mid-IR features, when all features are included (we find similar correlations if we use individual features, without any one standing out). We only use \ofive and \oseven, as the other two pointings have poorly constrained resolved fits, even for the bright \htwo lines. 
We find a tight correlation between both quantities, with slopes of 0.88 and 0.98 for \ofive and \oseven, respectively.
Such a strong correlation implies that the very small grains and molecular gas must be co-spatial in these two regions.
These slopes indicate a ratio that is close to constant between the two species. This is somewhat similar to the linear relation \citet{Roussel2007} found in the SINGS targets using \textit{Spitzer}/IRS measurements, and reported that the mid-IR features and \htwo emission relation is the tightest they find compared with other gas tracers (discussed more in Section~\ref{sec:discussion}).

The bottom panel of Fig.~\ref{fig:pah_vs_gas} shows the relation between the total mid-IR emission, normalised to the best-fit dust continuum, and the \neiii/\neii ratio, considered a tracer of the ionising radiation field hardness \citep[][]{Madden2006, Lai2025}. We clearly recover a similar trend as the one seen in literature, that is the decrease of PAH emission with increasing radiation field hardness, in \ofive,\oseven, and \oeight.
The behaviour in \osix is less clear, likely linked to the low S/N. In \osix, the plateau-like trend could be a sensitivity limit in the detection of the features, at the edge of continuum, features, and noise confusion. 

The negative slope hints at a destruction mechanism where the grains do not survive in a hard radiation field environment. 
This could be due to photo-destruction, or the sputtering of these grains in a more collisionally excited hot gas (as traced by higher ionisation lines).

\subsection{Mid-IR features and continuum}
A recurrent question that is being brought up in the recent JWST literature is the contribution of starlight, hot dust continuum, and PAH features in the MIRI filters. Several works have already addressed this question \citep[e.g.,][L.~Hands et al. in prep]{Sutter2024, Donnelly2025}.
In the Orion PDR, \citet[][]{Chown2025PDRs}, found that the continuum contributes to the F770W up to 60\% in the ionised region, with a sharp drop to $\sim 45\%$ in the atomic phase, and an increase in the dissociation fronts ($\sim 50\%$).

\begin{figure}
    \centering
    \includegraphics[width=1\linewidth]{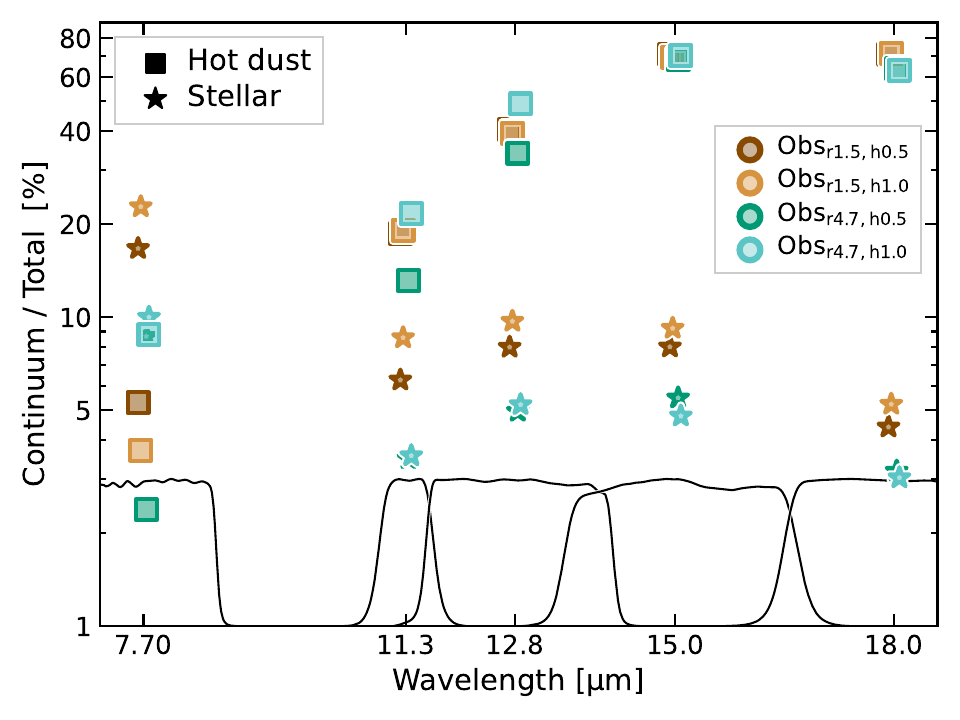}
    \caption{Contribution of the stellar and the hot dust continua in a few MIRI filters. For each 1D MRS spectrum we use the \pahfit best-fit of the stellar and hot dust components and integrate these continua in the F770W, F1130W, F1280W, F1500W, and F1800W transmission curves (black curves; arbitrary normalisation). Points are shown with horizontal offsets for clarity.
    }
    \label{fig:continuumRatios}
\end{figure}

In our case, besides being focused on an edge-on galaxy, the observations are pointed just off the mid-plane, which probes the ISM with a different kind of line of sight confusion, without including the thin disk component.
With \pahfit, we can extract the different best-fit components of the continuum in each observation, and measure the relative contribution of PAH features, stars, and hot dust emission. 
In Fig.~\ref{fig:continuumRatios}, we show the ratio of stellar-to-total and hot dust-to-total emission in the MIRI F770W, F1130W, F1280W, F1500W, and F1800W filters. 
Only in F770W does the starlight contribute to the continuum more than the three modified blackbodies accounting for hot dust. Although it covers the bright 7.7~\upmicron feature, the stellar emission is up to $\sim 20\%$ out to only 500~pc from the disk. Photometry analysis combining NIRCam and MIRI imaging of the disk however points at a much lower contribution in the disk itself, indicating a steep change in the contribution of the stellar continuum with distance from the disk, $h$ (I.~Faber et al., in prep).
As expected, this contribution decreases with wavelength, while the hot dust emission has the (also expected) opposite behaviour.
We also see that the observation with the lowest input from starlight is \oeight, the ``farthest'' both in galactocentric radius and distance from the mid-plane. \osix always shows the highest stellar contribution, which could be explained as an observation pointing at an ``in-between'' location: closer to the bulge than \oseven and \oeight, i.e. more stars, but further away from the mid-plane than \ofive, i.e. less dust.  

We also see that in F770W, F1130W, and F1280W (except for \oeight), mid-IR emission features contribute more than the continuum to the MIRI filters. The first two are expected results, as it is where we find the brightest mid-IR features. 
The relative contributions of the dust continuum in F1280W range between 35 and 50\% for all observations. This filter might be considered at the transition regime where (i) the hot dust continuum becomes more important as wavelength increases, and (ii) the mid-IR emission/dust continuum is at a balance between both components. That is, the moderate brightness of the 12 and 12.7~\upmicron features is supplemented by a relatively strong mid-IR-continuum between 10 and 15~\upmicron, identified in past work \citep[e.g.][]{Peeters2017, Maragkoudakis2023}. We do not model this plateau independently in this work (usually modelled with splines); as such, the ratio of PAH feature-to-dust continuum is likely increased in F1280W because this plateau is included in the numerator. 
As expected, the last two filters used here, F1500W and F1800W, show a much stronger dust continuum than PAH emission. The 17~\upmicron region, with several features, is mostly encompassed within F1800W but the 16.4~\upmicron sits in F1500W. The combined effects of splitting the complex in two features and a stronger dust continuum is likely why the continuum to total emission is close to 80\% in these filters.

Despite a strong stellar continuum in F770W, the main contribution in the MIRI filter remains that of mid-IR emission. This argues in favour of previous claims that the emission detected in F770W is mainly coming from the 7.7~\upmicron feature \citep[namely in][]{GO2180}. 
Previous works have found a stellar disk on average smaller than the dust disk \cite[][]{Xilouris1998, Bianchi2011, Fraternali2011, Bocchio2016}, outside of radii close to the bulge.
This supports the argument that the filamentary structures out to 4~kpc observed in the F770W deep observation of the CGM are mostly PAH emission features. Combined with the results in Section~\ref{sec:pah_gas}, this argues that the mid-IR features can be used as a high resolution tracer of cool material entrained in galactic winds \citep[][]{Girichidis2016, Richie2026}.

\section{Discussion}
\label{sec:discussion}
\subsection{Dust processing in the CGM}
We find variations in the properties of PAHs in the four positions studied in this paper. 
Several scenarios can induce the observed decrease in PAH emission with increasing distance from the disk (Obs$_{\rm r1.5,h0.5\to1.0}$ and Obs$_{\rm r4.7,h0.5\to1.0}$). 
First, it could be a decrease in overall surface density with increasing height. In that case, we would see a weakening of the PAH features even at fixed fractional abundance (with respect to the dust continuum) and radiation field. However, we find varying $I_{\rm \Sigma~PAH} / I_{\rm dust~continuum}$ between the four positions. Overall, \ofive and \osix show lower ratios than \oseven and \oeight. Additionally, we find a 20\% decrease from \oseven to \oeight (much lower between \ofive and \osix), indicating a preferential decrease of PAH emission compared to dust continuum for $h = 0.5$~kpc $\to h=1.0$~kpc, at this radius. 

Second, we could be probing a regime of PAH destruction.
The decreasing 17/7.7 ratio from $h\sim0.5 \to 1~$kpc potentially indicates shattering or sputtering of the larger grains emitting at 17~\upmicron (Fig.~\ref{fig:bandratios_comp}). 
Assuming that grains were launched from the disk and that two pointings at a similar radius at 0.5 and 1~kpc away from the plane are giving us snapshots of a time sequence, probing a decreasing 17/7.7 suggests grains farther from the disk have gotten smaller. Such a shift in the size distribution can be explained by shattering or sputtering, with both mechanisms affecting the grain size similarly. 
If we look at the overall PAH emission however, as the bottom panel in Fig.~\ref{fig:pah_vs_gas} shows, the mid-IR feature total intensity normalised to the continuum does not vary much between \oseven and \oeight despite a shift in \neiii/\neii, suggesting that the destruction of nanoparticles is not dominant, or that both PAHs and (non-PAHs) grains are subject to the same destruction rates (comparing \ofive and \osix is more difficult given the peculiar PAHs / dust~continuum values of \osix, potentially a sensitivity issue).
Related to size distribution, gas outflows and radiation pressure would also cause an effect on the 17/7.7 ratio similar to the one observed. In that scenario, it is not necessarily destruction of grains that lead to the varying size-tracing ratio. 

Finally, the lower emission with increasing height could be explained by the fact that the grains are exposed to a radiation field with decreasing intensity (following the vertical profile of a stellar disk), lowering mid-IR feature emission. This could also explain the varying 17/7.7 ratio, as shown in Fig.~6 of \citet{RichieHensley2025}, which shows that the ratio also depends on the radiation field properties and not (only) the grain properties.

The evolution of PAHs can be placed in a broader theoretical context by recent hydrodynamical simulations that explicitly follow the full grain size distribution of dust. In particular, \citet{Matsumoto2026} used the {\sc gadget4-osaka} framework to evolve a Milky~Way-like galaxy while tracking dust growth, destruction, shattering, and coagulation for grains with radius $3\times10^{-4}$~\upmicron$\leq a \leq 10$~\upmicron. 
They find that small grains ($a \leq 0.01$~\upmicron), including PAHs, are preferentially produced in low density, turbulent environments through grain shattering. In their study, this happens in these outer, diffuse regions of the disk; we can expect the same processes to affect the grain size distribution in extraplanar sightlines, where densities are low and turbulent velocities remain high. 
The simulations also indicate that the dust injected by supernovae into the CGM consist of large grains ($a \sim 0.1-0.3$~\upmicron), more resilient to sputtering. These large grains can survive transport out of the plane and subsequently undergo shattering. This ``survive then shatter'' pathway provides a plausible mechanism for maintaining a small grain population in the CGM. Alternatively, \citet{Richie2026} show that it is possible to transport PAH-sized grains formed in the ISM to large scales through cool clouds.

We find clearer signs of grain destruction as a function of galactocentric radius, as shown by the correlation between the radiation field hardness and feature emission for Obs$_{\rm r1.5\to4.7,h0.5}$. 
For example, \citet{Madden2006} shows a clear anti-correlation between the \neiii/\neii ratio and the ratio of PAH to small grains intensity in galaxies with varying morphologies \citep[][]{Madden2005}. That correlation is often considered a clue of grain destruction in a harder radiation field \citep[e.g.][]{Joblin1996, Chown2024, Lai2025}.
Signs of a varying global ionisation of the PAH population is more difficult to probe.
Indeed, we find a tentative positive trend of the 11.2/7.7 ratio with \neiii/\neii. This is contrary to the assumption that harder radiation fields would contribute to ionising the PAH population, increasing the emission at 7.7~\upmicron. 
However, this aligns with recent conclusions from \citet{Baron2024} who find a similar positive trend with  [\ion{S}{ii}]/\halpha in 19~PHANGS-JWST galaxies. This trend hints at the 7.7/11.2 ratio being a more effective tracer of the non-ionising radiation field spectrum rather than a tracer of the grains properties. 
In our case, however, the trend is very mild and only found considering all four observations together. Individual pointings (i.e. pixel-by-pixel fits) suggest trends that are difficult to interpret given their uncertainties.
The clear anti-correlation between radiation field hardness and the strength of mid-IR features is consistent with the physical picture emerging from the simulations of \citet{Matsumoto2026}, where the abundance of very small grains is found to be very sensitive to the local radiation and thermal environment. This would indicate the pointing closest to the bulge probe a region where the UV field is harder, consistent with the conditions where \citet{Matsumoto2026} predict rapid PAH destruction, or reduced contrast against the continuum. The modest increase of 17/7.7, if real, would indicate preferential destruction of the smaller PAH population near the bulge.

Finally, there are also indications of grain survival in the inner CGM.
The tight correlation between warm molecular gas and dust features seen in the top panel of Fig.~\ref{fig:pah_vs_gas} suggests that these two species are co-spatial in the (inner) CGM as they are in the ISM of evolved galaxies.
Several studies have found a tight correlation with gas. 
In external galaxies, other works have found (slightly) elevated PAH fractions---estimated from SED fitting (``$q_{\rm PAH}$'')---in molecular clouds \citep[e.g.][]{Sandstrom2010, Chastenet2019}, or provide scaling relation between CO and mid-IR emission, especially in the JWST era \citep[e.g.][]{Leroy2023, Chown2025Dwarfs, Chown2025PHANGS}.
The correlation of dust emission and molecular gas is expected even in the CGM based on simulation work \citep{Matsumoto2026, Richie2026}, and seen in M82 \citep{Villanueva2025, Lopez2026}, and we investigate this relation in the following section.

\subsection{The structure of dust emission in the CGM}
\begin{figure}
    \centering
    \includegraphics[width=1.0\linewidth]{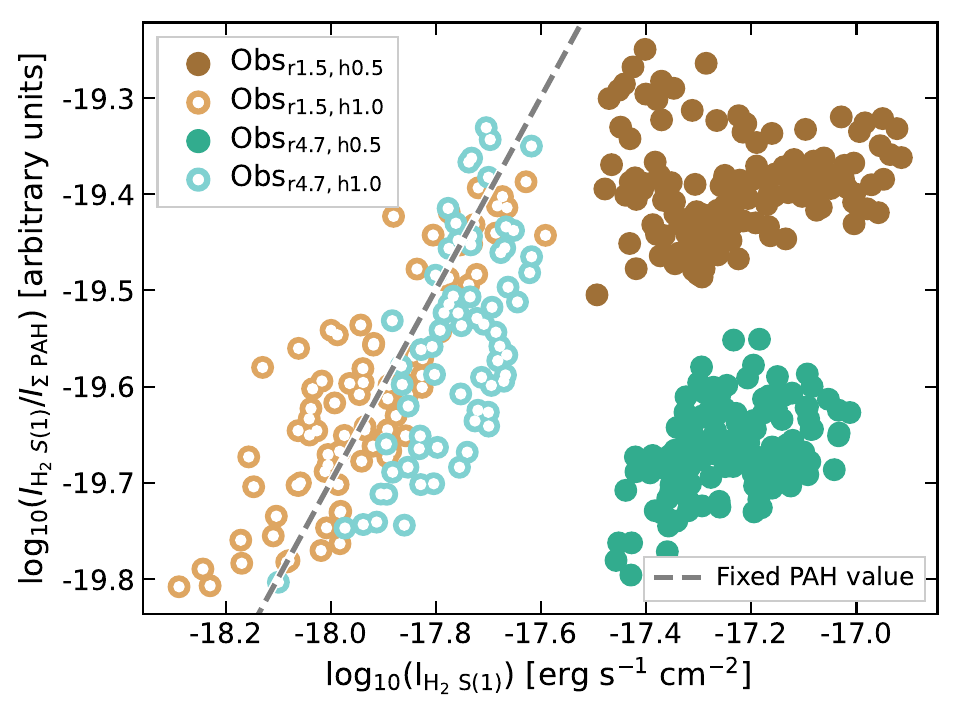}
    \caption{\htwo~$S(1)$ emission to total intensity in PAH features as a function of \htwo~$S(1)$ emission.}
    \label{fig:H2_H2-PAH}
\end{figure}

In the CGM in general, the spatial origin of the mid-IR features remains unclear. Simulations of cool outflows in the CGM suggest that the detected emission comes from a thin layer at the surface of dense, cold clumps \citep[e.g.][]{Richie2024}. It is broadly agreed that dust grains' survival is increased (and even only possible) in cold media \citep[e.g.][]{Richie2026},
and our results would therefore indicate a similar situation. This invokes a common heating mechanism where UV photons heat dust grains and the \htwo molecules, given the correlation in Fig.~\ref{fig:pah_vs_gas}.
In our case, one could consider the interface between a cold cloudlet and the hot gas of the outflow a similar region to that of a PDR in the disk.
Studying the warm molecular gas in the SINGS sample, \citet{Roussel2007} found that the tight correlation between \htwo and PAHs implies that the heating of \htwo must come from the PDRs in their targets. 
In Fig.~\ref{fig:H2_H2-PAH}, we present the ratio of \htwo~$S(1)$-to-$I_{\rm PAH}$ as a function of \htwo~$S(1)$. 
A flat trend similar to that seen by \citet[][see also \citealt{Haan2011}]{Roussel2007} implies a scaling of PAH content closely following that of warm \htwo. Here we see that we cannot quite consider the trends for \ofive or \oseven as flat, although there is a large scatter. This hints at a not-so-trivial co-evolution of the warm molecular gas and the small dust grains. 
The very clearly rising trend seen for \osix and \oeight implies a constant PAH distribution across the observed field-of-view, for a varying \htwo emission. 

We also point out the clear discrepancy in the top panel of Fig.~\ref{fig:pah_vs_gas}, where the mid-IR emission is higher in \oseven compared to \ofive at the same level of \htwo emission. Although \oseven is pointed towards a bright filament identified in mid-IR, this cannot explain the full picture, as we would expect a similar elevation of \htwo emission if the two species were truly co-spatial. 
The higher \neiii/\neii ratio in \ofive could suggest that grain destruction (sputtering, in a hot gas) leads to lower dust grain emission. However, \htwo molecules would photodissociate before dust grains, and the offset would not present itself as such.
Considering the features come from the surface layer of cold clouds, we could be looking at different stages of cloud evolution. In \oseven, the layer of \htwo would have started to dissociate, leaving grains more exposed and boosting their emission. 
To nuance this scenario, we note that \oseven is pointed towards a filament while \ofive is not (or much more diffuse). This could imply that \oseven has more cool gas in its environment, which could provide more opportunities for dust shielding than the diffuse pointing, favouring PAH survival \citep{Richie2026}. While the \htwo lines we use here probe warm gas, the cooler content would not be measured with our observations while still allowing for enhanced PAH emission within the cloud core.

Low-velocities shocks can also be an efficient heating mechanism for \htwo molecules, and affect the \htwo-to-grains ratio. 
The theoretical work by \citet{KaufmanNeufeld1996} suggest ratios \htwo~$S(1)/S(2) \sim 0.5-0.8$ in shock velocities spanning $20-40$~km~s$^{-1}$ and densities $10^4-10^6$~cm$^{-3}$. In our case, we see ratios of the \htwo~$S(1)/S(2)$ lines well above unity (easily noticeable in Fig.~\ref{fig:all1Dspectra}). This strongly discredits the shock-powered emission scenario as the main heating mechanism.
Diagnostics for X-ray heating of \htwo are limited, and mostly constrained to higher order \htwo lines, other cooling lines (e.g. [\ion{C}{ii}]), or H$^{+}_{3}$, not accessible in this work. 
However, we also point out that the survival rate of PAHs in a X-ray-filled medium is rather short \citep[e.g.][]{BoechatRoberty2009}, making this scenario also unlikely \citep[although note some correlation between PAHs and X-ray found in M82 by][]{Lopez2026}.

We can also consider turbulence as a heating mechanism. This was investigated by \citet{Stacey2010} in the disk of NGC~891 using \textit{Spitzer}/IRS data. In comparison, we lack the \htwo~$S(0)$ emission, although they use the $S(2)/S(1)$ ratio as an indicator of turbulence: in that scenario, the $S(2)$ emission is boosted compared to photoionisation heating. Comparing their measurements with turbulence models, they found that a non-negligible amount of the total emission could come from turbulence as well. We leave the detailed investigation of line emission to a future paper.

Overall, the relation between warm molecular gas and dust grains does not look as straightforward as it might be in the disks of galaxies, where the heating is largely dominated by PDRs. Here, we find that other mechanisms might be in tight competition, and even at different stages of the evolution of the ISM material entrained in the outflows.

\section{Conclusions}
We present new JWST/MIRI-MRS observations of the lower-CGM of NGC~891. These new data consist of four single-pointing positions, two south of the bulge $r \sim 1.5$~kpc and two at galactocentric radius $r \sim 4.7$~kpc, both in pair with a pointing at a distance from the mid-plane $h\sim 0.5$~kpc and one at $h \sim 1$~kpc (Fig.~\ref{fig:mosaic}). Each pointing was observed for $\sim 1.6$~hours.\footnote{The data products discussed in this paper are available upon request to the first or second author. All JWST level-1 products are available through the appropriate portal and the 1D spectrum extraction are available through the PDRs4All website.} 
The two pointings at shorter distance from the disk have higher S/N, and the 1D spectra extracted from the 3D cubes show several strong emission lines, \htwo rotational lines from \htwo~$S(1)$ to $S(5)$, and a few strong other emission lines (Fig.~\ref{fig:all1Dspectra}). In all observations, we identify the common mid-IR features attributed to very small carbonaceous dust grains, referred to as PAHs. 

The peak wavelengths of the identified features do not vary drastically compared to other works in the literature (albeit they have a lower S/N in our case), but we can identify some differences, mostly in the widths of some sub-features. These differences may be a hint of a different level of dust processing compared to the Orion PDR or M51 star-forming regions (Section~\ref{sec:qualitativedescription}, Figs.~\ref{fig:allObsSpectraContSub}, \ref{fig:specs_comp}).

Using \pahfit, we create a model that best reproduces all the features and complexes of the 1D spectra and 3D cubes to extract their relative intensities (Figs.~\ref{fig:pahfit_allObs}, \ref{fig:pahfitcube_allObs}). 
In that process, we report the detection of a new feature at 16.72~\upmicron, that seem to be correlated with the 7.7, 8.6, and/or 17~\upmicron (Fig.~\ref{fig:16.72corr}).
We find that the pointing at higher $r$ and smaller $h$ (\oseven) shows the strongest features (Fig.~\ref{fig:variations_r_h_1d}), although that observation aligns with an identified outflow, probing a dustier environment, and therefore limiting radius-driven variations. 
Variations relative to distance from the mid-plane ($h$) are more tentative, given the lower S/N of the two pointings at $h \sim 1$~kpc, although we can expect that the local environment becomes harsher for PAHs as $h$ increases.

We find some evidence of dust processing in the CGM.
Through band ratios, we find hints of a preferential survival of large grains near the bulge and a more ionised population at outer radius (Fig.~\ref{fig:bandratios_comp}). 
Combined with Ne line measurements, we find a clear anti-correlation between radiation field hardness and mid-IR feature emission (Fig.~\ref{fig:pah_vs_gas}, bottom panel), suggesting the destruction of dust grains with increasing radiation field hardness, even on the limited spatial scales of our observations. This trend does not appear directly linked to a global ionisation of the grain population through the 7.7/11.2 ratio in our data.

We find a tight correlation between the \htwo intensity and the total intensity from PAHs (Fig.~\ref{fig:pah_vs_gas}), indicating that they likely come from similar regions, and are subject to similar heating mechanisms.
However, an offset in that correlation raises questions about these mechanisms and/or destruction processes at play. 
This offset adds nuance to the assumption that the emission features come from the surface layers of molecular cloudlets, and can be used as a direct tracer of warm gas. It could be evidence of different prevalent mechanisms affecting the \htwo gas and the processing of very small carbonaceous grains.

\begin{acknowledgements}
We thank the referee for their detailed overview of the paper and concurrent suggestions to improve the manuscript clarity.
JC thanks David Law for his advice in running the pipeline for MIRI MRS data. JC thanks Jessica Sutter for sending distracting cat and puffin videos during this project.\newline
This work is based on observations made with the NASA/ESA/CSA James Webb Space Telescope. The data were obtained from the Mikulski Archive for Space Telescopes at the Space Telescope Science Institute, which is operated by the Association of Universities for Research in Astronomy, Inc., under NASA contract NAS 5-03127 for JWST. These observations are associated with the JWST Cycle~1 program GO~2180.
This work used observations associated with the JWST Cycle~2 GO~3435.\newline
JC, IDL and LP acknowledge funding from the Belgian Science Policy Office (BELSPO) through the PRODEX project ``JWST/MIRI Science exploitation'' (C4000142239).
FG acknowledges support by the French National Research Agency under the contracts WIDENING (ANR-23-ESDIR-0004) and REDEEMING (ANR-24-CE31-2530), as well as by the Actions Thématiques ``Physique et Chimie du Milieu Interstellaire'' (PCMI) of CNRS/INSU, with INC and INP, and ``Cosmologie et Galaxies'' (ATCG) of CNRS/INSU, with INP and IN2P3, both programs being co-funded by CEA and CNES.
IDL and FK acknowledges funding from the European Research Council (ERC) under the European Union’s Horizon 2020 research and innovation programme DustOrigin (ERC-2019-StG-851622). 
RSK acknowledges financial support from the ERC via Synergy Grant ``ECOGAL'' (project ID 855130) and from the German Excellence Strategy via the Heidelberg Cluster ``STRUCTURES'' (EXC 2181 - 390900948). In addition RSK is grateful for funding from the German Ministry for Economic Affairs and Climate Action in project ``MAINN'' (funding ID 50OO2206), and from DFG and ANR for project ``STARCLUSTERS'' (funding ID KL 1358/22-1). 
SW acknowledges financial support from the German Excellence Strategy via the Cologne-Bonn Cluster of Excellence ``Our Dynamic Unvierse'' (EXC 3037) as well as from the German Science Foundation (DFG) via the Collaborative Research Center 1601 (SFB 1601, project B6). 
JF acknowledges financial support from the DGAPA-PAPIIT project IN102226, Mexico.
\newline
This research has made use of the SIMBAD database, CDS, Strasbourg Astronomical Observatory, France \citep{CDS}.
This research made use of Astropy, a community-developed core Python package for Astronomy \citep{2013A&A...558A..33A,2018AJ....156..123A}.
This research made use of matplotlib, a Python library for publication quality graphics \citep{Hunter:2007}.
This research made use of NumPy \citep{harris2020array}.

\end{acknowledgements}

\bibliographystyle{aa} % style aa.bst
\bibliography{ngc891mrs} % your references Yourfile.bib

@ARTICLE{GO2180,
       author = {{Chastenet}, J{\'e}r{\'e}my and {De Looze}, Ilse and {Rela{\~n}o}, Monica and {Dale}, Daniel A. and {Williams}, Thomas G. and {Bianchi}, Simone and {Xilouris}, Emmanuel M. and {Baes}, Maarten and {Bolatto}, Alberto D. and {Boyer}, Martha L. and {Casasola}, Viviana and {Clark}, Christopher J.~R. and {Fraternali}, Filippo and {Fritz}, Jacopo and {Galliano}, Fr{\'e}d{\'e}ric and {Glover}, Simon C.~O. and {Gordon}, Karl D. and {Hirashita}, Hiroyuki and {Kennicutt}, Robert and {Nagamine}, Kentaro and {Kirchschlager}, Florian and {Klessen}, Ralf S. and {Koch}, Eric W. and {Levy}, Rebecca C. and {McCallum}, Lewis and {Madden}, Suzanne C. and {McLeod}, Anna F. and {Meidt}, Sharon E. and {Mosenkov}, Aleksandr V. and {Richie}, Helena M. and {Saintonge}, Am{\'e}lie and {Sandstrom}, Karin M. and {Schneider}, Evan E. and {Sivkova}, Evgenia E. and {Smith}, J.~D.~T. and {Smith}, Matthew W.~L. and {van der Wel}, Arjen and {Walch}, Stefanie and {Walter}, Fabian and {Wood}, Kenneth},
        title = "{JWST MIRI and NIRCam observations of NGC 891 and its circumgalactic medium}",
      journal = {\aap},
         year = 2024,
        month = oct,
       volume = {690},
          eid = {A348},
        pages = {A348},
          doi = {10.1051/0004-6361/202451033},
archivePrefix = {arXiv},
       eprint = {2408.08026},
 primaryClass = {astro-ph.GA},
       adsurl = {https://ui.adsabs.harvard.edu/abs/2024A&A...690A.348C}
}

@ARTICLE{Smith2007,
       author = {{Smith}, J.~D.~T. and {Draine}, B.~T. and {Dale}, D.~A. and {Moustakas}, J. and {Kennicutt}, R.~C., Jr. and {Helou}, G. and {Armus}, L. and {Roussel}, H. and {Sheth}, K. and {Bendo}, G.~J. and {Buckalew}, B.~A. and {Calzetti}, D. and {Engelbracht}, C.~W. and {Gordon}, K.~D. and {Hollenbach}, D.~J. and {Li}, A. and {Malhotra}, S. and {Murphy}, E.~J. and {Walter}, F.},
        title = "{The Mid-Infrared Spectrum of Star-forming Galaxies: Global Properties of Polycyclic Aromatic Hydrocarbon Emission}",
      journal = {\apj},
         year = 2007,
        month = feb,
       volume = {656},
       number = {2},
        pages = {770-791},
          doi = {10.1086/510549},
archivePrefix = {arXiv},
       eprint = {astro-ph/0610913},
 primaryClass = {astro-ph},
       adsurl = {https://ui.adsabs.harvard.edu/abs/2007ApJ...656..770S}
}

@ARTICLE{Chown2024,
       author = {{Chown}, Ryan and {Sidhu}, Ameek and {Peeters}, Els and {Tielens}, Alexander G.~G.~M. and {Cami}, Jan and {Bern{\'e}}, Olivier and {Habart}, Emilie and {Alarc{\'o}n}, Felipe and {Canin}, Am{\'e}lie and {Schroetter}, Ilane and {Trahin}, Boris and {Van De Putte}, Dries and {Abergel}, Alain and {Bergin}, Edwin A. and {Bernard-Salas}, Jeronimo and {Boersma}, Christiaan and {Bron}, Emeric and {Cuadrado}, Sara and {Dartois}, Emmanuel and {Dicken}, Daniel and {El-Yajouri}, Meriem and {Fuente}, Asunci{\'o}n and {Goicoechea}, Javier R. and {Gordon}, Karl D. and {Issa}, Lina and {Joblin}, Christine and {Kannavou}, Olga and {Khan}, Baria and {Lacinbala}, Ozan and {Languignon}, David and {Le Gal}, Romane and {Maragkoudakis}, Alexandros and {Meshaka}, Raphael and {Okada}, Yoko and {Onaka}, Takashi and {Pasquini}, Sofia and {Pound}, Marc W. and {Robberto}, Massimo and {R{\"o}llig}, Markus and {Schefter}, Bethany and {Schirmer}, Thi{\'e}baut and {Vicente}, S{\'\i}lvia and {Wolfire}, Mark G. and {Zannese}, Marion and {Aleman}, Isabel and {Allamandola}, Louis and {Auchettl}, Rebecca and {Baratta}, Giuseppe Antonio and {Bejaoui}, Salma and {Bera}, Partha P. and {Black}, John H. and {Boulanger}, Fran{\c{c}}ois and {Bouwman}, Jordy and {Brandl}, Bernhard and {Brechignac}, Philippe and {Br{\"u}nken}, Sandra and {Buragohain}, Mridusmita and {Burkhardt}, Andrew and {Candian}, Alessandra and {Cazaux}, St{\'e}phanie and {Cernicharo}, Jose and {Chabot}, Marin and {Chakraborty}, Shubhadip and {Champion}, Jason and {Colgan}, Sean W.~J. and {Cooke}, Ilsa R. and {Coutens}, Audrey and {Cox}, Nick L.~J. and {Demyk}, Karine and {Meyer}, Jennifer Donovan and {Foschino}, Sacha and {Garc{\'\i}a-Lario}, Pedro and {Gavilan}, Lisseth and {Gerin}, Maryvonne and {Gottlieb}, Carl A. and {Guillard}, Pierre and {Gusdorf}, Antoine and {Hartigan}, Patrick and {He}, Jinhua and {Herbst}, Eric and {Hornekaer}, Liv and {J{\"a}ger}, Cornelia and {Janot-Pacheco}, Eduardo and {Kaufman}, Michael and {Kemper}, Francisca and {Kendrew}, Sarah and {Kirsanova}, Maria S. and {Klaassen}, Pamela and {Kwok}, Sun and {Labiano}, {\'A}lvaro and {Lai}, Thomas S. -Y. and {Lee}, Timothy J. and {Lefloch}, Bertrand and {Le Petit}, Franck and {Li}, Aigen and {Linz}, Hendrik and {Mackie}, Cameron J. and {Madden}, Suzanne C. and {Mascetti}, Jo{\"e}lle and {McGuire}, Brett A. and {Merino}, Pablo and {Micelotta}, Elisabetta R. and {Misselt}, Karl and {Morse}, Jon A. and {Mulas}, Giacomo and {Neelamkodan}, Naslim and {Ohsawa}, Ryou and {Omont}, Alain and {Paladini}, Roberta and {Palumbo}, Maria Elisabetta and {Pathak}, Amit and {Pendleton}, Yvonne J. and {Petrignani}, Annemieke and {Pino}, Thomas and {Puga}, Elena and {Rangwala}, Naseem and {Rapacioli}, Mathias and {Ricca}, Alessandra and {Roman-Duval}, Julia and {Roser}, Joseph and {Roueff}, Evelyne and {Rouill{\'e}}, Ga{\"e}l and {Salama}, Farid and {Sales}, Dinalva A. and {Sandstrom}, Karin and {Sarre}, Peter and {Sciamma-O'Brien}, Ella and {Sellgren}, Kris and {Shenoy}, Sachindev S. and {Teyssier}, David and {Thomas}, Richard D. and {Togi}, Aditya and {Verstraete}, Laurent and {Witt}, Adolf N. and {Wootten}, Alwyn and {Zettergren}, Henning and {Zhang}, Yong and {Zhang}, Ziwei E. and {Zhen}, Junfeng},
        title = "{PDRs4All. IV. An embarrassment of riches: Aromatic infrared bands in the Orion Bar}",
      journal = {\aap},
         year = 2024,
        month = may,
       volume = {685},
          eid = {A75},
        pages = {A75},
          doi = {10.1051/0004-6361/202346662},
archivePrefix = {arXiv},
       eprint = {2308.16733},
 primaryClass = {astro-ph.GA},
       adsurl = {https://ui.adsabs.harvard.edu/abs/2024A&A...685A..75C}
}

@ARTICLE{vdPutte2024,
       author = {{Van De Putte}, Dries and {Meshaka}, Raphael and {Trahin}, Boris and {Habart}, Emilie and {Peeters}, Els and {Bern{\'e}}, Olivier and {Alarc{\'o}n}, Felipe and {Canin}, Am{\'e}lie and {Chown}, Ryan and {Schroetter}, Ilane and {Sidhu}, Ameek and {Boersma}, Christiaan and {Bron}, Emeric and {Dartois}, Emmanuel and {Goicoechea}, Javier R. and {Gordon}, Karl D. and {Onaka}, Takashi and {Tielens}, Alexander G.~G.~M. and {Verstraete}, Laurent and {Wolfire}, Mark G. and {Abergel}, Alain and {Bergin}, Edwin A. and {Bernard-Salas}, Jeronimo and {Cami}, Jan and {Cuadrado}, Sara and {Dicken}, Daniel and {Elyajouri}, Meriem and {Fuente}, Asunci{\'o}n and {Joblin}, Christine and {Khan}, Baria and {Lacinbala}, Ozan and {Languignon}, David and {Le Gal}, Romane and {Maragkoudakis}, Alexandros and {Okada}, Yoko and {Pasquini}, Sofia and {Pound}, Marc W. and {Robberto}, Massimo and {R{\"o}llig}, Markus and {Schefter}, Bethany and {Schirmer}, Thi{\'e}baut and {Tabone}, Benoit and {Vicente}, S{\'\i}lvia and {Zannese}, Marion and {Colgan}, Sean W.~J. and {He}, Jinhua and {Rouill{\'e}}, Ga{\"e}l and {Togi}, Aditya and {Aleman}, Isabel and {Auchettl}, Rebecca and {Baratta}, Giuseppe Antonio and {Bejaoui}, Salma and {Bera}, Partha P. and {Black}, John H. and {Boulanger}, Francois and {Bouwman}, Jordy and {Brandl}, Bernhard and {Brechignac}, Philippe and {Br{\"u}nken}, Sandra and {Buragohain}, Mridusmita and {Burkhardt}, Andrew and {Candian}, Alessandra and {Cazaux}, St{\'e}phanie and {Cernicharo}, Jose and {Chabot}, Marin and {Chakraborty}, Shubhadip and {Champion}, Jason and {Cooke}, Ilsa R. and {Coutens}, Audrey and {Cox}, Nick L.~J. and {Demyk}, Karine and {Meyer}, Jennifer Donovan and {Foschino}, Sacha and {Garc{\'\i}a-Lario}, Pedro and {Gerin}, Maryvonne and {Gottlieb}, Carl A. and {Guillard}, Pierre and {Gusdorf}, Antoine and {Hartigan}, Patrick and {Herbst}, Eric and {Hornekaer}, Liv and {Issa}, Lina and {J{\"a}ger}, Cornelia and {Janot-Pacheco}, Eduardo and {Kannavou}, Olga and {Kaufman}, Michael and {Kemper}, Francisca and {Kendrew}, Sarah and {Kirsanova}, Maria S. and {Klaassen}, Pamela and {Kwok}, Sun and {Labiano}, {\'A}lvaro and {Lai}, Thomas S. -Y. and {Le Floch}, Bertrand and {Le Petit}, Franck and {Li}, Aigen and {Linz}, Hendrik and {Mackie}, Cameron J. and {Madden}, Suzanne C. and {Mascetti}, Jo{\"e}lle and {McGuire}, Brett A. and {Merino}, Pablo and {Micelotta}, Elisabetta R. and {Morse}, Jon A. and {Mulas}, Giacomo and {Neelamkodan}, Naslim and {Ohsawa}, Ryou and {Omont}, Alain and {Paladini}, Roberta and {Palumbo}, Maria Elisabetta and {Pathak}, Amit and {Pendleton}, Yvonne J. and {Petrignani}, Annemieke and {Pino}, Thomas and {Puga}, Elena and {Rangwala}, Naseem and {Rapacioli}, Mathias and {Rho}, Jeonghee and {Ricca}, Alessandra and {Roman-Duval}, Julia and {Roser}, Joseph and {Roueff}, Evelyne and {Salama}, Farid and {Sales}, Dinalva A. and {Sandstrom}, Karin and {Sarre}, Peter and {Sciamma-O'Brien}, Ella and {Sellgren}, Kris and {Shenoy}, Sachindev S. and {Teyssier}, David and {Thomas}, Richard D. and {Witt}, Adolf N. and {Wootten}, Alwyn and {Ysard}, Nathalie and {Zettergren}, Henning and {Zhang}, Yong and {Zhang}, Ziwei E. and {Zhen}, Junfeng},
        title = "{PDRs4All. VIII. Mid-infrared emission line inventory of the Orion Bar}",
      journal = {\aap},
         year = 2024,
        month = jul,
       volume = {687},
          eid = {A86},
        pages = {A86},
          doi = {10.1051/0004-6361/202449295},
archivePrefix = {arXiv},
       eprint = {2404.03111},
 primaryClass = {astro-ph.GA},
       adsurl = {https://ui.adsabs.harvard.edu/abs/2024A&A...687A..86V}
}

@ARTICLE{vdPutte2025,
       author = {{Van De Putte}, Dries and {Peeters}, Els and {Gordon}, Karl D. and {Smith}, John-David T. and {Lai}, Thomas S.-Y. and {Maragkoudakis}, Alexandros and {Schefter}, Bethany and {Sidhu}, Ameek and {Doshi}, Dhruvil and {Bern{\'e}}, Olivier and {Cami}, Jan and {Boersma}, Christiaan and {Dartois}, Emmanuel and {Habart}, Emilie and {Onaka}, Takashi and {Tielens}, Alexander G.~G.~M.},
        title = "{PDRs4All: XVI. Tracing aromatic infrared band characteristics in photodissociation region spectra with PAHFIT in the JWST era}",
      journal = {\aap},
         year = 2025,
        month = sep,
       volume = {701},
          eid = {A111},
        pages = {A111},
          doi = {10.1051/0004-6361/202554991},
archivePrefix = {arXiv},
       eprint = {2507.05848},
 primaryClass = {astro-ph.GA},
       adsurl = {https://ui.adsabs.harvard.edu/abs/2025A&A...701A.111V}
}

@ARTICLE{IrwinMadden2006,
       author = {{Irwin}, J.~A. and {Madden}, S.~C.},
        title = "{Discovery of PAHs in the halo of NGC 5907}",
      journal = {\aap},
         year = 2006,
        month = jan,
       volume = {445},
       number = {1},
        pages = {123-141},
          doi = {10.1051/0004-6361:20053233},
archivePrefix = {arXiv},
       eprint = {astro-ph/0509726},
 primaryClass = {astro-ph},
       adsurl = {https://ui.adsabs.harvard.edu/abs/2006A&A...445..123I}
}

@ARTICLE{Irwin2007,
       author = {{Irwin}, J.~A. and {Kennedy}, H. and {Parkin}, T. and {Madden}, S.},
        title = "{PAHs in the halo of NGC 5529}",
      journal = {\aap},
         year = 2007,
        month = nov,
       volume = {474},
       number = {2},
        pages = {461-472},
          doi = {10.1051/0004-6361:20077729},
archivePrefix = {arXiv},
       eprint = {0708.3808},
 primaryClass = {astro-ph},
       adsurl = {https://ui.adsabs.harvard.edu/abs/2007A&A...474..461I}
}

@ARTICLE{Lee2001,
       author = {{Lee}, S. -W. and {Irwin}, J.~A. and {Dettmar}, R. -J. and {Cunningham}, C.~T. and {Golla}, G. and {Wang}, Q.~D.},
        title = "{NGC 5775: Anatomy of a disk-halo interface}",
      journal = {\aap},
         year = 2001,
        month = oct,
       volume = {377},
        pages = {759-777},
          doi = {10.1051/0004-6361:20011046},
archivePrefix = {arXiv},
       eprint = {astro-ph/0108510},
 primaryClass = {astro-ph},
       adsurl = {https://ui.adsabs.harvard.edu/abs/2001A&A...377..759L}
}

@ARTICLE{Bolatto2024,
       author = {{Bolatto}, Alberto D. and {Levy}, Rebecca C. and {Tarantino}, Elizabeth and {Boyer}, Martha L. and {Fisher}, Deanne B. and {Cronin}, Serena A. and {Leroy}, Adam K. and {Klessen}, Ralf S. and {Smith}, J.~D. and {Berg}, Danielle A. and {B{\"o}ker}, Torsten and {Boogaard}, Leindert A. and {Ostriker}, Eve C. and {Thompson}, Todd A. and {Ott}, Juergen and {Lenki{\'c}}, Laura and {Lopez}, Laura A. and {Dale}, Daniel A. and {Veilleux}, Sylvain and {van der Werf}, Paul P. and {Glover}, Simon C.~O. and {Sandstrom}, Karin M. and {Skillman}, Evan D. and {Chisholm}, John and {Villanueva}, Vicente and {Lai}, Thomas S. -Y. and {Lopez}, Sebastian and {Mills}, Elisabeth A.~C. and {Emig}, Kimberly L. and {Armus}, Lee and {Mayya}, Divakara and {Meier}, David S. and {De Looze}, Ilse and {Herrera-Camus}, Rodrigo and {Walter}, Fabian and {Rela{\~n}o}, M{\'o}nica and {Koziol}, Hannah B. and {Marvil}, Joshua and {Jim{\'e}nez-Donaire}, Mar{\'\i}a J. and {Martini}, Paul},
        title = "{JWST Observations of Starbursts: Polycyclic Aromatic Hydrocarbon Emission at the Base of the M82 Galactic Wind}",
      journal = {\apj},
         year = 2024,
        month = may,
       volume = {967},
       number = {1},
          eid = {63},
        pages = {63},
          doi = {10.3847/1538-4357/ad33c8},
       adsurl = {https://ui.adsabs.harvard.edu/abs/2024ApJ...967...63B}
}

@ARTICLE{Fisher2025,
       author = {{Fisher}, Deanne B. and {Bolatto}, Alberto D. and {Chisholm}, John and {Fielding}, Drummond and {Levy}, Rebecca C. and {Tarantino}, Elizabeth and {Boyer}, Martha L. and {Cronin}, Serena A. and {Lopez}, Laura A. and {Smith}, J.~D. and {Berg}, Danielle A. and {Lopez}, Sebastian and {Veilleux}, Sylvain and {van der Werf}, Paul P. and {B{\"o}ker}, Torsten and {Boogaard}, Leindert A. and {Lenki{\'c}}, Laura and {Glover}, Simon C.~O. and {Villanueva}, Vicente and {Mayya}, Divakara and {Lai}, Thomas S. -Y. and {Dale}, Daniel A. and {Emig}, Kimberly L. and {Walter}, Fabian and {Rela{\~n}o}, Monica and {De Looze}, Ilse and {Mills}, Elisabeth A.~C. and {Leroy}, Adam K. and {Meier}, David S. and {Herrera-Camus}, Rodrigo and {Klessen}, Ralf S.},
        title = "{JWST observations of starbursts: cold clouds and plumes launching in the M 82 outflow}",
      journal = {\mnras},
         year = 2025,
        month = apr,
       volume = {538},
       number = {4},
        pages = {3068-3083},
          doi = {10.1093/mnras/staf363},
archivePrefix = {arXiv},
       eprint = {2405.03686},
 primaryClass = {astro-ph.GA},
       adsurl = {https://ui.adsabs.harvard.edu/abs/2025MNRAS.538.3068F}
}

@ARTICLE{THEMIS,
       author = {{Jones}, A.~P. and {K{\"o}hler}, M. and {Ysard}, N. and {Bocchio}, M. and {Verstraete}, L.},
        title = "{The global dust modelling framework THEMIS}",
      journal = {\aap},
         year = 2017,
        month = jun,
       volume = {602},
          eid = {A46},
        pages = {A46},
          doi = {10.1051/0004-6361/201630225},
archivePrefix = {arXiv},
       eprint = {1703.00775},
 primaryClass = {astro-ph.GA},
       adsurl = {https://ui.adsabs.harvard.edu/abs/2017A&A...602A..46J}
}

@ARTICLE{Menard2010,
       author = {{M{\'e}nard}, Brice and {Scranton}, Ryan and {Fukugita}, Masataka and {Richards}, Gordon},
        title = "{Measuring the galaxy-mass and galaxy-dust correlations through magnification and reddening}",
      journal = {\mnras},
         year = 2010,
        month = jun,
       volume = {405},
       number = {2},
        pages = {1025-1039},
          doi = {10.1111/j.1365-2966.2010.16486.x},
archivePrefix = {arXiv},
       eprint = {0902.4240},
 primaryClass = {astro-ph.CO},
       adsurl = {https://ui.adsabs.harvard.edu/abs/2010MNRAS.405.1025M}
}

@ARTICLE{HD2023,
       author = {{Hensley}, Brandon S. and {Draine}, B.~T.},
        title = "{The Astrodust+PAH Model: A Unified Description of the Extinction, Emission, and Polarization from Dust in the Diffuse Interstellar Medium}",
      journal = {\apj},
         year = 2023,
        month = may,
       volume = {948},
       number = {1},
          eid = {55},
        pages = {55},
          doi = {10.3847/1538-4357/acc4c2},
archivePrefix = {arXiv},
       eprint = {2208.12365},
 primaryClass = {astro-ph.GA},
       adsurl = {https://ui.adsabs.harvard.edu/abs/2023ApJ...948...55H}
}

@ARTICLE{DL2007,
       author = {{Draine}, B.~T. and {Li}, Aigen},
        title = "{Infrared Emission from Interstellar Dust. IV. The Silicate-Graphite-PAH Model in the Post-Spitzer Era}",
      journal = {\apj},
         year = 2007,
        month = mar,
       volume = {657},
       number = {2},
        pages = {810-837},
          doi = {10.1086/511055},
archivePrefix = {arXiv},
       eprint = {astro-ph/0608003},
 primaryClass = {astro-ph},
       adsurl = {https://ui.adsabs.harvard.edu/abs/2007ApJ...657..810D}
}

@ARTICLE{Chastenet2019,
       author = {{Chastenet}, J{\'e}r{\'e}my and {Sandstrom}, Karin and {Chiang}, I-Da and {Leroy}, Adam K. and {Utomo}, Dyas and {Bot}, Caroline and {Gordon}, Karl D. and {Draine}, Bruce T. and {Fukui}, Yasuo and {Onishi}, Toshikazu and {Tsuge}, Kisetsu},
        title = "{The Polycyclic Aromatic Hydrocarbon Mass Fraction on a 10 pc Scale in the Magellanic Clouds}",
      journal = {\apj},
         year = 2019,
        month = may,
       volume = {876},
       number = {1},
          eid = {62},
        pages = {62},
          doi = {10.3847/1538-4357/ab16cf},
archivePrefix = {arXiv},
       eprint = {1904.02705},
 primaryClass = {astro-ph.GA},
       adsurl = {https://ui.adsabs.harvard.edu/abs/2019ApJ...876...62C}
}

@ARTICLE{Sutter2024,
       author = {{Sutter}, Jessica and {Sandstrom}, Karin and {Chastenet}, J{\'e}r{\'e}my and {Leroy}, Adam K. and {Koch}, Eric W. and {Williams}, Thomas G. and {Chown}, Ryan and {Belfiore}, Francesco and {Bigiel}, Frank and {Boquien}, M{\'e}d{\'e}ric and {Cao}, Yixian and {Chevance}, M{\'e}lanie and {Dale}, Daniel A. and {Egorov}, Oleg V. and {Glover}, Simon C.~O. and {Groves}, Brent and {Klessen}, Ralf S. and {Kreckel}, Kathryn and {Larson}, Kirsten L. and {Oakes}, Elias K. and {Pathak}, Debosmita and {Ramambason}, Lise and {Rosolowsky}, Erik and {Watkins}, Elizabeth J.},
        title = "{The Fraction of Dust Mass in the Form of Polycyclic Aromatic Hydrocarbons on 10{\textendash}50 pc Scales in Nearby Galaxies}",
      journal = {\apj},
         year = 2024,
        month = aug,
       volume = {971},
       number = {2},
          eid = {178},
        pages = {178},
          doi = {10.3847/1538-4357/ad54bd},
archivePrefix = {arXiv},
       eprint = {2405.15102},
 primaryClass = {astro-ph.GA},
       adsurl = {https://ui.adsabs.harvard.edu/abs/2024ApJ...971..178S}
}

@ARTICLE{Egorov2023,
       author = {{Egorov}, Oleg V. and {Kreckel}, Kathryn and {Sandstrom}, Karin M. and {Leroy}, Adam K. and {Glover}, Simon C.~O. and {Groves}, Brent and {Kruijssen}, J.~M. Diederik and {Barnes}, Ashley. T. and {Belfiore}, Francesco and {Bigiel}, F. and {Blanc}, Guillermo A. and {Boquien}, M{\'e}d{\'e}ric and {Cao}, Yixian and {Chastenet}, J{\'e}r{\'e}my and {Chevance}, M{\'e}lanie and {Congiu}, Enrico and {Dale}, Daniel A. and {Emsellem}, Eric and {Grasha}, Kathryn and {Klessen}, Ralf S. and {Larson}, Kirsten L. and {Liu}, Daizhong and {Murphy}, Eric J. and {Pan}, Hsi-An and {Pessa}, Ismael and {Pety}, J{\'e}r{\^o}me and {Rosolowsky}, Erik and {Scheuermann}, Fabian and {Schinnerer}, Eva and {Sutter}, Jessica and {Thilker}, David A. and {Watkins}, Elizabeth J. and {Williams}, Thomas G.},
        title = "{PHANGS-JWST First Results: Destruction of the PAH Molecules in H II Regions Probed by JWST and MUSE}",
      journal = {\apjl},
         year = 2023,
        month = feb,
       volume = {944},
       number = {2},
          eid = {L16},
        pages = {L16},
          doi = {10.3847/2041-8213/acac92},
archivePrefix = {arXiv},
       eprint = {2212.09159},
 primaryClass = {astro-ph.GA},
       adsurl = {https://ui.adsabs.harvard.edu/abs/2023ApJ...944L..16E}
}

@ARTICLE{Madden2006,
       author = {{Madden}, S.~C. and {Galliano}, F. and {Jones}, A.~P. and {Sauvage}, M.},
        title = "{ISM properties in low-metallicity environments}",
      journal = {\aap},
         year = 2006,
        month = feb,
       volume = {446},
       number = {3},
        pages = {877-896},
          doi = {10.1051/0004-6361:20053890},
archivePrefix = {arXiv},
       eprint = {astro-ph/0510086},
 primaryClass = {astro-ph},
       adsurl = {https://ui.adsabs.harvard.edu/abs/2006A&A...446..877M}
}

@ARTICLE{Chastenet2025,
       author = {{Chastenet}, J{\'e}r{\'e}my and {Sandstrom}, Karin and {Leroy}, Adam K. and {Bot}, Caroline and {Chiang}, I-Da and {Chown}, Ryan and {Gordon}, Karl D. and {Koch}, Eric W. and {Roussel}, H{\'e}l{\`e}ne and {Sutter}, Jessica and {Williams}, Thomas G.},
        title = "{The Resolved Behavior of Dust Mass, Polycyclic Aromatic Hydrocarbon Fraction, and Radiation Field in {\ensuremath{\sim}}800 Nearby Galaxies}",
      journal = {\apjs},
         year = 2025,
        month = jan,
       volume = {276},
       number = {1},
          eid = {2},
        pages = {2},
          doi = {10.3847/1538-4365/ad8a5c},
archivePrefix = {arXiv},
       eprint = {2410.03835},
 primaryClass = {astro-ph.GA},
       adsurl = {https://ui.adsabs.harvard.edu/abs/2025ApJS..276....2C}
}

@ARTICLE{Gordon2008,
       author = {{Gordon}, Karl D. and {Engelbracht}, Charles W. and {Rieke}, George H. and {Misselt}, K.~A. and {Smith}, J. -D.~T. and {Kennicutt}, Jr., Robert C.},
        title = "{The Behavior of the Aromatic Features in M101 H II Regions: Evidence for Dust Processing}",
      journal = {\apj},
         year = 2008,
        month = jul,
       volume = {682},
       number = {1},
        pages = {336-354},
          doi = {10.1086/589567},
archivePrefix = {arXiv},
       eprint = {0804.3223},
 primaryClass = {astro-ph},
       adsurl = {https://ui.adsabs.harvard.edu/abs/2008ApJ...682..336G}
}

@ARTICLE{RemyRuyer2015,
   author = {{R{\'e}my-Ruyer}, A. and {Madden}, S.~C. and {Galliano}, F. and 
	{Lebouteiller}, V. and {Baes}, M. and {Bendo}, G.~J. and {Boselli}, A. and 
	{Ciesla}, L. and {Cormier}, D. and {Cooray}, A. and {Cortese}, L. and 
	{De Looze}, I. and {Doublier-Pritchard}, V. and {Galametz}, M. and 
	{Jones}, A.~P. and {Karczewski}, O.~{\L}. and {Lu}, N. and {Spinoglio}, L.
	},
    title = "{Linking dust emission to fundamental properties in galaxies: the low-metallicity picture}",
  journal = {\aap},
archivePrefix = "arXiv",
   eprint = {1507.05432},
     year = 2015,
    month = oct,
   volume = 582,
      eid = {A121},
    pages = {A121},
      doi = {10.1051/0004-6361/201526067},
   adsurl = {http://adsabs.harvard.edu/abs/2015A%26A...582A.121R}
}

@ARTICLE{Engelbracht2005,
   author = {{Engelbracht}, C.~W. and {Gordon}, K.~D. and {Rieke}, G.~H. and 
	{Werner}, M.~W. and {Dale}, D.~A. and {Latter}, W.~B.},
    title = "{Metallicity Effects on Mid-Infrared Colors and the 8 {$\mu$}m PAH Emission in Galaxies}",
  journal = {\apjl},
   eprint = {astro-ph/0506214},
     year = 2005,
    month = jul,
   volume = 628,
    pages = {L29-L32},
      doi = {10.1086/432613},
   adsurl = {http://adsabs.harvard.edu/abs/2005ApJ...628L..29E}
}

@ARTICLE{Galliano2008,
   author = {{Galliano}, F. and {Dwek}, E. and {Chanial}, P.},
    title = "{Stellar Evolutionary Effects on the Abundances of Polycyclic Aromatic Hydrocarbons and Supernova-Condensed Dust in Galaxies}",
  journal = {\apj},
archivePrefix = "arXiv",
   eprint = {0708.0790},
     year = 2008,
    month = jan,
   volume = 672,
    pages = {214-243},
      doi = {10.1086/523621},
   adsurl = {http://adsabs.harvard.edu/abs/2008ApJ...672..214G}
}

@ARTICLE{Draine2021,
       author = {{Draine}, B.~T. and {Li}, Aigen and {Hensley}, Brandon S. and {Hunt}, L.~K. and {Sandstrom}, K. and {Smith}, J. -D.~T.},
        title = "{Excitation of Polycyclic Aromatic Hydrocarbon Emission: Dependence on Size Distribution, Ionization, and Starlight Spectrum and Intensity}",
      journal = {\apj},
         year = 2021,
        month = aug,
       volume = {917},
       number = {1},
          eid = {3},
        pages = {3},
          doi = {10.3847/1538-4357/abff51},
archivePrefix = {arXiv},
       eprint = {2011.07046},
 primaryClass = {astro-ph.GA},
       adsurl = {https://ui.adsabs.harvard.edu/abs/2021ApJ...917....3D}
}

@ARTICLE{Micelotta2010shocks,
   author = {{Micelotta}, E.~R. and {Jones}, A.~P. and {Tielens}, A.~G.~G.~M.
	},
    title = "{Polycyclic aromatic hydrocarbon processing in interstellar shocks}",
  journal = {\aap},
archivePrefix = "arXiv",
   eprint = {0910.2461},
     year = 2010,
    month = feb,
   volume = 510,
      eid = {A36},
    pages = {A36},
      doi = {10.1051/0004-6361/200911682},
   adsurl = {http://adsabs.harvard.edu/abs/2010A%26A...510A..36M}
}

@ARTICLE{Micelotta2010gas,
   author = {{Micelotta}, E.~R. and {Jones}, A.~P. and {Tielens}, A.~G.~G.~M.
	},
    title = "{Polycyclic aromatic hydrocarbon processing in a hot gas}",
  journal = {\aap},
archivePrefix = "arXiv",
   eprint = {0912.1595},
     year = 2010,
    month = feb,
   volume = 510,
      eid = {A37},
    pages = {A37},
      doi = {10.1051/0004-6361/200911683},
   adsurl = {http://adsabs.harvard.edu/abs/2010A%26A...510A..37M}
}

@ARTICLE{Bocchio2012,
   author = {{Bocchio}, M. and {Micelotta}, E.~R. and {Gautier}, A.-L. and 
	{Jones}, A.~P.},
    title = "{Small hydrocarbon particle erosion in a hot gas. A comparative study}",
  journal = {\aap},
     year = 2012,
    month = sep,
   volume = 545,
      eid = {A124},
    pages = {A124},
      doi = {10.1051/0004-6361/201219705},
   adsurl = {http://adsabs.harvard.edu/abs/2012A%26A...545A.124B}
}

@ARTICLE{Montillaud2013,
   author = {{Montillaud}, J. and {Joblin}, C. and {Toublanc}, D.},
    title = "{Evolution of polycyclic aromatic hydrocarbons in photodissociation regions. Hydrogenation and charge states}",
  journal = {\aap},
archivePrefix = "arXiv",
   eprint = {1301.6507},
     year = 2013,
    month = apr,
   volume = 552,
      eid = {A15},
    pages = {A15},
      doi = {10.1051/0004-6361/201220757},
   adsurl = {http://adsabs.harvard.edu/abs/2013A%26A...552A..15M}
}

@ARTICLE{Chown2025Dwarfs,
       author = {{Chown}, Ryan and {Leroy}, Adam K. and {Bolatto}, Alberto D. and {Chastenet}, J{\'e}r{\'e}my and {Glover}, Simon C.~O. and {Indebetouw}, R{\'e}my and {Koch}, Eric W. and {Donovan Meyer}, Jennifer and {Pingel}, Nickolas M. and {Rosolowsky}, Erik and {Sandstrom}, Karin and {Sutter}, Jessica and {Tarantino}, Elizabeth and {Bigiel}, Frank and {Boquien}, M{\'e}d{\'e}ric and {Chiang}, I. -Da and {Dale}, Daniel A. and {Dalcanton}, Julianne J. and {Egorov}, Oleg V. and {Eibensteiner}, Cosima and {Grasha}, Kathryn and {Hassani}, Hamid and {He}, Hao and {Kim}, Jaeyeon and {Meidt}, Sharon and {Pathak}, Debosmita and {Sarbadhicary}, Sumit K. and {Stanimirovic}, Snezana and {Villanueva}, Vicente and {Williams}, Thomas G.},
        title = "{Relationships between Polycyclic Aromatic Hydrocarbons, Small Dust Grains, H$_{2}$, and H I in Local Group Dwarf Galaxies NGC 6822 and WLM Using JWST, ALMA, and the VLA}",
      journal = {\apj},
         year = 2025,
        month = jul,
       volume = {987},
       number = {1},
          eid = {91},
        pages = {91},
          doi = {10.3847/1538-4357/add73a},
archivePrefix = {arXiv},
       eprint = {2504.08069},
 primaryClass = {astro-ph.GA},
       adsurl = {https://ui.adsabs.harvard.edu/abs/2025ApJ...987...91C}
}

@ARTICLE{Chown2025PHANGS,
       author = {{Chown}, Ryan and {Leroy}, Adam K. and {Sandstrom}, Karin and {Chastenet}, J{\'e}r{\'e}my and {Sutter}, Jessica and {Koch}, Eric W. and {Koziol}, Hannah B. and {Neumann}, Lukas and {Sun}, Jiayi and {Williams}, Thomas G. and {Baron}, Dalya and {Anand}, Gagandeep S. and {Barnes}, Ashley. T. and {Bazzi}, Zein and {Belfiore}, Francesco and {Bigiel}, Frank and {Bolatto}, Alberto and {Boquien}, M{\'e}d{\'e}ric and {Cao}, Yixian and {Chevance}, M{\'e}lanie and {Colombo}, Dario and {Dale}, Daniel A. and {den Brok}, Jakob and {Egorov}, Oleg V. and {Eibensteiner}, Cosima and {Emsellem}, Eric and {Hassani}, Hamid and {Henshaw}, Jonathan D. and {He}, Hao and {Kim}, Jaeyeon and {Klessen}, Ralf S. and {Kreckel}, Kathryn and {Larson}, Kirsten L. and {Lee}, Janice C. and {Meidt}, Sharon E. and {Murphy}, Eric J. and {Oakes}, Elias K. and {Ostriker}, Eve C. and {Pan}, Hsi-An and {Pathak}, Debosmita and {Rosolowsky}, Erik and {Sarbadhicary}, Sumit K. and {Schinnerer}, Eva and {Teng}, Yu-Hsuan and {Thilker}, David A. and {Weinbeck}, Tony D. and {Watkins}, Elizabeth J.},
        title = "{Polycyclic Aromatic Hydrocarbon and CO(2{\textendash}1) Emission at 50{\textendash}150 pc Scales in 70 Nearby Galaxies}",
      journal = {\apj},
         year = 2025,
        month = apr,
       volume = {983},
       number = {1},
          eid = {64},
        pages = {64},
          doi = {10.3847/1538-4357/adbd40},
archivePrefix = {arXiv},
       eprint = {2410.05397},
 primaryClass = {astro-ph.GA},
       adsurl = {https://ui.adsabs.harvard.edu/abs/2025ApJ...983...64C}
}

@ARTICLE{Cortzen2019,
   author = {{Cortzen}, I. and {Garrett}, J. and {Magdis}, G. and {Rigopoulou}, D. and 
	{Valentino}, F. and {Pereira-Santaella}, M. and {Combes}, F. and 
	{Alonso-Herrero}, A. and {Toft}, S. and {Daddi}, E. and {Elbaz}, D. and 
	{G{\'o}mez-Guijarro}, C. and {Stockmann}, M. and {Huang}, J. and 
	{Kramer}, C.},
    title = "{PAHs as tracers of the molecular gas in star-forming galaxies}",
  journal = {\mnras},
archivePrefix = "arXiv",
   eprint = {1810.05178},
     year = 2019,
    month = jan,
   volume = 482,
    pages = {1618-1633},
      doi = {10.1093/mnras/sty2777},
   adsurl = {http://adsabs.harvard.edu/abs/2019MNRAS.482.1618C}
}

@ARTICLE{Leroy2023,
       author = {{Leroy}, Adam K. and {Sandstrom}, Karin and {Rosolowsky}, Erik and {Belfiore}, Francesco and {Bolatto}, Alberto D. and {Cao}, Yixian and {Koch}, Eric W. and {Schinnerer}, Eva and {Barnes}, Ashley. T. and {Be{\v{s}}li{\'c}}, Ivana and {Bigiel}, F. and {Blanc}, Guillermo A. and {Chastenet}, J{\'e}r{\'e}my and {Chen}, Ness Mayker and {Chevance}, M{\'e}lanie and {Chown}, Ryan and {Congiu}, Enrico and {Dale}, Daniel A. and {Egorov}, Oleg V. and {Emsellem}, Eric and {Eibensteiner}, Cosima and {Faesi}, Christopher M. and {Glover}, Simon C.~O. and {Grasha}, Kathryn and {Groves}, Brent and {Hassani}, Hamid and {Henshaw}, Jonathan D. and {Hughes}, Annie and {Jim{\'e}nez-Donaire}, Mar{\'\i}a J. and {Kim}, Jaeyeon and {Klessen}, Ralf S. and {Kreckel}, Kathryn and {Kruijssen}, J.~M. Diederik and {Larson}, Kirsten L. and {Lee}, Janice C. and {Levy}, Rebecca C. and {Liu}, Daizhong and {Lopez}, Laura A. and {Meidt}, Sharon E. and {Murphy}, Eric J. and {Neumann}, Justus and {Pessa}, Ismael and {Pety}, J{\'e}r{\^o}me and {Saito}, Toshiki and {Sardone}, Amy and {Sun}, Jiayi and {Thilker}, David A. and {Usero}, Antonio and {Watkins}, Elizabeth J. and {Whitcomb}, Cory M. and {Williams}, Thomas G.},
        title = "{PHANGS-JWST First Results: Mid-infrared Emission Traces Both Gas Column Density and Heating at 100 pc Scales}",
      journal = {\apjl},
         year = 2023,
        month = feb,
       volume = {944},
       number = {2},
          eid = {L9},
        pages = {L9},
          doi = {10.3847/2041-8213/acaf85},
archivePrefix = {arXiv},
       eprint = {2212.10574},
 primaryClass = {astro-ph.GA},
       adsurl = {https://ui.adsabs.harvard.edu/abs/2023ApJ...944L...9L}
}

@ARTICLE{Rand2008,
       author = {{Rand}, Richard J. and {Wood}, Kenneth and {Benjamin}, Robert A.},
        title = "{Infrared Spectroscopy of the Diffuse Ionized Halo of NGC 891}",
      journal = {\apj},
         year = 2008,
        month = jun,
       volume = {680},
       number = {1},
        pages = {263-275},
          doi = {10.1086/587779},
archivePrefix = {arXiv},
       eprint = {0802.3156},
 primaryClass = {astro-ph},
       adsurl = {https://ui.adsabs.harvard.edu/abs/2008ApJ...680..263R}
}

@ARTICLE{Bregman2013,
       author = {{Bregman}, Joel N. and {Miller}, Eric D. and {Seitzer}, Patrick and {Cowley}, C.~R. and {Miller}, Matthew J.},
        title = "{Outflow versus Infall in Spiral Galaxies: Metal Absorption in the Halo of NGC 891}",
      journal = {\apj},
         year = 2013,
        month = mar,
       volume = {766},
       number = {1},
          eid = {57},
        pages = {57},
          doi = {10.1088/0004-637X/766/1/57},
archivePrefix = {arXiv},
       eprint = {1304.0795},
 primaryClass = {astro-ph.CO},
       adsurl = {https://ui.adsabs.harvard.edu/abs/2013ApJ...766...57B}
}

@ARTICLE{Walch2015,
       author = {{Walch}, S. and {Girichidis}, P. and {Naab}, T. and {Gatto}, A. and {Glover}, S.~C.~O. and {W{\"u}nsch}, R. and {Klessen}, R.~S. and {Clark}, P.~C. and {Peters}, T. and {Derigs}, D. and {Baczynski}, C.},
        title = "{The SILCC (SImulating the LifeCycle of molecular Clouds) project - I. Chemical evolution of the supernova-driven ISM}",
      journal = {\mnras},
         year = 2015,
        month = nov,
       volume = {454},
       number = {1},
        pages = {238-268},
          doi = {10.1093/mnras/stv1975},
archivePrefix = {arXiv},
       eprint = {1412.2749},
 primaryClass = {astro-ph.GA},
       adsurl = {https://ui.adsabs.harvard.edu/abs/2015MNRAS.454..238W}
}

@ARTICLE{TanFielding2024,
       author = {{Tan}, Brent and {Fielding}, Drummond B.},
        title = "{Cloud atlas: navigating the multiphase landscape of tempestuous galactic winds}",
      journal = {\mnras},
         year = 2024,
        month = feb,
       volume = {527},
       number = {4},
        pages = {9683-9714},
          doi = {10.1093/mnras/stad3793},
archivePrefix = {arXiv},
       eprint = {2305.14424},
 primaryClass = {astro-ph.GA},
       adsurl = {https://ui.adsabs.harvard.edu/abs/2024MNRAS.527.9683T}
}

@ARTICLE{McCallum2024,
       author = {{McCallum}, Lewis and {Wood}, Kenneth and {Benjamin}, Robert and {Pe{\~n}aloza}, Camilo and {Krishnarao}, Dhanesh and {Smith}, Rowan and {Vandenbroucke}, Bert},
        title = "{The persistence of high altitude non-equilibrium diffuse ionized gas in simulations of star-forming galaxies}",
      journal = {\mnras},
         year = 2024,
        month = may,
       volume = {530},
       number = {3},
        pages = {2548-2564},
          doi = {10.1093/mnras/stae988},
archivePrefix = {arXiv},
       eprint = {2404.05651},
 primaryClass = {astro-ph.GA},
       adsurl = {https://ui.adsabs.harvard.edu/abs/2024MNRAS.530.2548M}
}

@ARTICLE{Schneider2020,
       author = {{Schneider}, Evan E. and {Ostriker}, Eve C. and {Robertson}, Brant E. and {Thompson}, Todd A.},
        title = "{The Physical Nature of Starburst-driven Galactic Outflows}",
      journal = {\apj},
         year = 2020,
        month = may,
       volume = {895},
       number = {1},
          eid = {43},
        pages = {43},
          doi = {10.3847/1538-4357/ab8ae8},
archivePrefix = {arXiv},
       eprint = {2002.10468},
 primaryClass = {astro-ph.GA},
       adsurl = {https://ui.adsabs.harvard.edu/abs/2020ApJ...895...43S}
}

@ARTICLE{Girichidis2016,
       author = {{Girichidis}, Philipp and {Walch}, Stefanie and {Naab}, Thorsten and {Gatto}, Andrea and {W{\"u}nsch}, Richard and {Glover}, Simon C.~O. and {Klessen}, Ralf S. and {Clark}, Paul C. and {Peters}, Thomas and {Derigs}, Dominik and {Baczynski}, Christian},
        title = "{The SILCC (SImulating the LifeCycle of molecular Clouds) project - II. Dynamical evolution of the supernova-driven ISM and the launching of outflows}",
      journal = {\mnras},
         year = 2016,
        month = mar,
       volume = {456},
       number = {4},
        pages = {3432-3455},
          doi = {10.1093/mnras/stv2742},
archivePrefix = {arXiv},
       eprint = {1508.06646},
 primaryClass = {astro-ph.GA},
       adsurl = {https://ui.adsabs.harvard.edu/abs/2016MNRAS.456.3432G}
}

@ARTICLE{Emerick2019,
       author = {{Emerick}, Andrew and {Bryan}, Greg L. and {Mac Low}, Mordecai-Mark},
        title = "{Simulating an isolated dwarf galaxy with multichannel feedback and chemical yields from individual stars}",
      journal = {\mnras},
         year = 2019,
        month = jan,
       volume = {482},
       number = {1},
        pages = {1304-1329},
          doi = {10.1093/mnras/sty2689},
archivePrefix = {arXiv},
       eprint = {1807.07182},
 primaryClass = {astro-ph.GA},
       adsurl = {https://ui.adsabs.harvard.edu/abs/2019MNRAS.482.1304E}
}

@ARTICLE{Yoon2021,
       author = {{Yoon}, J.~H. and {Martin}, Crystal L. and {Veilleux}, S. and {Mel{\'e}ndez}, M. and {Mueller}, T. and {Gordon}, K.~D. and {Cecil}, G. and {Bland-Hawthorn}, J. and {Engelbracht}, C.},
        title = "{Exploring the dust content of galactic haloes with Herschel III. NGC 891}",
      journal = {\mnras},
         year = 2021,
        month = mar,
       volume = {502},
       number = {1},
        pages = {969-984},
          doi = {10.1093/mnras/staa3583},
archivePrefix = {arXiv},
       eprint = {2012.08686},
 primaryClass = {astro-ph.GA},
       adsurl = {https://ui.adsabs.harvard.edu/abs/2021MNRAS.502..969Y}
}

@INPROCEEDINGS{Heckman2002,
       author = {{Heckman}, T.~M.},
        title = "{Galactic Superwinds Circa 2001}",
    booktitle = {Extragalactic Gas at Low Redshift},
         year = 2002,
       editor = {{Mulchaey}, John S. and {Stocke}, John T.},
       series = {Astronomical Society of the Pacific Conference Series},
       volume = {254},
        month = jan,
        pages = {292},
          doi = {10.48550/arXiv.astro-ph/0107438},
archivePrefix = {arXiv},
       eprint = {astro-ph/0107438},
 primaryClass = {astro-ph},
       adsurl = {https://ui.adsabs.harvard.edu/abs/2002ASPC..254..292H}
}

@ARTICLE{Bocchio2016,
       author = {{Bocchio}, M. and {Bianchi}, S. and {Hunt}, L.~K. and {Schneider}, R.},
        title = "{Halo dust detection around NGC 891}",
      journal = {\aap},
         year = 2016,
        month = feb,
       volume = {586},
          eid = {A8},
        pages = {A8},
          doi = {10.1051/0004-6361/201526950},
archivePrefix = {arXiv},
       eprint = {1509.07677},
 primaryClass = {astro-ph.GA},
       adsurl = {https://ui.adsabs.harvard.edu/abs/2016A&A...586A...8B}
}

@ARTICLE{Whaley2009,
       author = {{Whaley}, C.~H. and {Irwin}, J.~A. and {Madden}, S.~C. and {Galliano}, F. and {Bendo}, G.~J.},
        title = "{A multiwavelength infrared study of NGC 891}",
      journal = {\mnras},
         year = 2009,
        month = may,
       volume = {395},
       number = {1},
        pages = {97-113},
          doi = {10.1111/j.1365-2966.2009.14532.x},
archivePrefix = {arXiv},
       eprint = {0901.4351},
 primaryClass = {astro-ph.CO},
       adsurl = {https://ui.adsabs.harvard.edu/abs/2009MNRAS.395...97W}
}

@ARTICLE{Maragkoudakis2022,
       author = {{Maragkoudakis}, A. and {Boersma}, C. and {Temi}, P. and {Bregman}, J.~D. and {Allamandola}, L.~J.},
        title = "{Linking Characteristics of the Polycyclic Aromatic Hydrocarbon Population with Galaxy Properties: A Quantitative Approach Using the NASA Ames PAH IR Spectroscopic Database}",
      journal = {\apj},
         year = 2022,
        month = may,
       volume = {931},
       number = {1},
          eid = {38},
        pages = {38},
          doi = {10.3847/1538-4357/ac666f},
archivePrefix = {arXiv},
       eprint = {2204.05292},
 primaryClass = {astro-ph.GA},
       adsurl = {https://ui.adsabs.harvard.edu/abs/2022ApJ...931...38M}
}

@ARTICLE{Maragkoudakis2020,
       author = {{Maragkoudakis}, A. and {Peeters}, E. and {Ricca}, A.},
        title = "{Probing the size and charge of polycyclic aromatic hydrocarbons}",
      journal = {\mnras},
         year = 2020,
        month = may,
       volume = {494},
       number = {1},
        pages = {642-664},
          doi = {10.1093/mnras/staa681},
archivePrefix = {arXiv},
       eprint = {2003.02823},
 primaryClass = {astro-ph.GA},
       adsurl = {https://ui.adsabs.harvard.edu/abs/2020MNRAS.494..642M}
}

@ARTICLE{Maragkoudakis2025,
       author = {{Maragkoudakis}, A. and {Boersma}, C. and {Temi}, P. and {Bregman}, J.~D. and {Allamandola}, L.~J. and {Esposito}, V.~J. and {Ricca}, A. and {Peeters}, E.},
        title = "{A Sensitivity Analysis of the Modeling of Polycyclic Aromatic Hydrocarbon Emission in Galaxies}",
      journal = {\apj},
         year = 2025,
        month = jan,
       volume = {979},
       number = {1},
          eid = {90},
        pages = {90},
          doi = {10.3847/1538-4357/ad9918},
archivePrefix = {arXiv},
       eprint = {2412.01875},
 primaryClass = {astro-ph.GA},
       adsurl = {https://ui.adsabs.harvard.edu/abs/2025ApJ...979...90M}
}

@ARTICLE{Boersma2010,
       author = {{Boersma}, C. and {Bauschlicher}, C.~W. and {Allamandola}, L.~J. and {Ricca}, A. and {Peeters}, E. and {Tielens}, A.~G.~G.~M.},
        title = "{The 15-20 {\ensuremath{\mu}}m PAH emission features: probes of individual PAHs?}",
      journal = {\aap},
         year = 2010,
        month = feb,
       volume = {511},
          eid = {A32},
        pages = {A32},
          doi = {10.1051/0004-6361/200912714},
       adsurl = {https://ui.adsabs.harvard.edu/abs/2010A&A...511A..32B}
}

@ARTICLE{Peeters2012,
       author = {{Peeters}, Els and {Tielens}, Alexander G.~G.~M. and {Allamandola}, Louis J. and {Wolfire}, Mark G.},
        title = "{The 15-20 {\ensuremath{\mu}}m Emission in the Reflection Nebula NGC 2023}",
      journal = {\apj},
         year = 2012,
        month = mar,
       volume = {747},
       number = {1},
          eid = {44},
        pages = {44},
          doi = {10.1088/0004-637X/747/1/44},
archivePrefix = {arXiv},
       eprint = {1112.3386},
 primaryClass = {astro-ph.GA},
       adsurl = {https://ui.adsabs.harvard.edu/abs/2012ApJ...747...44P}
}

@ARTICLE{Shannon2015,
       author = {{Shannon}, M.~J. and {Stock}, D.~J. and {Peeters}, E.},
        title = "{Probing the Ionization States of Polycyclic Aromatic Hydrocarbons via the 15-20 {\ensuremath{\mu}}m Emission Bands}",
      journal = {\apj},
         year = 2015,
        month = oct,
       volume = {811},
       number = {2},
          eid = {153},
        pages = {153},
          doi = {10.1088/0004-637X/811/2/153},
archivePrefix = {arXiv},
       eprint = {1508.04766},
 primaryClass = {astro-ph.GA},
       adsurl = {https://ui.adsabs.harvard.edu/abs/2015ApJ...811..153S}
}

@ARTICLE{Chown2025PDRs,
       author = {{Chown}, Ryan and {Okada}, Yoko and {Peeters}, Els and {Sidhu}, Ameek and {Khan}, Baria and {Schefter}, Bethany and {Trahin}, Boris and {Canin}, Am{\'e}lie and {Van De Putte}, Dries and {Alarc{\'o}n}, Felipe and {Schroetter}, Ilane and {Kannavou}, Olga and {Habart}, Emilie and {Bern{\'e}}, Olivier and {Boersma}, Christiaan and {Cami}, Jan and {Dartois}, Emmanuel and {Goicoechea}, Javier and {Gordon}, Karl and {Onaka}, Takashi},
        title = "{PDRs4All: XIII. Empirical prescriptions for the interpretation of JWST imaging observations of star-forming regions}",
      journal = {\aap},
         year = 2025,
        month = jun,
       volume = {698},
          eid = {A86},
        pages = {A86},
          doi = {10.1051/0004-6361/202452940},
archivePrefix = {arXiv},
       eprint = {2411.06061},
 primaryClass = {astro-ph.GA},
       adsurl = {https://ui.adsabs.harvard.edu/abs/2025A&A...698A..86C}
}

@ARTICLE{Donnelly2025,
       author = {{Donnelly}, Grant P. and {Lai}, Thomas S. -Y. and {Armus}, Lee and {D{\'\i}az-Santos}, Tanio and {Larson}, Kirsten L. and {Barcos-Mu{\~n}oz}, Loreto and {Bianchin}, Marina and {Bohn}, Thomas and {B{\"o}ker}, Torsten and {Buiten}, Victorine A. and {Charmandaris}, Vassilis and {Evans}, Aaron S. and {Howell}, Justin and {Inami}, Hanae and {Kakkad}, Darshan and {Lenki{\'c}}, Laura and {Linden}, Sean T. and {Lofaro}, Cristina M. and {Malkan}, Matthew A. and {Medling}, Anne M. and {Privon}, George C. and {Ricci}, Claudio and {Smith}, J.~D.~T. and {Song}, Yiqing and {Stierwalt}, Sabrina and {van der Werf}, Paul P. and {U}, Vivian},
        title = "{A Spectroscopically Calibrated Prescription for Extracting Polycyclic Aromatic Hydrocarbon Flux from JWST MIRI Imaging}",
      journal = {\apj},
         year = 2025,
        month = apr,
       volume = {983},
       number = {1},
          eid = {79},
        pages = {79},
          doi = {10.3847/1538-4357/adb97f},
archivePrefix = {arXiv},
       eprint = {2501.19397},
 primaryClass = {astro-ph.GA},
       adsurl = {https://ui.adsabs.harvard.edu/abs/2025ApJ...983...79D}
}

@ARTICLE{Whitcomb2023,
       author = {{Whitcomb}, C.~M. and {Sandstrom}, K. and {Leroy}, A. and {Smith}, J.-D.~T.},
        title = "{Star Formation and Molecular Gas Diagnostics with Mid- and Far-infrared Emission}",
      journal = {\apj},
         year = 2023,
        month = may,
       volume = {948},
       number = {2},
          eid = {88},
        pages = {88},
          doi = {10.3847/1538-4357/acc316},
archivePrefix = {arXiv},
       eprint = {2212.00180},
 primaryClass = {astro-ph.GA},
       adsurl = {https://ui.adsabs.harvard.edu/abs/2023ApJ...948...88W}
}

@ARTICLE{Maragkoudakis2023,
       author = {{Maragkoudakis}, A. and {Peeters}, E. and {Ricca}, A.},
        title = "{Spectral variations among different scenarios of PAH processing or formation}",
      journal = {\mnras},
         year = 2023,
        month = apr,
       volume = {520},
       number = {4},
        pages = {5354-5372},
          doi = {10.1093/mnras/stad465},
archivePrefix = {arXiv},
       eprint = {2302.03678},
 primaryClass = {astro-ph.GA},
       adsurl = {https://ui.adsabs.harvard.edu/abs/2023MNRAS.520.5354M}
}

@ARTICLE{Peeters2017,
       author = {{Peeters}, Els and {Bauschlicher}, Jr., Charles W. and {Allamandola}, Louis J. and {Tielens}, Alexander G.~G.~M. and {Ricca}, Alessandra and {Wolfire}, Mark G.},
        title = "{The PAH Emission Characteristics of the Reflection Nebula NGC 2023}",
      journal = {\apj},
         year = 2017,
        month = feb,
       volume = {836},
       number = {2},
          eid = {198},
        pages = {198},
          doi = {10.3847/1538-4357/836/2/198},
archivePrefix = {arXiv},
       eprint = {1701.06585},
 primaryClass = {astro-ph.GA},
       adsurl = {https://ui.adsabs.harvard.edu/abs/2017ApJ...836..198P}
}

@ARTICLE{Hensley2022,
       author = {{Hensley}, Brandon S. and {Murray}, Claire E. and {Dodici}, Mark},
        title = "{Polycyclic Aromatic Hydrocarbons, Anomalous Microwave Emission, and their Connection to the Cold Neutral Medium}",
      journal = {\apj},
         year = 2022,
        month = apr,
       volume = {929},
       number = {1},
          eid = {23},
        pages = {23},
          doi = {10.3847/1538-4357/ac5cbd},
archivePrefix = {arXiv},
       eprint = {2111.03067},
 primaryClass = {astro-ph.GA},
       adsurl = {https://ui.adsabs.harvard.edu/abs/2022ApJ...929...23H}
}

@ARTICLE{Sandstrom2010,
       author = {{Sandstrom}, Karin M. and {Bolatto}, Alberto D. and {Draine}, B.~T. and {Bot}, Caroline and {Stanimirovi{\'c}}, Sne{\v{z}}ana},
        title = "{The Spitzer Survey of the Small Magellanic Cloud (S$^{3}$MC): Insights into the Life Cycle of Polycyclic Aromatic Hydrocarbons}",
      journal = {\apj},
         year = 2010,
        month = jun,
       volume = {715},
       number = {2},
        pages = {701-723},
          doi = {10.1088/0004-637X/715/2/701},
archivePrefix = {arXiv},
       eprint = {1003.4516},
 primaryClass = {astro-ph.CO},
       adsurl = {https://ui.adsabs.harvard.edu/abs/2010ApJ...715..701S}
}

@ARTICLE{Hsieh2017,
       author = {{Hsieh}, B.~C. and {Lin}, Lihwai and {Lin}, J.~H. and {Pan}, H.~A. and {Hsu}, C.~H. and {S{\'a}nchez}, S.~F. and {Cano-D{\'\i}az}, M. and {Zhang}, K. and {Yan}, R. and {Barrera-Ballesteros}, J.~K. and {Boquien}, M. and {Riffel}, R. and {Brownstein}, J. and {Cruz-Gonz{\'a}lez}, I. and {Hagen}, A. and {Ibarra}, H. and {Pan}, K. and {Bizyaev}, D. and {Oravetz}, D. and {Simmons}, A.},
        title = "{SDSS-IV MaNGA: Spatially Resolved Star Formation Main Sequence and LI(N)ER Sequence}",
      journal = {\apjl},
         year = 2017,
        month = dec,
       volume = {851},
       number = {2},
          eid = {L24},
        pages = {L24},
          doi = {10.3847/2041-8213/aa9d80},
archivePrefix = {arXiv},
       eprint = {1711.09162},
 primaryClass = {astro-ph.GA},
       adsurl = {https://ui.adsabs.harvard.edu/abs/2017ApJ...851L..24H}
}

@ARTICLE{Leroy2008,
       author = {{Leroy}, Adam K. and {Walter}, Fabian and {Brinks}, Elias and {Bigiel}, Frank and {de Blok}, W.~J.~G. and {Madore}, Barry and {Thornley}, M.~D.},
        title = "{The Star Formation Efficiency in Nearby Galaxies: Measuring Where Gas Forms Stars Effectively}",
      journal = {\aj},
         year = 2008,
        month = dec,
       volume = {136},
       number = {6},
        pages = {2782-2845},
          doi = {10.1088/0004-6256/136/6/2782},
archivePrefix = {arXiv},
       eprint = {0810.2556},
 primaryClass = {astro-ph},
       adsurl = {https://ui.adsabs.harvard.edu/abs/2008AJ....136.2782L}
}

@ARTICLE{Lin2019,
       author = {{Lin}, Lihwai and {Pan}, Hsi-An and {Ellison}, Sara L. and {Belfiore}, Francesco and {Shi}, Yong and {S{\'a}nchez}, Sebasti{\'a}n F. and {Hsieh}, Bau-Ching and {Rowlands}, Kate and {Ramya}, S. and {Thorp}, Mallory D. and {Li}, Cheng and {Maiolino}, Roberto},
        title = "{The ALMaQUEST Survey: The Molecular Gas Main Sequence and the Origin of the Star-forming Main Sequence}",
      journal = {\apjl},
         year = 2019,
        month = oct,
       volume = {884},
       number = {2},
          eid = {L33},
        pages = {L33},
          doi = {10.3847/2041-8213/ab4815},
archivePrefix = {arXiv},
       eprint = {1909.11243},
 primaryClass = {astro-ph.GA},
       adsurl = {https://ui.adsabs.harvard.edu/abs/2019ApJ...884L..33L}
}

@ARTICLE{DeLooze2020,
       author = {{De Looze}, I. and {Lamperti}, I. and {Saintonge}, A. and {Rela{\~n}o}, M. and {Smith}, M.~W.~L. and {Clark}, C.~J.~R. and {Wilson}, C.~D. and {Decleir}, M. and {Jones}, A.~P. and {Kennicutt}, R.~C. and {Accurso}, G. and {Brinks}, E. and {Bureau}, M. and {Cigan}, P. and {Clements}, D.~L. and {De Vis}, P. and {Fanciullo}, L. and {Gao}, Y. and {Gear}, W.~K. and {Ho}, L.~C. and {Hwang}, H.~S. and {Micha{\l}owski}, M.~J. and {Lee}, J.~C. and {Li}, C. and {Lin}, L. and {Liu}, T. and {Lomaeva}, M. and {Pan}, H. -A. and {Sargent}, M. and {Williams}, T. and {Xiao}, T. and {Zhu}, M.},
        title = "{JINGLE - IV. Dust, H I gas, and metal scaling laws in the local Universe}",
      journal = {\mnras},
         year = 2020,
        month = aug,
       volume = {496},
       number = {3},
        pages = {3668-3687},
          doi = {10.1093/mnras/staa1496},
archivePrefix = {arXiv},
       eprint = {2006.01856},
 primaryClass = {astro-ph.GA},
       adsurl = {https://ui.adsabs.harvard.edu/abs/2020MNRAS.496.3668D}
}

@ARTICLE{Saintonge2022,
       author = {{Saintonge}, Am{\'e}lie and {Catinella}, Barbara},
        title = "{The Cold Interstellar Medium of Galaxies in the Local Universe}",
      journal = {\araa},
         year = 2022,
        month = aug,
       volume = {60},
        pages = {319-361},
          doi = {10.1146/annurev-astro-021022-043545},
archivePrefix = {arXiv},
       eprint = {2202.00690},
 primaryClass = {astro-ph.GA},
       adsurl = {https://ui.adsabs.harvard.edu/abs/2022ARA&A..60..319S}
}

@ARTICLE{Casasola2022,
       author = {{Casasola}, Viviana and {Bianchi}, Simone and {Magrini}, Laura and {Mosenkov}, Aleksandr V. and {Salvestrini}, Francesco and {Baes}, Maarten and {Calura}, Francesco and {Cassar{\`a}}, Letizia P. and {Clark}, Christopher J.~R. and {Corbelli}, Edvige and {Fritz}, Jacopo and {Galliano}, Fr{\'e}d{\'e}ric and {Liuzzo}, Elisabetta and {Madden}, Suzanne and {Nersesian}, Angelos and {Pozzi}, Francesca and {Roychowdhury}, Sambit and {Baronchelli}, Ivano and {Bonato}, Matteo and {Gruppioni}, Carlotta and {Pantoni}, Lara},
        title = "{The resolved scaling relations in DustPedia: Zooming in on the local Universe}",
      journal = {\aap},
         year = 2022,
        month = dec,
       volume = {668},
          eid = {A130},
        pages = {A130},
          doi = {10.1051/0004-6361/202245043},
archivePrefix = {arXiv},
       eprint = {2210.15993},
 primaryClass = {astro-ph.GA},
       adsurl = {https://ui.adsabs.harvard.edu/abs/2022A&A...668A.130C}
}

@ARTICLE{Cortese2012,
       author = {{Cortese}, L. and {Ciesla}, L. and {Boselli}, A. and {Bianchi}, S. and {Gomez}, H. and {Smith}, M.~W.~L. and {Bendo}, G.~J. and {Eales}, S. and {Pohlen}, M. and {Baes}, M. and {Corbelli}, E. and {Davies}, J.~I. and {Hughes}, T.~M. and {Hunt}, L.~K. and {Madden}, S.~C. and {Pierini}, D. and {di Serego Alighieri}, S. and {Zibetti}, S. and {Boquien}, M. and {Clements}, D.~L. and {Cooray}, A. and {Galametz}, M. and {Magrini}, L. and {Pappalardo}, C. and {Spinoglio}, L. and {Vlahakis}, C.},
        title = "{The dust scaling relations of the Herschel Reference Survey}",
      journal = {\aap},
         year = 2012,
        month = apr,
       volume = {540},
          eid = {A52},
        pages = {A52},
          doi = {10.1051/0004-6361/201118499},
archivePrefix = {arXiv},
       eprint = {1201.2762},
 primaryClass = {astro-ph.CO},
       adsurl = {https://ui.adsabs.harvard.edu/abs/2012A&A...540A..52C}
}

@ARTICLE{Peroux2020,
       author = {{P{\'e}roux}, C{\'e}line and {Howk}, J. Christopher},
        title = "{The Cosmic Baryon and Metal Cycles}",
      journal = {\araa},
         year = 2020,
        month = aug,
       volume = {58},
        pages = {363-406},
          doi = {10.1146/annurev-astro-021820-120014},
archivePrefix = {arXiv},
       eprint = {2011.01935},
 primaryClass = {astro-ph.GA},
       adsurl = {https://ui.adsabs.harvard.edu/abs/2020ARA&A..58..363P}
}

@ARTICLE{KennicuttEvans12KSReview,
       author = {{Kennicutt}, Robert C. and {Evans}, Neal J.},
        title = "{Star Formation in the Milky Way and Nearby Galaxies}",
      journal = {\araa},
         year = 2012,
        month = sep,
       volume = {50},
        pages = {531-608},
          doi = {10.1146/annurev-astro-081811-125610},
archivePrefix = {arXiv},
       eprint = {1204.3552},
 primaryClass = {astro-ph.GA},
       adsurl = {https://ui.adsabs.harvard.edu/abs/2012ARA&A..50..531K}
}

@ARTICLE{Maiolino2019,
       author = {{Maiolino}, R. and {Mannucci}, F.},
        title = "{De re metallica: the cosmic chemical evolution of galaxies}",
      journal = {\aapr},
         year = 2019,
        month = feb,
       volume = {27},
       number = {1},
          eid = {3},
        pages = {3},
          doi = {10.1007/s00159-018-0112-2},
archivePrefix = {arXiv},
       eprint = {1811.09642},
 primaryClass = {astro-ph.GA},
       adsurl = {https://ui.adsabs.harvard.edu/abs/2019A&ARv..27....3M}
}

@ARTICLE{Allamandola1999,
       author = {{Allamandola}, L.~J. and {Hudgins}, D.~M. and {Sandford}, S.~A.},
        title = "{Modeling the Unidentified Infrared Emission with Combinations of Polycyclic Aromatic Hydrocarbons}",
      journal = {\apjl},
         year = 1999,
        month = feb,
       volume = {511},
       number = {2},
        pages = {L115-L119},
          doi = {10.1086/311843},
       adsurl = {https://ui.adsabs.harvard.edu/abs/1999ApJ...511L.115A}
}

@ARTICLE{Allamandola1989b,
       author = {{Allamandola}, L.~J. and {Tielens}, A.~G.~G.~M. and {Barker}, J.~R.},
        title = "{Interstellar Polycyclic Aromatic Hydrocarbons: The Infrared Emission Bands, the Excitation/Emission Mechanism, and the Astrophysical Implications}",
      journal = {\apjs},
         year = 1989,
        month = dec,
       volume = {71},
        pages = {733},
          doi = {10.1086/191396},
       adsurl = {https://ui.adsabs.harvard.edu/abs/1989ApJS...71..733A}
}

@ARTICLE{Roussel2007,
       author = {{Roussel}, H. and {Helou}, G. and {Hollenbach}, D.~J. and {Draine}, B.~T. and {Smith}, J.~D. and {Armus}, L. and {Schinnerer}, E. and {Walter}, F. and {Engelbracht}, C.~W. and {Thornley}, M.~D. and {Kennicutt}, R.~C. and {Calzetti}, D. and {Dale}, D.~A. and {Murphy}, E.~J. and {Bot}, C.},
        title = "{Warm Molecular Hydrogen in the Spitzer SINGS Galaxy Sample}",
      journal = {\apj},
         year = 2007,
        month = nov,
       volume = {669},
       number = {2},
        pages = {959-981},
          doi = {10.1086/521667},
archivePrefix = {arXiv},
       eprint = {0707.0395},
 primaryClass = {astro-ph},
       adsurl = {https://ui.adsabs.harvard.edu/abs/2007ApJ...669..959R}
}

@ARTICLE{Galliano2021,
       author = {{Galliano}, Fr{\'e}d{\'e}ric and {Nersesian}, Angelos and {Bianchi}, Simone and {De Looze}, Ilse and {Roychowdhury}, Sambit and {Baes}, Maarten and {Casasola}, Viviana and {Cassar{\'a}}, Letizia P. and {Dobbels}, Wouter and {Fritz}, Jacopo and {Galametz}, Maud and {Jones}, Anthony P. and {Madden}, Suzanne C. and {Mosenkov}, Aleksandr and {Xilouris}, Emmanuel M. and {Ysard}, Nathalie},
        title = "{A nearby galaxy perspective on dust evolution. Scaling relations and constraints on the dust build-up in galaxies with the DustPedia and DGS samples}",
      journal = {\aap},
         year = 2021,
        month = may,
       volume = {649},
          eid = {A18},
        pages = {A18},
          doi = {10.1051/0004-6361/202039701},
archivePrefix = {arXiv},
       eprint = {2101.00456},
 primaryClass = {astro-ph.GA},
       adsurl = {https://ui.adsabs.harvard.edu/abs/2021A&A...649A..18G}
}

@ARTICLE{Reichardt2025,
       author = {{Reichardt Chu}, Bronwyn and {Fisher}, Deanne B. and {Chisholm}, John and {Berg}, Danielle and {Bolatto}, Alberto and {Cameron}, Alex J. and {Fielding}, Drummond B. and {Herrera-Camus}, Rodrigo and {Kacprzak}, Glenn G. and {Li}, Miao and {McLeod}, Anna F. and {McPherson}, Daniel K. and {Nielsen}, Nikole M. and {Rickards Vaught}, Ryan J. and {Ridolfo}, Sophia G. and {Sandstrom}, Karin},
        title = "{DUVET: sub-kiloparsec resolved star formation driven outflows in a sample of local starbursting disc galaxies}",
      journal = {\mnras},
         year = 2025,
        month = jan,
       volume = {536},
       number = {2},
        pages = {1799-1821},
          doi = {10.1093/mnras/stae2705},
archivePrefix = {arXiv},
       eprint = {2402.17830},
 primaryClass = {astro-ph.GA},
       adsurl = {https://ui.adsabs.harvard.edu/abs/2025MNRAS.536.1799R}
}

@ARTICLE{Richie2024,
       author = {{Richie}, Helena M. and {Schneider}, Evan E. and {Abruzzo}, Matthew W. and {Torrey}, Paul},
        title = "{Dust Survival in Galactic Winds}",
      journal = {\apj},
         year = 2024,
        month = oct,
       volume = {974},
       number = {1},
          eid = {81},
        pages = {81},
          doi = {10.3847/1538-4357/ad6a1c},
archivePrefix = {arXiv},
       eprint = {2403.03711},
 primaryClass = {astro-ph.GA},
       adsurl = {https://ui.adsabs.harvard.edu/abs/2024ApJ...974...81R}
}

@ARTICLE{Richie2026,
       author = {{Richie}, Helena M. and {Schneider}, Evan E.},
        title = "{Dust Evolution in Simulated Multiphase Galactic Outflows}",
      journal = {\apj},
         year = 2026,
        month = jan,
       volume = {996},
       number = {1},
          eid = {17},
        pages = {17},
          doi = {10.3847/1538-4357/ae12e8},
archivePrefix = {arXiv},
       eprint = {2505.11734},
 primaryClass = {astro-ph.GA},
       adsurl = {https://ui.adsabs.harvard.edu/abs/2026ApJ...996...17R}
}

@ARTICLE{Fraternali2011,
       author = {{Fraternali}, F. and {Sancisi}, R. and {Kamphuis}, P.},
        title = "{A tale of two galaxies: light and mass in NGC 891 and NGC 7814}",
      journal = {\aap},
         year = 2011,
        month = jul,
       volume = {531},
          eid = {A64},
        pages = {A64},
          doi = {10.1051/0004-6361/201116634},
archivePrefix = {arXiv},
       eprint = {1105.3867},
 primaryClass = {astro-ph.CO},
       adsurl = {https://ui.adsabs.harvard.edu/abs/2011A&A...531A..64F}
}

@ARTICLE{Xilouris1998,
       author = {{Xilouris}, E.~M. and {Alton}, P.~B. and {Davies}, J.~I. and {Kylafis}, N.~D. and {Papamastorakis}, J. and {Trewhella}, M.},
        title = "{Optical and NIR modelling of NGC 891}",
      journal = {\aap},
         year = 1998,
        month = mar,
       volume = {331},
        pages = {894-900},
       adsurl = {https://ui.adsabs.harvard.edu/abs/1998A&A...331..894X}
}

@ARTICLE{Bianchi2011,
       author = {{Bianchi}, S. and {Xilouris}, E.~M.},
        title = "{The extent of dust in NGC 891 from Herschel/SPIRE images}",
      journal = {\aap},
         year = 2011,
        month = jul,
       volume = {531},
          eid = {L11},
        pages = {L11},
          doi = {10.1051/0004-6361/201116772},
archivePrefix = {arXiv},
       eprint = {1106.1518},
 primaryClass = {astro-ph.CO},
       adsurl = {https://ui.adsabs.harvard.edu/abs/2011A&A...531L..11B}
}

@ARTICLE{Mathis1983,
       author = {{Mathis}, J.~S. and {Mezger}, P.~G. and {Panagia}, N.},
        title = "{Interstellar radiation field and dust temperatures in the diffuse interstellar medium and in giant molecular clouds}",
      journal = {\aap},
         year = 1983,
        month = nov,
       volume = {128},
        pages = {212-229},
       adsurl = {https://ui.adsabs.harvard.edu/abs/1983A&A...128..212M}
}

@ARTICLE{Lai2025,
       author = {{Lai}, Thomas S.-Y. and {Duval}, Sara and {Smith}, J.~D.~T. and {Armus}, Lee and {Witt}, Adolf N. and {Sandstrom}, Karin and {Tarantino}, Elizabeth and {Baba}, Shunsuke and {Bolatto}, Alberto and {Donnelly}, Grant P. and {Hensley}, Brandon S. and {Imanishi}, Masatoshi and {Lenkic}, Laura and {Linden}, Sean and {Nakagawa}, Takao and {Spoon}, Henrik W.~W. and {Togi}, Aditya and {Whitcomb}, Cory M.},
        title = "{Resolving Emission from Small Dust Grains in the Blue Compact Dwarf II Zw 40 with JWST}",
      journal = {\apjl},
         year = 2025,
        month = oct,
       volume = {991},
       number = {2},
          eid = {L56},
        pages = {L56},
          doi = {10.3847/2041-8213/ae0467},
archivePrefix = {arXiv},
       eprint = {2509.04662},
 primaryClass = {astro-ph.GA},
       adsurl = {https://ui.adsabs.harvard.edu/abs/2025ApJ...991L..56L}
}

@ARTICLE{KaufmanNeufeld1996,
       author = {{Kaufman}, Michael J. and {Neufeld}, David A.},
        title = "{Far-Infrared Water Emission from Magnetohydrodynamic Shock Waves}",
      journal = {\apj},
         year = 1996,
        month = jan,
       volume = {456},
        pages = {611},
          doi = {10.1086/176683},
       adsurl = {https://ui.adsabs.harvard.edu/abs/1996ApJ...456..611K}
}

@ARTICLE{BoechatRoberty2009,
       author = {{Boechat-Roberty}, H.~M. and {Neves}, R. and {Pilling}, S. and {Lago}, A.~F. and {de Souza}, G.~G.~B.},
        title = "{Dissociation of the benzene molecule by ultraviolet and soft X-rays in circumstellar environment}",
      journal = {\mnras},
         year = 2009,
        month = apr,
       volume = {394},
       number = {2},
        pages = {810-817},
          doi = {10.1111/j.1365-2966.2008.14368.x},
archivePrefix = {arXiv},
       eprint = {0812.0116},
 primaryClass = {astro-ph},
       adsurl = {https://ui.adsabs.harvard.edu/abs/2009MNRAS.394..810B}
}

@ARTICLE{Laine2010,
       author = {{Laine}, Seppo and {Appleton}, Philip N. and {Gottesman}, Stephen T. and {Ashby}, Matthew L.~N. and {Garland}, Catherine A.},
        title = "{Warm Molecular Hydrogen Emission in Normal Edge-on Galaxies NGC 4565 and NGC 5907}",
      journal = {\aj},
         year = 2010,
        month = sep,
       volume = {140},
       number = {3},
        pages = {753-769},
          doi = {10.1088/0004-6256/140/3/753},
archivePrefix = {arXiv},
       eprint = {1007.4194},
 primaryClass = {astro-ph.CO},
       adsurl = {https://ui.adsabs.harvard.edu/abs/2010AJ....140..753L}
}

@INPROCEEDINGS{Madden2005,
       author = {{Madden}, Suzanne C.},
        title = "{Modeling the Dust Spectral Energy Distributions of Dwarf Galaxies}",
    booktitle = {The Spectral Energy Distributions of Gas-Rich Galaxies: Confronting Models with Data},
         year = 2005,
       editor = {{Popescu}, Cristina C. and {Tuffs}, Richard J.},
       series = {American Institute of Physics Conference Series},
       volume = {761},
        month = apr,
        pages = {223-230},
          doi = {10.1063/1.1913932},
archivePrefix = {arXiv},
       eprint = {astro-ph/0501666},
 primaryClass = {astro-ph},
       adsurl = {https://ui.adsabs.harvard.edu/abs/2005AIPC..761..223M}
}

@ARTICLE{Joblin1996,
       author = {{Joblin}, C. and {Tielens}, A.~G.~G.~M. and {Allamandola}, L.~J. and {Geballe}, T.~R.},
        title = "{Spatial Variation of the 3.29 and 3.40 Micron Emission Bands within Reflection Nebulae and the Photochemical Evolution of Methylated Polycyclic Aromatic Hydrocarbons}",
      journal = {\apj},
         year = 1996,
        month = feb,
       volume = {458},
        pages = {610},
          doi = {10.1086/176843},
       adsurl = {https://ui.adsabs.harvard.edu/abs/1996ApJ...458..610J}
}

@ARTICLE{Baron2024,
       author = {{Baron}, Dalya and {Sandstrom}, Karin M. and {Rosolowsky}, Erik and {Egorov}, Oleg V. and {Klessen}, Ralf S. and {Leroy}, Adam K. and {Boquien}, M{\'e}d{\'e}ric and {Schinnerer}, Eva and {Belfiore}, Francesco and {Groves}, Brent and {Chastenet}, J{\'e}r{\'e}my and {Dale}, Daniel A. and {Blanc}, Guillermo A. and {M{\'e}ndez-Delgado}, Jos{\'e} E. and {Koch}, Eric W. and {Grasha}, Kathryn and {Chevance}, M{\'e}lanie and {Thilker}, David A. and {Colombo}, Dario and {Williams}, Thomas G. and {Pathak}, Debosmita and {Sutter}, Jessica and {Brown}, Toby and {Wu}, John F. and {Peek}, Josh E.~G. and {Emsellem}, Eric and {Larson}, Kirsten L. and {Neumann}, Justus},
        title = "{PHANGS-ML: Dissecting Multiphase Gas and Dust in Nearby Galaxies Using Machine Learning}",
      journal = {\apj},
         year = 2024,
        month = jun,
       volume = {968},
       number = {1},
          eid = {24},
        pages = {24},
          doi = {10.3847/1538-4357/ad39e5},
archivePrefix = {arXiv},
       eprint = {2402.04330},
 primaryClass = {astro-ph.GA},
       adsurl = {https://ui.adsabs.harvard.edu/abs/2024ApJ...968...24B}
}

@ARTICLE{Haan2011,
       author = {{Haan}, S. and {Armus}, L. and {Laine}, S. and {Charmandaris}, V. and {Smith}, J.~D. and {Schweizer}, F. and {Brandl}, B. and {Evans}, A.~S. and {Surace}, J.~A. and {Diaz-Santos}, T. and {Beir{\~a}o}, P. and {Murphy}, E.~J. and {Stierwalt}, S. and {Hibbard}, J.~E. and {Yun}, M. and {Jarrett}, T.~H.},
        title = "{Spitzer IRS Spectral Mapping of the Toomre Sequence: Spatial Variations of PAH, Gas, and Dust Properties in nearby Major Mergers}",
      journal = {\apjs},
         year = 2011,
        month = dec,
       volume = {197},
       number = {2},
          eid = {27},
        pages = {27},
          doi = {10.1088/0067-0049/197/2/27},
archivePrefix = {arXiv},
       eprint = {1110.3046},
 primaryClass = {astro-ph.CO},
       adsurl = {https://ui.adsabs.harvard.edu/abs/2011ApJS..197...27H}
}

@ARTICLE{Stacey2010,
       author = {{Stacey}, G.~J. and {Charmandaris}, V. and {Boulanger}, F. and {Wu}, Yanling and {Combes}, F. and {Higdon}, S.~J.~U. and {Smith}, J.~D.~T. and {Nikola}, T.},
        title = "{The Energetics of Molecular Gas in NGC 891 from H$_{2}$ and Far-infrared Spectroscopy}",
      journal = {\apj},
         year = 2010,
        month = sep,
       volume = {721},
       number = {1},
        pages = {59-73},
          doi = {10.1088/0004-637X/721/1/59},
archivePrefix = {arXiv},
       eprint = {1007.4701},
 primaryClass = {astro-ph.CO},
       adsurl = {https://ui.adsabs.harvard.edu/abs/2010ApJ...721...59S}
}

@ARTICLE{Pendleton2025,
       author = {{Pendleton}, Yvonne J. and {Geballe}, T.~R. and {Chu}, Laurie E.~U. and {Decleir}, Marjorie and {Gordon}, Karl D. and {Tielens}, A.~G.~G.~M. and {Allamandola}, Louis J. and {Bouwman}, Jeroen and {Chiar}, J.~E. and {Dewitt}, Curtis and et al.},
        title = "{A Tale of Two Sightlines: Comparison of Hydrocarbon Dust Absorption Bands toward Cygnus OB2-12 and the Galactic Center}",
      journal = {\apj},
         year = 2025,
        month = oct,
       volume = {992},
       number = {1},
          eid = {8},
        pages = {8},
          doi = {10.3847/1538-4357/adfc3d},
archivePrefix = {arXiv},
       eprint = {2508.12601},
 primaryClass = {astro-ph.GA},
       adsurl = {https://ui.adsabs.harvard.edu/abs/2025ApJ...992....8P}
}

@ARTICLE{Bregman1989,
       author = {{Bregman}, J.~D. and {Allamandola}, L.~J. and {Tielens}, A.~G.~G.~M. and {Geballe}, T.~R. and {Witteborn}, F.~C.},
        title = "{The Infrared Emission Bands. II. A Spatial and Spectral Study of the Orion Bar}",
      journal = {\apj},
         year = 1989,
        month = sep,
       volume = {344},
        pages = {791},
          doi = {10.1086/167844},
       adsurl = {https://ui.adsabs.harvard.edu/abs/1989ApJ...344..791B}
}

@ARTICLE{Peeters2002,
       author = {{Peeters}, E. and {Hony}, S. and {Van Kerckhoven}, C. and {Tielens}, A.~G.~G.~M. and {Allamandola}, L.~J. and {Hudgins}, D.~M. and {Bauschlicher}, C.~W.},
        title = "{The rich 6 to 9 vec mu m spectrum of interstellar PAHs}",
      journal = {\aap},
         year = 2002,
        month = aug,
       volume = {390},
        pages = {1089-1113},
          doi = {10.1051/0004-6361:20020773},
archivePrefix = {arXiv},
       eprint = {astro-ph/0205400},
 primaryClass = {astro-ph},
       adsurl = {https://ui.adsabs.harvard.edu/abs/2002A&A...390.1089P}
}

@ARTICLE{Bauschlicher2008,
       author = {{Bauschlicher}, Jr., Charles W. and {Peeters}, Els and {Allamandola}, Louis J.},
        title = "{The Infrared Spectra of Very Large, Compact, Highly Symmetric, Polycyclic Aromatic Hydrocarbons (PAHs)}",
      journal = {\apj},
         year = 2008,
        month = may,
       volume = {678},
       number = {1},
        pages = {316-327},
          doi = {10.1086/533424},
archivePrefix = {arXiv},
       eprint = {0802.1071},
 primaryClass = {astro-ph},
       adsurl = {https://ui.adsabs.harvard.edu/abs/2008ApJ...678..316B}
}

@ARTICLE{Hony2001,
       author = {{Hony}, S. and {Van Kerckhoven}, C. and {Peeters}, E. and {Tielens}, A.~G.~G.~M. and {Hudgins}, D.~M. and {Allamandola}, L.~J.},
        title = "{The CH out-of-plane bending modes of PAH molecules in astrophysical environments}",
      journal = {\aap},
         year = 2001,
        month = may,
       volume = {370},
        pages = {1030-1043},
          doi = {10.1051/0004-6361:20010242},
archivePrefix = {arXiv},
       eprint = {astro-ph/0103035},
 primaryClass = {astro-ph},
       adsurl = {https://ui.adsabs.harvard.edu/abs/2001A&A...370.1030H}
}

@ARTICLE{Shannon2019,
       author = {{Shannon}, Matthew J. and {Boersma}, Christiaan},
        title = "{Examining the Class B to A Shift of the 7.7 {\ensuremath{\mu}}m PAH Band with the NASA Ames PAH IR Spectroscopic Database}",
      journal = {\apj},
         year = 2019,
        month = jan,
       volume = {871},
       number = {1},
          eid = {124},
        pages = {124},
          doi = {10.3847/1538-4357/aaf562},
archivePrefix = {arXiv},
       eprint = {1812.02178},
 primaryClass = {astro-ph.GA},
       adsurl = {https://ui.adsabs.harvard.edu/abs/2019ApJ...871..124S}
}

@ARTICLE{Peeters2024,
       author = {{Peeters}, Els and {Habart}, Emilie and {Bern{\'e}}, Olivier and {Sidhu}, Ameek and {Chown}, Ryan and {Van De Putte}, Dries and {Trahin}, Boris and {Schroetter}, Ilane and {Canin}, Am{\'e}lie and {Alarc{\'o}n}, Felipe and {Schefter}, Bethany and {Khan}, Baria and {Pasquini}, Sofia and {Tielens}, Alexander G.~G.~M. and {Wolfire}, Mark G. and {Dartois}, Emmanuel and {Goicoechea}, Javier R. and {Maragkoudakis}, Alexandros and {Onaka}, Takashi and {Pound}, Marc W. and {Vicente}, S{\'\i}lvia and {Abergel}, Alain and {Bergin}, Edwin A. and {Bernard-Salas}, Jeronimo and {Boersma}, Christiaan and {Bron}, Emeric and {Cami}, Jan and {Cuadrado}, Sara and {Dicken}, Daniel and {Elyajouri}, Meriem and {Fuente}, Asunci{\'o}n and {Gordon}, Karl D. and {Issa}, Lina and {Joblin}, Christine and {Kannavou}, Olga and {Lacinbala}, Ozan and {Languignon}, David and {Le Gal}, Romane and {Meshaka}, Raphael and {Okada}, Yoko and {Robberto}, Massimo and {R{\"o}llig}, Markus and {Schirmer}, Thi{\'e}baut and {Tabone}, Benoit and {Zannese}, Marion and {Aleman}, Isabel and {Allamandola}, Louis and {Auchettl}, Rebecca and {Baratta}, Giuseppe Antonio and {Bejaoui}, Salma and {Bera}, Partha P. and {Black}, John H. and {Boulanger}, Francois and {Bouwman}, Jordy and {Brandl}, Bernhard and {Brechignac}, Philippe and {Br{\"u}nken}, Sandra and {Buragohain}, Mridusmita and {Burkhardt}, Andrew and {Candian}, Alessandra and {Cazaux}, St{\'e}phanie and {Cernicharo}, Jose and {Chabot}, Marin and {Chakraborty}, Shubhadip and {Champion}, Jason and {Colgan}, Sean W.~J. and {Cooke}, Ilsa R. and {Coutens}, Audrey and {Cox}, Nick L.~J. and {Demyk}, Karine and {Meyer}, Jennifer Donovan and {Foschino}, Sacha and {Garc{\'\i}a-Lario}, Pedro and {Gerin}, Maryvonne and {Gottlieb}, Carl A. and {Guillard}, Pierre and {Gusdorf}, Antoine and {Hartigan}, Patrick and {He}, Jinhua and {Herbst}, Eric and {Hornekaer}, Liv and {J{\"a}ger}, Cornelia and {Janot-Pacheco}, Eduardo and {Kaufman}, Michael and {Kendrew}, Sarah and {Kirsanova}, Maria S. and {Klaassen}, Pamela and {Kwok}, Sun and {Labiano}, {\'A}lvaro and {Lai}, Thomas S.-Y. and {Lee}, Timothy J. and {Lefloch}, Bertrand and {Le Petit}, Franck and {Li}, Aigen and {Linz}, Hendrik and {Mackie}, Cameron J. and {Madden}, Suzanne C. and {Mascetti}, Jo{\"e}lle and {McGuire}, Brett A. and {Merino}, Pablo and {Micelotta}, Elisabetta R. and {Misselt}, Karl and {Morse}, Jon A. and {Mulas}, Giacomo and {Neelamkodan}, Naslim and {Ohsawa}, Ryou and {Paladini}, Roberta and {Palumbo}, Maria Elisabetta and {Pathak}, Amit and {Pendleton}, Yvonne J. and {Petrignani}, Annemieke and {Pino}, Thomas and {Puga}, Elena and {Rangwala}, Naseem and {Rapacioli}, Mathias and {Ricca}, Alessandra and {Roman-Duval}, Julia and {Roser}, Joseph and {Roueff}, Evelyne and {Rouill{\'e}}, Ga{\"e}l and {Salama}, Farid and {Sales}, Dinalva A. and {Sandstrom}, Karin and {Sarre}, Peter and {Sciamma-O'Brien}, Ella and {Sellgren}, Kris and {Shenoy}, Sachindev S. and {Teyssier}, David and {Thomas}, Richard D. and {Togi}, Aditya and {Verstraete}, Laurent and {Witt}, Adolf N. and {Wootten}, Alwyn and {Ysard}, Nathalie and {Zettergren}, Henning and {Zhang}, Yong and {Zhang}, Ziwei E. and {Zhen}, Junfeng},
        title = "{PDRs4All: III. JWST's NIR spectroscopic view of the Orion Bar}",
      journal = {\aap},
         year = 2024,
        month = may,
       volume = {685},
          eid = {A74},
        pages = {A74},
          doi = {10.1051/0004-6361/202348244},
archivePrefix = {arXiv},
       eprint = {2310.08720},
 primaryClass = {astro-ph.GA},
       adsurl = {https://ui.adsabs.harvard.edu/abs/2024A&A...685A..74P}
}

@ARTICLE{Peeters2004,
       author = {{Peeters}, E. and {Mattioda}, A.~L. and {Hudgins}, D.~M. and {Allamandola}, L.~J.},
        title = "{Polycyclic Aromatic Hydrocarbon Emission in the 15-21 Micron Region}",
      journal = {\apjl},
         year = 2004,
        month = dec,
       volume = {617},
       number = {1},
        pages = {L65-L68},
          doi = {10.1086/427186},
       adsurl = {https://ui.adsabs.harvard.edu/abs/2004ApJ...617L..65P}
}

@ARTICLE{CDS,
       author = {{Wenger}, M. and {Ochsenbein}, F. and {Egret}, D. and {Dubois}, P. and {Bonnarel}, F. and {Borde}, S. and {Genova}, F. and {Jasniewicz}, G. and {Lalo{\"e}}, S. and {Lesteven}, S. and {Monier}, R.},
        title = "{The SIMBAD astronomical database. The CDS reference database for astronomical objects}",
      journal = {\aaps},
         year = 2000,
        month = apr,
       volume = {143},
        pages = {9-22},
          doi = {10.1051/aas:2000332},
archivePrefix = {arXiv},
       eprint = {astro-ph/0002110},
 primaryClass = {astro-ph},
       adsurl = {https://ui.adsabs.harvard.edu/abs/2000A&AS..143....9W}
}

@ARTICLE{Chown2021,
       author = {{Chown}, Ryan and {Li}, Cheng and {Parker}, Laura and {Wilson}, Christine D. and {Li}, Niu and {Gao}, Yang},
        title = "{A new estimator of resolved molecular gas in nearby galaxies}",
      journal = {\mnras},
         year = 2021,
        month = jan,
       volume = {500},
       number = {1},
        pages = {1261-1278},
          doi = {10.1093/mnras/staa3288},
archivePrefix = {arXiv},
       eprint = {2007.00174},
 primaryClass = {astro-ph.GA},
       adsurl = {https://ui.adsabs.harvard.edu/abs/2021MNRAS.500.1261C}
}

@ARTICLE{Kerkeni2022,
       author = {{Kerkeni}, B. and {Garc{\'\i}a-Bernete}, I. and {Rigopoulou}, D. and {Tew}, D.~P. and {Roche}, P.~F. and {Clary}, D.~C.},
        title = "{Probing computational methodologies in predicting mid-infrared spectra for large polycyclic aromatic hydrocarbons}",
      journal = {\mnras},
         year = 2022,
        month = jul,
       volume = {513},
       number = {3},
        pages = {3663-3681},
          doi = {10.1093/mnras/stac976},
archivePrefix = {arXiv},
       eprint = {2210.00955},
 primaryClass = {astro-ph.GA},
       adsurl = {https://ui.adsabs.harvard.edu/abs/2022MNRAS.513.3663K}
}

@ARTICLE{Bauschlicher2009,
       author = {{Bauschlicher}, Jr., Charles W. and {Peeters}, Els and {Allamandola}, Louis J.},
        title = "{The Infrared Spectra of Very Large Irregular Polycyclic Aromatic Hydrocarbons (PAHs): Observational Probes of Astronomical PAH Geometry, Size, and Charge}",
      journal = {\apj},
         year = 2009,
        month = may,
       volume = {697},
       number = {1},
        pages = {311-327},
          doi = {10.1088/0004-637X/697/1/311},
archivePrefix = {arXiv},
       eprint = {0903.0412},
 primaryClass = {astro-ph.GA},
       adsurl = {https://ui.adsabs.harvard.edu/abs/2009ApJ...697..311B}
}

@ARTICLE{Draine2007,
       author = {{Draine}, B.~T. and {Dale}, D.~A. and {Bendo}, G. and {Gordon}, K.~D. and {Smith}, J.~D.~T. and {Armus}, L. and {Engelbracht}, C.~W. and {Helou}, G. and {Kennicutt}, Jr., R.~C. and {Li}, A. and {Roussel}, H. and {Walter}, F. and {Calzetti}, D. and {Moustakas}, J. and {Murphy}, E.~J. and {Rieke}, G.~H. and {Bot}, C. and {Hollenbach}, D.~J. and {Sheth}, K. and {Teplitz}, H.~I.},
        title = "{Dust Masses, PAH Abundances, and Starlight Intensities in the SINGS Galaxy Sample}",
      journal = {\apj},
         year = 2007,
        month = jul,
       volume = {663},
       number = {2},
        pages = {866-894},
          doi = {10.1086/518306},
archivePrefix = {arXiv},
       eprint = {astro-ph/0703213},
 primaryClass = {astro-ph},
       adsurl = {https://ui.adsabs.harvard.edu/abs/2007ApJ...663..866D}
}

@ARTICLE{RemyRuyer2013,
       author = {{R{\'e}my-Ruyer}, A. and {Madden}, S.~C. and {Galliano}, F. and {Hony}, S. and {Sauvage}, M. and {Bendo}, G.~J. and {Roussel}, H. and {Pohlen}, M. and {Smith}, M.~W.~L. and {Galametz}, M. and {Cormier}, D. and {Lebouteiller}, V. and {Wu}, R. and {Baes}, M. and {Barlow}, M.~J. and {Boquien}, M. and {Boselli}, A. and {Ciesla}, L. and {De Looze}, I. and {Karczewski}, O. {\L}. and {Panuzzo}, P. and {Spinoglio}, L. and {Vaccari}, M. and {Wilson}, C.~D.},
        title = "{Revealing the cold dust in low-metallicity environments. I. Photometry analysis of the Dwarf Galaxy Survey with Herschel}",
      journal = {\aap},
         year = 2013,
        month = sep,
       volume = {557},
          eid = {A95},
        pages = {A95},
          doi = {10.1051/0004-6361/201321602},
archivePrefix = {arXiv},
       eprint = {1309.1371},
 primaryClass = {astro-ph.CO},
       adsurl = {https://ui.adsabs.harvard.edu/abs/2013A&A...557A..95R}
}

@ARTICLE{Davies2017,
       author = {{Davies}, J.~I. and {Baes}, M. and {Bianchi}, S. and {Jones}, A. and {Madden}, S. and {Xilouris}, M. and {Bocchio}, M. and {Casasola}, V. and {Cassara}, L. and {Clark}, C. and {De Looze}, I. and {Evans}, R. and {Fritz}, J. and {Galametz}, M. and {Galliano}, F. and {Lianou}, S. and {Mosenkov}, A.~V. and {Smith}, M. and {Verstocken}, S. and {Viaene}, S. and {Vika}, M. and {Wagle}, G. and {Ysard}, N.},
        title = "{DustPedia: A Definitive Study of Cosmic Dust in the Local Universe}",
      journal = {\pasp},
         year = 2017,
        month = apr,
       volume = {129},
       number = {974},
        pages = {044102},
          doi = {10.1088/1538-3873/129/974/044102},
archivePrefix = {arXiv},
       eprint = {1609.06138},
 primaryClass = {astro-ph.GA},
       adsurl = {https://ui.adsabs.harvard.edu/abs/2017PASP..129d4102D}
}

@ARTICLE{Aniano2020,
       author = {{Aniano}, G. and {Draine}, B.~T. and {Hunt}, L.~K. and {Sandstrom}, K. and {Calzetti}, D. and {Kennicutt}, R.~C. and {Dale}, D.~A. and {Galametz}, M. and {Gordon}, K.~D. and {Leroy}, A.~K. and {Smith}, J.-D.~T. and {Roussel}, H. and {Sauvage}, M. and {Walter}, F. and {Armus}, L. and {Bolatto}, A.~D. and {Boquien}, M. and {Crocker}, A. and {De Looze}, I. and {Donovan Meyer}, J. and {Helou}, G. and {Hinz}, J. and {Johnson}, B.~D. and {Koda}, J. and {Miller}, A. and {Montiel}, E. and {Murphy}, E.~J. and {Rela{\~n}o}, M. and {Rix}, H.-W. and {Schinnerer}, E. and {Skibba}, R. and {Wolfire}, M.~G. and {Engelbracht}, C.~W.},
        title = "{Modeling Dust and Starlight in Galaxies Observed by Spitzer and Herschel: The KINGFISH Sample}",
      journal = {\apj},
         year = 2020,
        month = feb,
       volume = {889},
       number = {2},
          eid = {150},
        pages = {150},
          doi = {10.3847/1538-4357/ab5fdb},
archivePrefix = {arXiv},
       eprint = {1912.04914},
 primaryClass = {astro-ph.GA},
       adsurl = {https://ui.adsabs.harvard.edu/abs/2020ApJ...889..150A}
}

@ARTICLE{Khan2025,
       author = {{Khan}, Baria and {Abbott}, Benjamin and {Peeters}, Els and {Tielens}, Alexander G.~G.~M. and {Onaka}, Takashi and {Cami}, Jan and {Schefter}, Bethany and {Boersma}, Christiaan and {Dartois}, Emmanuel and {Goicoechea}, Javier R. and {Maragkoudakis}, Alexandros and {Van De Putte}, Dries and {Buragohain}, Mridusmita and {Candian}, Alessandra and {Labiano}, {\'A}lvaro and {Lai}, Thomas S.-Y. and {Ricca}, Alessandra and {Sales}, Dinalva A. and {Zhang}, Yong and {Sidhu}, Ameek and {Chown}, Ryan and {Canin}, Am{\'e}lie and {Trahin}, Boris and {Schroetter}, Ilane and {Kannavou}, Olga and {Alarc{\'o}n}, Felipe and {Bern{\'e}}, Olivier and {Habart}, Emilie},
        title = "{PDRs4All: XIV. Probing CH out-of-plane bending modes of PAH molecules in the Orion Bar with JWST}",
      journal = {\aap},
         year = 2025,
        month = jul,
       volume = {699},
          eid = {A133},
        pages = {A133},
          doi = {10.1051/0004-6361/202554096},
       adsurl = {https://ui.adsabs.harvard.edu/abs/2025A&A...699A.133K}
}

@ARTICLE{Boersma2014,
       author = {{Boersma}, C. and {Bregman}, J. and {Allamandola}, L.~J.},
        title = "{Properties of Polycyclic Aromatic Hydrocarbons in the Northwest Photon Dominated Region of NGC 7023. II. Traditional PAH Analysis Using k-means as a Visualization Tool}",
      journal = {\apj},
         year = 2014,
        month = nov,
       volume = {795},
       number = {2},
          eid = {110},
        pages = {110},
          doi = {10.1088/0004-637X/795/2/110},
       adsurl = {https://ui.adsabs.harvard.edu/abs/2014ApJ...795..110B}
}

@ARTICLE{THEMIS2,
       author = {{Ysard}, N. and {Jones}, A.~P. and {Guillet}, V. and {Demyk}, K. and {Decleir}, M. and {Verstraete}, L. and {Choubani}, I. and {Miville-Desch{\^e}nes}, M.-A. and {Fanciullo}, L.},
        title = "{THEMIS 2.0: A self-consistent model for dust extinction, emission, and polarisation}",
      journal = {\aap},
         year = 2024,
        month = apr,
       volume = {684},
          eid = {A34},
        pages = {A34},
          doi = {10.1051/0004-6361/202348391},
archivePrefix = {arXiv},
       eprint = {2401.07739},
 primaryClass = {astro-ph.GA},
       adsurl = {https://ui.adsabs.harvard.edu/abs/2024A&A...684A..34Y}
}

@ARTICLE{RichieHensley2025,
       author = {{Richie}, Helena M. and {Hensley}, Brandon S.},
        title = "{PAH Emission Spectra and Band Ratios for Arbitrary Radiation Fields with the Single Photon Approximation}",
      journal = {arXiv e-prints},
         year = 2025,
        month = oct,
          eid = {arXiv:2510.16861},
        pages = {arXiv:2510.16861},
          doi = {10.48550/arXiv.2510.16861},
archivePrefix = {arXiv},
       eprint = {2510.16861},
 primaryClass = {astro-ph.GA},
       adsurl = {https://ui.adsabs.harvard.edu/abs/2025arXiv251016861R}
}

@ARTICLE{ShivaeiBoogaard2024,
       author = {{Shivaei}, Irene and {Boogaard}, Leindert A.},
        title = "{The tight correlation between PAH and CO emission from z {\ensuremath{\sim}} 0 to 4}",
      journal = {\aap},
         year = 2024,
        month = nov,
       volume = {691},
          eid = {L2},
        pages = {L2},
          doi = {10.1051/0004-6361/202451826},
archivePrefix = {arXiv},
       eprint = {2409.05710},
 primaryClass = {astro-ph.GA},
       adsurl = {https://ui.adsabs.harvard.edu/abs/2024A&A...691L...2S}
}

@ARTICLE{Misselt2025,
       author = {{Misselt}, K. and {Witt}, A.~N. and {Gordon}, K.~D. and {Van De Putte}, D. and {Trahin}, B. and {Abergel}, A. and {Noriega-Crespo}, A. and {Guillard}, P. and {Zannese}, M. and {Dell'ova}, P. and {Baes}, M. and {Klaassen}, P. and {Ysard}, N.},
        title = "{JWST observations of photodissociation regions: I. Aliphatic and aromatic carbonaceous dust, ices, and gas-phase spectral line inventory}",
      journal = {\aap},
         year = 2025,
        month = aug,
       volume = {700},
          eid = {A158},
        pages = {A158},
          doi = {10.1051/0004-6361/202554851},
archivePrefix = {arXiv},
       eprint = {2506.20468},
 primaryClass = {astro-ph.GA},
       adsurl = {https://ui.adsabs.harvard.edu/abs/2025A&A...700A.158M}
}

@ARTICLE{Casasola2020,
       author = {{Casasola}, V. and {Bianchi}, S. and {De Vis}, P. and {Magrini}, L. and {Corbelli}, E. and {Clark}, C.~J.~R. and {Fritz}, J. and {Nersesian}, A. and {Viaene}, S. and {Baes}, M. and {Cassar{\`a}}, L.~P. and {Davies}, J. and {De Looze}, I. and {Dobbels}, W. and {Galametz}, M. and {Galliano}, F. and {Jones}, A.~P. and {Madden}, S.~C. and {Mosenkov}, A.~V. and {Tr{\v{c}}ka}, A. and {Xilouris}, E.},
        title = "{The ISM scaling relations in DustPedia late-type galaxies: A benchmark study for the Local Universe}",
      journal = {\aap},
         year = 2020,
        month = jan,
       volume = {633},
          eid = {A100},
        pages = {A100},
          doi = {10.1051/0004-6361/201936665},
archivePrefix = {arXiv},
       eprint = {1911.09187},
 primaryClass = {astro-ph.GA},
       adsurl = {https://ui.adsabs.harvard.edu/abs/2020A&A...633A.100C}
}

@ARTICLE{Peek2015,
       author = {{Peek}, J.~E.~G. and {M{\'e}nard}, Brice and {Corrales}, Lia},
        title = "{Dust in the Circumgalactic Medium of Low-redshift Galaxies}",
      journal = {\apj},
         year = 2015,
        month = nov,
       volume = {813},
       number = {1},
          eid = {7},
        pages = {7},
          doi = {10.1088/0004-637X/813/1/7},
archivePrefix = {arXiv},
       eprint = {1411.3333},
 primaryClass = {astro-ph.GA},
       adsurl = {https://ui.adsabs.harvard.edu/abs/2015ApJ...813....7P}
}

@ARTICLE{Lu2024,
       author = {{Lu}, Li-Yuan and {Li}, Jiang-Tao and {Vargas}, Carlos J. and {Fang}, Taotao and {Benjamin}, Robert A. and {Bregman}, Joel N. and {Dettmar}, Ralf-J{\"u}rgen and {English}, Jayanne and {Heald}, George H. and {Jiang}, Yan and {Daniel Wang}, Q. and {Yang}, Yang},
        title = "{eDIG-CHANGES: III. The lagging eDIG revealed by multi-slit spectroscopy of NGC 891}",
      journal = {\aap},
         year = 2024,
        month = nov,
       volume = {691},
          eid = {A217},
        pages = {A217},
          doi = {10.1051/0004-6361/202451783},
archivePrefix = {arXiv},
       eprint = {2410.02347},
 primaryClass = {astro-ph.GA},
       adsurl = {https://ui.adsabs.harvard.edu/abs/2024A&A...691A.217L}
}

@ARTICLE{Levy2018,
       author = {{Levy}, Rebecca C. and {Bolatto}, Alberto D. and {Teuben}, Peter and {S{\'a}nchez}, Sebasti{\'a}n F. and {Barrera-Ballesteros}, Jorge K. and {Blitz}, Leo and {Colombo}, Dario and {Garc{\'\i}a-Benito}, Rub{\'e}n and {Herrera-Camus}, Rodrigo and {Husemann}, Bernd and {Kalinova}, Veselina and {Lan}, Tian and {Leung}, Gigi Y.~C. and {Mast}, Dami{\'a}n and {Utomo}, Dyas and {van de Ven}, Glenn and {Vogel}, Stuart N. and {Wong}, Tony},
        title = "{The EDGE-CALIFA Survey: Molecular and Ionized Gas Kinematics in Nearby Galaxies}",
      journal = {\apj},
         year = 2018,
        month = jun,
       volume = {860},
       number = {2},
          eid = {92},
        pages = {92},
          doi = {10.3847/1538-4357/aac2e5},
archivePrefix = {arXiv},
       eprint = {1804.05853},
 primaryClass = {astro-ph.GA},
       adsurl = {https://ui.adsabs.harvard.edu/abs/2018ApJ...860...92L}
}

@ARTICLE{Levy2019,
       author = {{Levy}, Rebecca C. and {Bolatto}, Alberto D. and {S{\'a}nchez}, Sebasti{\'a}n F. and {Blitz}, Leo and {Colombo}, Dario and {Kalinova}, Veselina and {L{\'o}pez-Cob{\'a}}, Carlos and {Ostriker}, Eve C. and {Teuben}, Peter and {Utomo}, Dyas and {Vogel}, Stuart N. and {Wong}, Tony},
        title = "{The EDGE-CALIFA Survey: Evidence for Pervasive Extraplanar Diffuse Ionized Gas in Nearby Edge-on Galaxies}",
      journal = {\apj},
         year = 2019,
        month = sep,
       volume = {882},
       number = {2},
          eid = {84},
        pages = {84},
          doi = {10.3847/1538-4357/ab2ed4},
archivePrefix = {arXiv},
       eprint = {1905.05196},
 primaryClass = {astro-ph.GA},
       adsurl = {https://ui.adsabs.harvard.edu/abs/2019ApJ...882...84L}
}

@ARTICLE{Villanueva2025,
       author = {{Villanueva}, V. and {Bolatto}, A.~D. and {Herrera-Camus}, R. and {Leroy}, A. and {Fisher}, D.~B. and {Levy}, R.~C. and {B{\"o}ker}, T. and {Boogaard}, L. and {Cronin}, S.~A. and {Dale}, D.~A. and {Emig}, K. and {De Looze}, I. and {Donnelly}, G.~P. and {Lai}, T.~S.-Y. and {Lenkic}, L. and {Lopez}, L.~A. and {Lopez}, S. and {Meier}, D.~S. and {Ott}, J. and {Relano}, M. and {Smith}, J.~D. and {Tarantino}, E. and {Veilleux}, S. and {Walter}, F. and {van der Werf}, P.},
        title = "{JWST Observations of Starbursts: Relations between PAH features and CO clouds in the starburst galaxy M 82}",
      journal = {\aap},
         year = 2025,
        month = mar,
       volume = {695},
          eid = {A202},
        pages = {A202},
          doi = {10.1051/0004-6361/202553891},
archivePrefix = {arXiv},
       eprint = {2501.14893},
 primaryClass = {astro-ph.GA},
       adsurl = {https://ui.adsabs.harvard.edu/abs/2025A&A...695A.202V}
}

@ARTICLE{Lopez2026,
       author = {{Lopez}, Sebastian and {Ring}, Colton and {Leroy}, Adam K. and {Cronin}, Serena A. and {Bolatto}, Alberto D. and {Lopez}, Laura A. and {Villanueva}, Vicente and {Fisher}, Deanne B. and {Thompson}, Todd A. and {Donnelly}, Grant P. and {Armus}, Lee and {B{\"o}ker}, Torsten and {Boogaard}, Leindert A. and {Boyer}, Martha L. and {Chown}, Ryan and {Dale}, Daniel A. and {Donaghue}, Keaton and {Emig}, Kimberly and {Glover}, Simon C.~O. and {Herrera-Camus}, Rodrigo and {Klessen}, Ralf S. and {Lai}, Thomas S.-Y. and {Lenki{\'c}}, Laura and {Levy}, Rebecca C. and {Meier}, David S. and {Mills}, Elisabeth and {Ott}, Juergen and {Skillman}, Evan D. and {Smith}, J.~D.~T. and {Tarantino}, Elizabeth J. and {Veilleux}, Sylvain and {Walter}, Fabian and {van der Werf}, Paul P.},
        title = "{JWST Observations of Starbursts: Polycyclic Aromatic Hydrocarbons Closely Trace the Cool Phase of M82's Galactic Wind}",
      journal = {\apjl},
         year = 2026,
        month = mar,
       volume = {999},
       number = {1},
          eid = {L7},
        pages = {L7},
          doi = {10.3847/2041-8213/ae4508},
archivePrefix = {arXiv},
       eprint = {2510.01314},
 primaryClass = {astro-ph.GA},
       adsurl = {https://ui.adsabs.harvard.edu/abs/2026ApJ...999L...7L}
}

@ARTICLE{Matsumoto2026,
       author = {{Matsumoto}, Kosei and {Sommovigo}, Laura and {Gebek}, Andrea and {Nagamine}, Kentaro and {Nersesian}, Angelos and {Baes}, Maarten and {De Looze}, Ilse and {van der Wel}, Arjen and {Somerville}, Rachel and {Romano}, Leonard E.~C. and {Cochrane}, Rachel K.},
        title = "{Evolution of galaxy attenuation curves driven by evolving dust mass and grain size distributions}",
      journal = {\aap},
         year = 2026,
        month = jan,
       volume = {705},
          eid = {A75},
        pages = {A75},
          doi = {10.1051/0004-6361/202555658},
archivePrefix = {arXiv},
       eprint = {2508.21157},
 primaryClass = {astro-ph.GA},
       adsurl = {https://ui.adsabs.harvard.edu/abs/2026A&A...705A..75M}
}

@ARTICLE{McCleary2026,
       author = {{McCleary}, Jacqueline E. and {Huff}, Eric M. and {Bartlett}, James G. and {Hensley}, Brandon S.},
        title = "{A Detection of Circumgalactic Dust at Megaparsec Scales with Maximum Likelihood Estimation}",
      journal = {\apj},
         year = 2026,
        month = apr,
       volume = {1000},
       number = {2},
          eid = {313},
        pages = {313},
          doi = {10.3847/1538-4357/ae4c3d},
archivePrefix = {arXiv},
       eprint = {2503.04098},
 primaryClass = {astro-ph.GA},
       adsurl = {https://ui.adsabs.harvard.edu/abs/2026ApJ..1000..313M}
}

@ARTICLE{RossaDettmar2003,
       author = {{Rossa}, J. and {Dettmar}, R.-J.},
        title = "{An H{\ensuremath{\alpha}}  survey aiming at the detection of extraplanar diffuse ionized gas in halos of edge-on spiral galaxies.  I. How common are gaseous halos among non-starburst galaxies?}",
      journal = {\aap},
         year = 2003,
        month = aug,
       volume = {406},
        pages = {493-503},
          doi = {10.1051/0004-6361:20030615},
archivePrefix = {arXiv},
       eprint = {astro-ph/0304452},
 primaryClass = {astro-ph},
       adsurl = {https://ui.adsabs.harvard.edu/abs/2003A&A...406..493R}
}

@ARTICLE{Strickland2004,
       author = {{Strickland}, David K. and {Heckman}, Timothy M. and {Colbert}, Edward J.~M. and {Hoopes}, Charles G. and {Weaver}, Kimberly A.},
        title = "{A High Spatial Resolution X-Ray and H{\ensuremath{\alpha}} Study of Hot Gas in the Halos of Star-forming Disk Galaxies. II. Quantifying Supernova Feedback}",
      journal = {\apj},
         year = 2004,
        month = may,
       volume = {606},
       number = {2},
        pages = {829-852},
          doi = {10.1086/383136},
archivePrefix = {arXiv},
       eprint = {astro-ph/0306598},
 primaryClass = {astro-ph},
       adsurl = {https://ui.adsabs.harvard.edu/abs/2004ApJ...606..829S}
}

@ARTICLE{Rand2011,
       author = {{Rand}, Richard J. and {Wood}, Kenneth and {Benjamin}, Robert A. and {Meidt}, Sharon E.},
        title = "{Infrared Spectroscopy of the Diffuse Ionized Halos of Edge-on Galaxies}",
      journal = {\apj},
         year = 2011,
        month = feb,
       volume = {728},
       number = {2},
          eid = {163},
        pages = {163},
          doi = {10.1088/0004-637X/728/2/163},
archivePrefix = {arXiv},
       eprint = {1101.1491},
 primaryClass = {astro-ph.CO},
       adsurl = {https://ui.adsabs.harvard.edu/abs/2011ApJ...728..163R}
}

@ARTICLE{BernardSalas2009,
       author = {{Bernard-Salas}, J. and {Spoon}, H.~W.~W. and {Charmandaris}, V. and {Lebouteiller}, V. and {Farrah}, D. and {Devost}, D. and {Brandl}, B.~R. and {Wu}, Yanling and {Armus}, L. and {Hao}, L. and {Sloan}, G.~C. and {Weedman}, D. and {Houck}, J.~R.},
        title = "{A Spitzer High-resolution Mid-Infrared Spectral Atlas of Starburst Galaxies}",
      journal = {\apjs},
         year = 2009,
        month = oct,
       volume = {184},
       number = {2},
        pages = {230-247},
          doi = {10.1088/0067-0049/184/2/230},
archivePrefix = {arXiv},
       eprint = {0908.2812},
 primaryClass = {astro-ph.CO},
       adsurl = {https://ui.adsabs.harvard.edu/abs/2009ApJS..184..230B}
}

@ARTICLE{Kemper2004,
       author = {{Kemper}, F. and {Vriend}, W.~J. and {Tielens}, A.~G.~G.~M.},
        title = "{The Absence of Crystalline Silicates in the Diffuse Interstellar Medium}",
      journal = {\apj},
         year = 2004,
        month = jul,
       volume = {609},
       number = {2},
        pages = {826-837},
          doi = {10.1086/421339},
archivePrefix = {arXiv},
       eprint = {astro-ph/0403609},
 primaryClass = {astro-ph},
       adsurl = {https://ui.adsabs.harvard.edu/abs/2004ApJ...609..826K}
}

@INCOLLECTION{Field1975,
       author = {{Field}, G.~B.},
        title = "{The composition of interstellar dust.}",
    booktitle = {The Dusty Universe},
         year = 1975,
       editor = {{Field}, G.~B. and {Cameron}, A.~G.~W.},
        pages = {89-112},
       adsurl = {https://ui.adsabs.harvard.edu/abs/1975duun.book...89F}
}

@ARTICLE{Min2008,
       author = {{Min}, M. and {Waters}, L.~B.~F.~M. and {de Koter}, A. and {Hovenier}, J.~W. and {Keller}, L.~P. and {Markwick-Kemper}, F.},
        title = "{The shape and composition of interstellar silicate grains}",
      journal = {\aap},
         year = 2008,
        month = aug,
       volume = {486},
       number = {3},
        pages = {779-780},
          doi = {10.1051/0004-6361:20065436e},
       adsurl = {https://ui.adsabs.harvard.edu/abs/2008A&A...486..779M}
}

@ARTICLE{WD2001,
       author = {{Weingartner}, Joseph C. and {Draine}, B.~T.},
        title = "{Dust Grain-Size Distributions and Extinction in the Milky Way, Large Magellanic Cloud, and Small Magellanic Cloud}",
      journal = {\apj},
         year = 2001,
        month = feb,
       volume = {548},
       number = {1},
        pages = {296-309},
          doi = {10.1086/318651},
archivePrefix = {arXiv},
       eprint = {astro-ph/0008146},
 primaryClass = {astro-ph},
       adsurl = {https://ui.adsabs.harvard.edu/abs/2001ApJ...548..296W}
}

@ARTICLE{Mannucci2010,
       author = {{Mannucci}, F. and {Cresci}, G. and {Maiolino}, R. and {Marconi}, A. and {Gnerucci}, A.},
        title = "{A fundamental relation between mass, star formation rate and metallicity in local and high-redshift galaxies}",
      journal = {\mnras},
         year = 2010,
        month = nov,
       volume = {408},
       number = {4},
        pages = {2115-2127},
          doi = {10.1111/j.1365-2966.2010.17291.x},
archivePrefix = {arXiv},
       eprint = {1005.0006},
 primaryClass = {astro-ph.CO},
       adsurl = {https://ui.adsabs.harvard.edu/abs/2010MNRAS.408.2115M}
}

@ARTICLE{Popescu2011,
       author = {{Popescu}, C.~C. and {Tuffs}, R.~J. and {Dopita}, M.~A. and {Fischera}, J. and {Kylafis}, N.~D. and {Madore}, B.~F.},
        title = "{Modelling the spectral energy distribution of galaxies. V. The dust and PAH emission SEDs of disk galaxies}",
      journal = {\aap},
         year = 2011,
        month = mar,
       volume = {527},
          eid = {A109},
        pages = {A109},
          doi = {10.1051/0004-6361/201015217},
archivePrefix = {arXiv},
       eprint = {1011.2942},
 primaryClass = {astro-ph.CO},
       adsurl = {https://ui.adsabs.harvard.edu/abs/2011A&A...527A.109P}
}

@ARTICLE{2018AJ....156..123A, author = {{Astropy Collaboration} and {Price-Whelan}, A.~M. and {Sip{\H o}cz}, B.~M. and {G{\"u}nther}, H.~M. and {Lim}, P.~L. and {Crawford}, S.~M. and {Conseil}, S. and {Shupe}, D.~L. and {Craig}, M.~W. and {Dencheva}, N. and {Ginsburg}, A. and {VanderPlas}, J.~T. and {Bradley}, L.~D. and {P{\'e}rez-Su{\'a}rez}, D. and {de Val-Borro}, M. and {Aldcroft}, T.~L. and {Cruz}, K.~L. and {Robitaille}, T.~P. and {Tollerud}, E.~J. and {Ardelean}, C. and {Babej}, T. and {Bach}, Y.~P. and {Bachetti}, M. and {Bakanov}, A.~V. and {Bamford}, S.~P. and {Barentsen}, G. and {Barmby}, P. and {Baumbach}, A. and {Berry}, K.~L. and {Biscani}, F. and {Boquien}, M. and {Bostroem}, K.~A. and {Bouma}, L.~G. and {Brammer}, G.~B. and {Bray}, E.~M. and {Breytenbach}, H. and {Buddelmeijer}, H. and {Burke}, D.~J. and {Calderone}, G. and {Cano Rodr{\'{\i}}guez}, J.~L. and {Cara}, M. and {Cardoso}, J.~V.~M. and {Cheedella}, S. and {Copin}, Y. and {Corrales}, L. and {Crichton}, D. and {D'Avella}, D. and {Deil}, C. and {Depagne}, {\'E}. and {Dietrich}, J.~P. and {Donath}, A. and {Droettboom}, M. and {Earl}, N. and {Erben}, T. and {Fabbro}, S. and {Ferreira}, L.~A. and {Finethy}, T. and {Fox}, R.~T. and {Garrison}, L.~H. and {Gibbons}, S.~L.~J. and {Goldstein}, D.~A. and {Gommers}, R. and {Greco}, J.~P. and {Greenfield}, P. and {Groener}, A.~M. and {Grollier}, F. and {Hagen}, A. and {Hirst}, P. and {Homeier}, D. and {Horton}, A.~J. and {Hosseinzadeh}, G. and {Hu}, L. and {Hunkeler}, J.~S. and {Ivezi{\'c}}, {\v Z}. and {Jain}, A. and {Jenness}, T. and {Kanarek}, G. and {Kendrew}, S. and {Kern}, N.~S. and {Kerzendorf}, W.~E. and {Khvalko}, A. and {King}, J. and {Kirkby}, D. and {Kulkarni}, A.~M. and {Kumar}, A. and {Lee}, A. and {Lenz}, D. and {Littlefair}, S.~P. and {Ma}, Z. and {Macleod}, D.~M. and {Mastropietro}, M. and {McCully}, C. and {Montagnac}, S. and {Morris}, B.~M. and {Mueller}, M. and {Mumford}, S.~J. and {Muna}, D. and {Murphy}, N.~A. and {Nelson}, S. and {Nguyen}, G.~H. and {Ninan}, J.~P. and {N{\"o}the}, M. and {Ogaz}, S. and {Oh}, S. and {Parejko}, J.~K. and {Parley}, N. and {Pascual}, S. and {Patil}, R. and {Patil}, A.~A. and {Plunkett}, A.~L. and {Prochaska}, J.~X. and {Rastogi}, T. and {Reddy Janga}, V. and {Sabater}, J. and {Sakurikar}, P. and {Seifert}, M. and {Sherbert}, L.~E. and {Sherwood-Taylor}, H. and {Shih}, A.~Y. and {Sick}, J. and {Silbiger}, M.~T. and {Singanamalla}, S. and {Singer}, L.~P. and {Sladen}, P.~H. and {Sooley}, K.~A. and {Sornarajah}, S. and {Streicher}, O. and {Teuben}, P. and {Thomas}, S.~W. and {Tremblay}, G.~R. and {Turner}, J.~E.~H. and {Terr{\'o}n}, V. and {van Kerkwijk}, M.~H. and {de la Vega}, A. and {Watkins}, L.~L. and {Weaver}, B.~A. and {Whitmore}, J.~B. and {Woillez}, J. and {Zabalza}, V. and {Astropy Contributors}}, title = "{The Astropy Project: Building an Open-science Project and Status of the v2.0 Core Package}", journal = {\aj}, archivePrefix = "arXiv", eprint = {1801.02634}, primaryClass = "astro-ph.IM", year = 2018, month = sep, volume = 156, eid = {123}, pages = {123}, doi = {10.3847/1538-3881/aabc4f}, adsurl = {http://adsabs.harvard.edu/abs/2018AJ....156..123A} }

@ARTICLE{2013A&A...558A..33A, author = {{Astropy Collaboration} and {Robitaille}, T.~P. and {Tollerud}, E.~J. and {Greenfield}, P. and {Droettboom}, M. and {Bray}, E. and {Aldcroft}, T. and {Davis}, M. and {Ginsburg}, A. and {Price-Whelan}, A.~M. and {Kerzendorf}, W.~E. and {Conley}, A. and {Crighton}, N. and {Barbary}, K. and {Muna}, D. and {Ferguson}, H. and {Grollier}, F. and {Parikh}, M.~M. and {Nair}, P.~H. and {Unther}, H.~M. and {Deil}, C. and {Woillez}, J. and {Conseil}, S. and {Kramer}, R. and {Turner}, J.~E.~H. and {Singer}, L. and {Fox}, R. and {Weaver}, B.~A. and {Zabalza}, V. and {Edwards}, Z.~I. and {Azalee Bostroem}, K. and {Burke}, D.~J. and {Casey}, A.~R. and {Crawford}, S.~M. and {Dencheva}, N. and {Ely}, J. and {Jenness}, T. and {Labrie}, K. and {Lim}, P.~L. and {Pierfederici}, F. and {Pontzen}, A. and {Ptak}, A. and {Refsdal}, B. and {Servillat}, M. and {Streicher}, O.}, title = "{Astropy: A community Python package for astronomy}", journal = {\aap}, archivePrefix = "arXiv", eprint = {1307.6212}, primaryClass = "astro-ph.IM", year = 2013, month = oct, volume = 558, eid = {A33}, pages = {A33}, doi = {10.1051/0004-6361/201322068}, adsurl = {http://adsabs.harvard.edu/abs/2013A%26A...558A..33A} }

@Article{Hunter:2007, Author = {Hunter, J. D.}, Title = {Matplotlib: A 2D graphics environment}, Journal = {Computing In Science \& Engineering}, Volume = {9}, Number = {3}, Pages = {90--95}, publisher = {IEEE COMPUTER SOC}, year = 2007 }
\begin{appendix} %First appendix
\section{Convolved spectra}
\label{sec:app:convolvedSpectra}
The wavelength at which some features peak can be an indication of the primary underlying population of grains. For instance, that is the case for the 7.7~\upmicron, where variations of the peak wavelength between 7.6 and 7.8~\upmicron indicate a shift in the average grain size.

In an attempt to identify such shifts in our observations, we convolve our 1D spectra (using the {\sc astropy.convolution}'s {\sc Gaussian1DKernel} and {\sc convolve} functions) to minimise the noise and extract local maxima. 
We test several Gaussian widths to find the best value, but the impact of the convolution varies with the observations and the features considered.
Overall, the maximum shifts in the peak wavelength for a single feature and the same observation are relatively small (with a couple of exceptions) when varying $\sigma$:
$0.2-0.7\%$ for the 6.2~\upmicron, 
$0.03-0.07\%$ for the 7.7~\upmicron,
$0.04-0.2\%$ for the 8.6~\upmicron, and
$0.13-0.2\%$ for the 11.2~\upmicron features.
Aside from the $0.5-0.7\%$ difference in the 6.2~\upmicron peak wavelength for \ofive and \oeight, these numbers are small. 
The main issue is the presence of multiple peaks in some features that end up blended when decreasing the spectral resolution.
For example, as seen in Section~\ref{sec:qualitativedescriptionIntra}, the 11.2~\upmicron feature exhibits three peaks, and smoothing the spectra leads to blending them and shifting the peak wavelengths towards higher values, when $\sigma$ increases.
Similarly, if we consider real the double peaks in the 6.2~\upmicron features, a high $\sigma$ value will inevitably blend them. 
The 7.7~\upmicron feature has a high enough S/N that the standard deviations considered here do not blend the multiple peaks together.

\begin{figure*}
    \centering
    \includegraphics[width=1.0\linewidth]{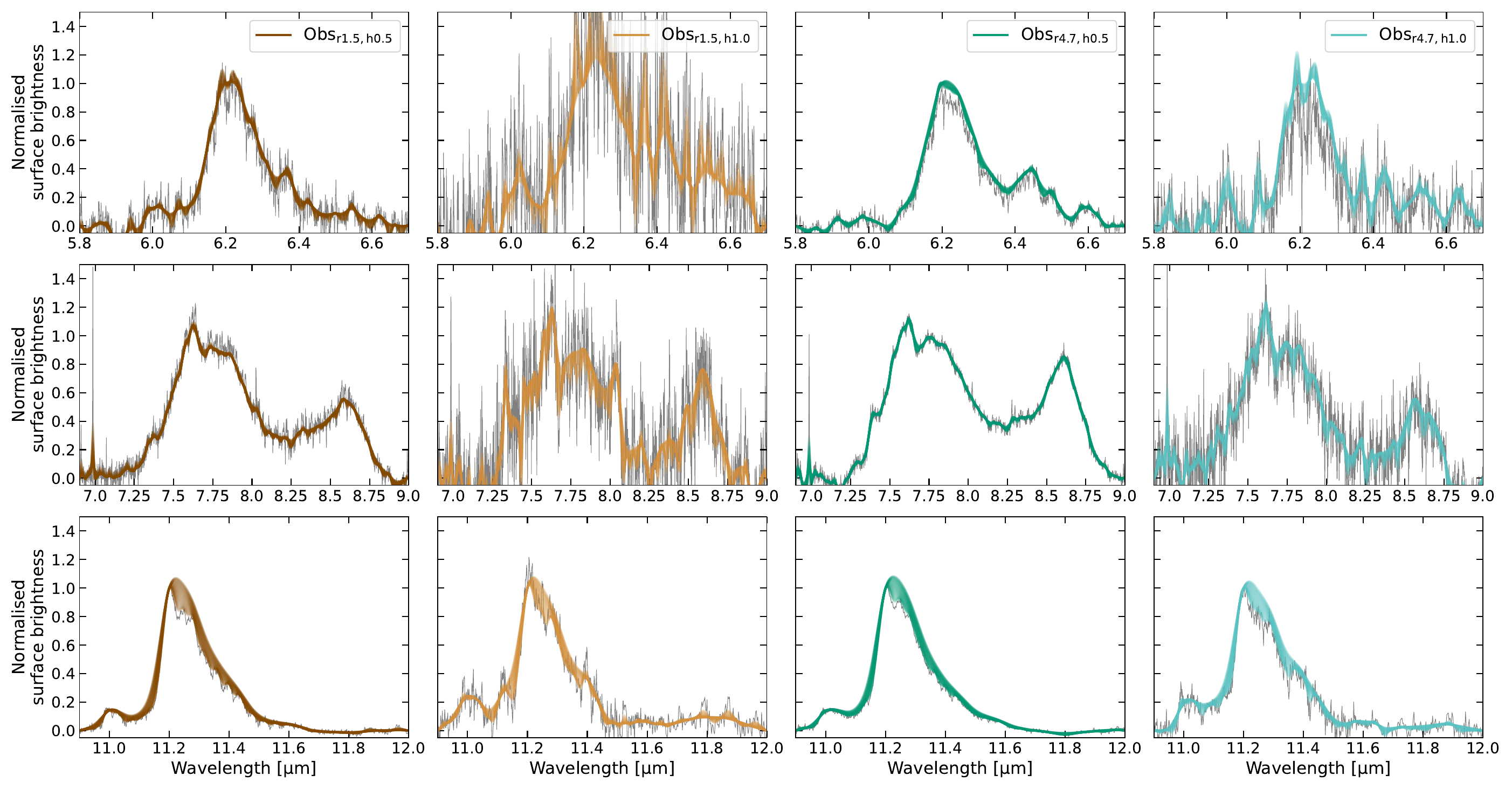}
    \caption{Convolved spectra for all observations, centred on a few PAH features ( normalised after local background subtracted). We use several Gaussian widths to show the effect of spectra degradations, namely peak shifts and multiple peaks blending. The native spectra are shown in grey.}
    \label{fig:app:convolved1Dspecs}
\end{figure*}

\section{NGC~891 and PDRs4All spectral comparison}
\label{app:compwithPDRs}
In Fig.~\ref{fig:app:wPDRs}, we show our average spectrum of \ofive and \oseven together with spectra from PDRs4All tracing the atomic, ionised, and dissociation front templates (the last one is an average of DF1, DF2, and DF3). 
In the 6.2~\upmicron feature region, all three PDR templates behave very similarly. 
In the 7.7~\upmicron region, with our normalisation, the DF templates seem closer to our data, but the atomic spectrum shows a more similar slope in the red wing of the broad 7.7~\upmicron feature. 
This is also true for the red wing of the 11.2~\upmicron features, where the \hii and DF templates display a steeper slope, as well as a more prominent 12.245~\upmicron bump compared to our data.

\begin{figure}
    \centering
    \includegraphics[width=1.0\linewidth]{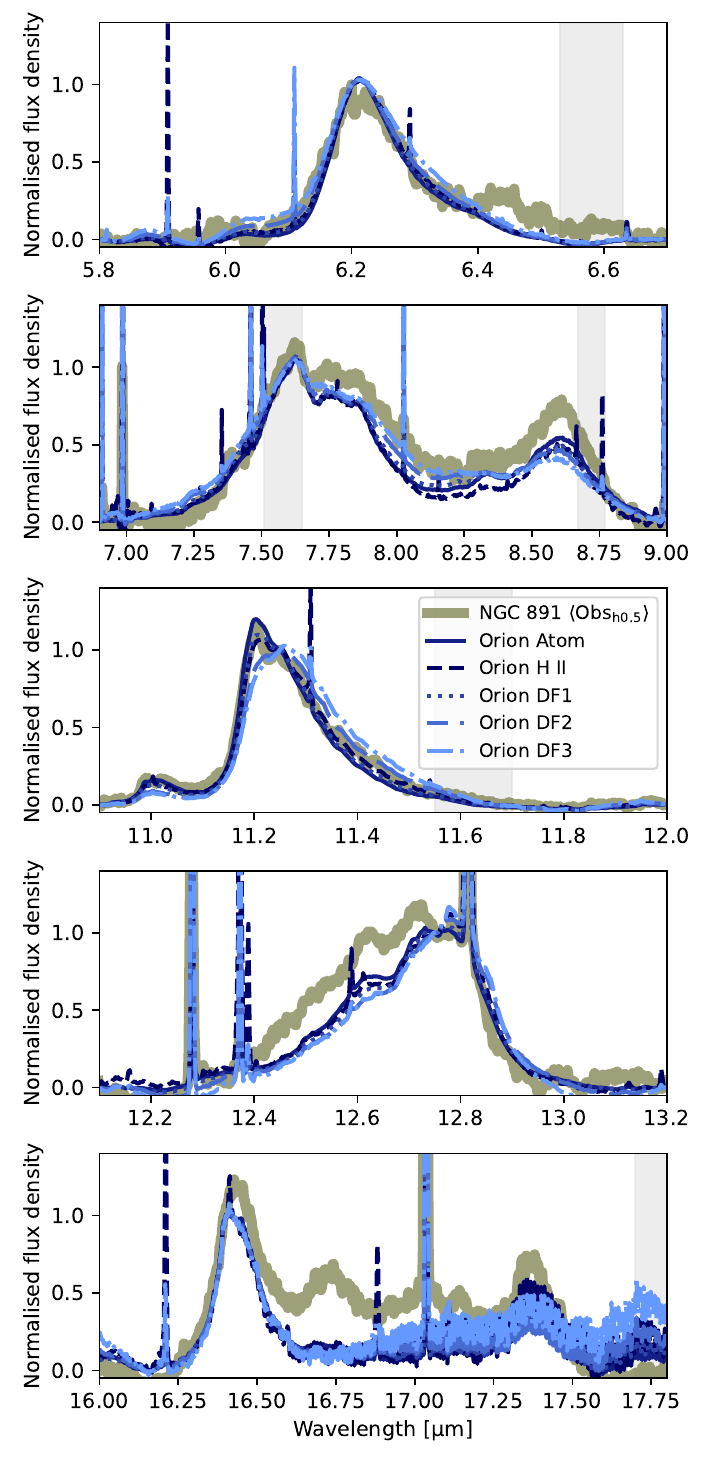}
    \caption{Average spectrum of \ofive and \oseven, and the templates of the PDRs4All observations.}
    \label{fig:app:wPDRs}
\end{figure}

\section{Final \pahfit model}
\label{sec:app:finalmodel}
The final model used in \pahfit includes:
\begin{itemize}
    \item a starlight component with $T=3000~$K;
    \item three dust continua with temperatures $=50,150,200~$K;
    \item emission lines for \htwo~$S(1)$ to $S(5)$, $[\ion{S}{iii}]$, $[\ion{Fe}{ii}]$, $[\ion{Ne}{ii}]$, $[\ion{Cl}{ii}]$, $[\ion{Ne}{ii}]$, $[\ion{Ar}{iii}]$, $[\ion{Ar}{ii}]$, $[\ion{Fe}{iii}]$, modelled as Gaussians, with central wavelength referenced in \citet[][see also \citealt{Peeters2024}]{vdPutte2024};
    \item a series of dust features listed in Table~\ref{tab:app:features}.

\end{itemize}
\begin{table}[]
    \centering
    \caption{Carbonaceous dust features in the final \pahfit model, modelled as Drude profiles. }
    \begin{tabular}{c|c}
    \textbf{$\lambda_0$ [\upmicron]} & \textbf{FWHM [\upmicron]} \\
    % \hline
    \multicolumn{2}{c}{Individual features} \\
    5.99 & 0.03 \\
    6.91 & 0.15 \\
    7.04 & 0.08 \\
    8.30 & 0.25 \\
    8.60 & 0.25 \\
    10.60 & 0.40 \\
    11.01 & 0.10 \\
    11.96 & 0.50 \\
    13.30 & 0.60 \\
    13.57 & 0.15 \\
    14.20 & 0.15 \\
    16.72 & 0.30 \\
    \multicolumn{2}{c}{6.2~\upmicron complex} \\
     6.19 & 0.05 \\
     6.26 & 0.15 \\
     6.44 & 0.13 \\
     6.58 & 0.05 \\
    \multicolumn{2}{c}{7.7~\upmicron complex} \\
     7.39 & 0.05 \\
     7.55 & 0.30 \\
     7.61 & 0.08 \\
     7.65 & 0.70\\
     7.84 & 0.40 \\
    \multicolumn{2}{c}{11.2~\upmicron complex} \\
    11.20 & 0.06 \\
    11.24 & 0.04 \\
    11.27 & 0.08 \\
    11.34 & 0.10 \\
    11.41 & 0.10 \\
    11.49 & 0.30 \\
    \multicolumn{2}{c}{12.7~\upmicron complex} \\
    12.51 & 0.53 \\
    12.61 & 0.10 \\
    12.71 & 0.160 \\
    12.80 & 0.12 \\
    \multicolumn{2}{c}{16.4~\upmicron feature} \\
    16.42 & 0.10 \\
    16.45 & 0.10 \\
    16.50 & 0.15 \\
    \multicolumn{2}{c}{17.0~\upmicron complex} \\
    17.09 & 0.40 \\
    17.37 & 0.18 \\
    17.79 & 0.30 \\
    \hline
    \end{tabular}
    \label{tab:app:features}
\end{table}

\section{Spatial variations of the 16.72~$\mu$m feature}
\label{app:16.72feature}
In Fig.~\ref{fig:app:16.72maps}, we show some of the carbon feature maps from \ofive and \oseven, with a contour of the extracted map of the 16.72~\upmicron features, at 3.6 and 6.4~W~m$^{-2}$~sr$^{-1}$, respectively. The contours could indicate a closer co-spatiality of the new feature with the 7.7 and/or 8.6~\upmicron features. Some observation-dependent can be seen, e.g. with the 17~\upmicron feature that shows little similarity in \oseven but a good spatial agreement in \ofive and a consistently high Pearson coefficient, $\rho$, with the 16.72~\upmicron feature (see Section~\ref{sec:16.72bump}).

\begin{figure}
    \centering
    \includegraphics[width=0.9\linewidth]{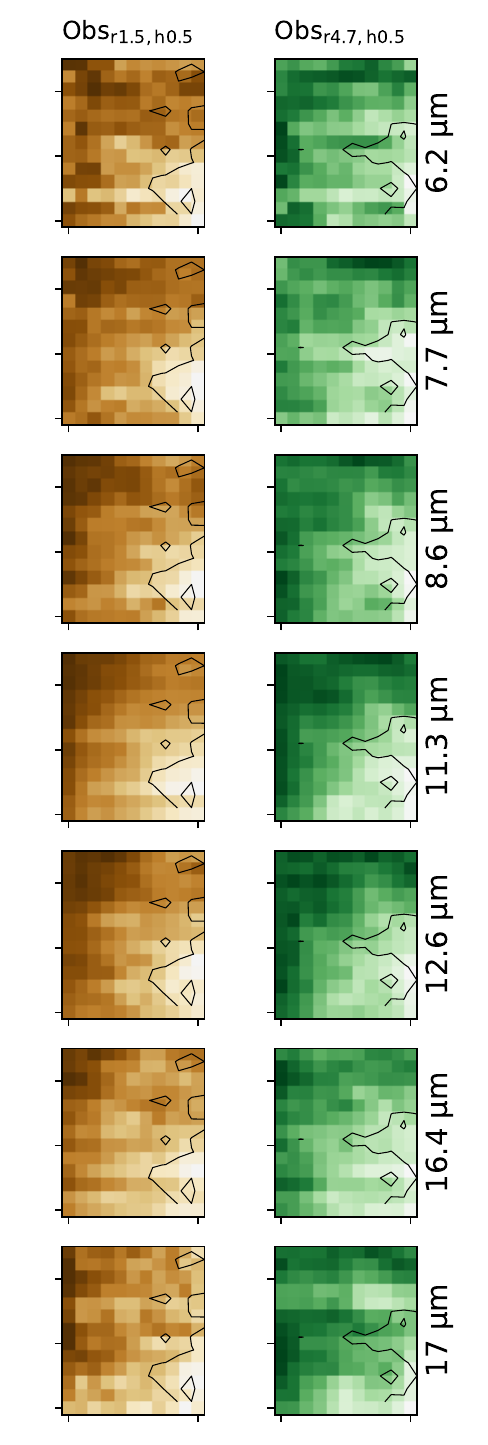}
    \caption{Total intensity in the carbon features in \ofive and \oseven extracted with \pahfit, with 3.6 and 6.4~W~m$^{-2}$~sr$^{-1}$ contours of the 16.72~\upmicron feature total intensity.}
    \label{fig:app:16.72maps}
\end{figure}

\end{appendix}
\end{document}